\documentclass[galaxies,preprints,review,accept,moreauthors,pdftex]{Definitions/mdpi}

\newcommand{\nat}{Nature}

\newcommand{\iaucirc}{IAU Circ.}

\newcommand{\fcp}{Fund. Cosmic Phys.}

\newcommand{\planss}{Planet. Space Sci.}
\newcommand{\ssr}{Space Sci. Rev.}

\newcommand{\memsai}[0]{Mem. Della Soc. Astron. Ital.}
\firstpage{1} 
\pubvolume{xx}
\issuenum{1}
\articlenumber{5}
\pubyear{2019}
\copyrightyear{2019}
\usepackage{longtable}

\Title{Interstellar dust production, destruction and effects of dust depletion in galaxies} 

\def\lesssim{\mathrel{\hbox{\rlap{\hbox{\lower4pt\hbox{$\sim$}}}\hbox{$<$}}}}
\def\gtrsim{\mathrel{\hbox{\rlap{\hbox{\lower4pt\hbox{$\sim$}}}\hbox{$>$}}}}

\Author{Francesco Calura $^{1}$\orcidA{}}

\AuthorNames{Firstname Lastname, Firstname Lastname and Firstname Lastname}

\address{%
$^{1}$ \quad INAF – Osservatorio di Astrofisica e Scienza dello Spazio di Bologna, via Gobetti 93/3, 40129 Bologna, Italy}

\corres{Correspondence: francesco.calura@inaf.it.}

\abstract{Despite its small mass fraction typically observed in the interstellar medium, dust plays a significant role as a key component of galaxies,
  affecting a wide range of properties. 
  This review focuses specifically on how dust grains influence interstellar chemical abundances and on the processes that regulate the evolution of the galactic dust budget. 
  I describe the main physical processes regulating dust evolution, including production by stars and other sources,  
  destruction in supernova shocks and interstellar growth and how they are included in galactic chemical evolution models
  and simulations.  
  I discuss the main effects of interstellar dust on the abundances measured in various high-redshift systems that include
  Damped Lyman $\alpha$ absorbers detected along the lines of sight of distant quasars and in the absorption spectra of Gamma Ray Burst
  afterglows.  
  I discuss the measure of the dust mass in galaxies and review its global budget, evaluated through the study of the evolution of the comoving dust mass density,
  for which I present an up-to-date compilation of data chosen from the literature. Interstellar dust growth plays a critical role in regulating the dust budget,
  for which I present a list of evidence both in favour of it and against.  The dust budget at high redshift is one aspect that requires attention to drive significant
  progress in the future, along with the investigation of the properties of dust in local, low-metallicity systems.
  Our poor theoretical knowledge of basic aspects related to dust evolution evidences the need for a new high-sensitivity space telescope operating
  in the far-infrared regime, still awaited by the community since the demise of Herschel.}


\begin{document}
\tableofcontents

\section{Introduction} 
The microscopic, solid particles commonly referred to as dust grains  
represent a fundamental component of the neutral and molecular interstellar matter of galaxies.    
Although interstellar dust 
does not dominate the cold mass budget, with very low typical dust-to-gas ratios $\lesssim$ 1\%  in local star-forming galaxies \cite{issa1990,devis2019}, 
dust has a dominant role 
in shaping some of the most fundamental observed galactic properties. 
First of all, the presence of dust strongly influences the spectral properties of galaxies and
of its sub-components, in both the form of stellar populations and active galactic nuclei. This
occurs on an extremely wide range of wavelengths, virtually on the entire electromagnetic spectrum. 
It is well known since a few decades that 
at optical and UV rest-frame wavelengths, the light emitted by astrophysical sources is scattered and
absorbed by dust grains, to be re-radiated at much larger wavelengths, i.e., in the submillimetric band.
The knowledge of the effects of dust on the properties of galaxies from the UV band to the far-infrared (and beyond)
dates back to the early 20th century, when it was invoked as one of the possible explanations of Olbers’ Paradox,
questioning how could the night sky be so dark if stars in the Universe were uniformly distributed.
In the most easily accessible band, i.e., the optical one, the recognised `official' discovery
of the effects of dust extinction occurred in the `30s of the 19th century and originated from the fact that 
more distant clusters appeared not only fainter, but also redder than expected and attributed
to a distribution of `{\it}{Fine cosmic dust particles of various sizes}' \cite{trumpler1930}. 
It is estimated that up to 30\% of the light emitted in the Universe has been reprocessed by dust grains \cite{bernstein2002,bianchi2018}.
However, dust influences galaxy spectra even beyond the UV-FIR interval, as the grains are known to become
nearly (but not completely) transparent at X-ray energies \cite{draine2004} 
and continuum emission from rotating grains occur in the radio band \cite{ferrara1995}.  
In this context, it is probably easier to identify the portion of the electromagnetic spectrum in which
dust grains have no influence at all, i.e., practically only at the two opposite ends of the electromagnetic spectrum,
in the spectral range of the $\gamma$-rays and of $>$meter-sized radio waves. 

Dust does not influence only the spectral properties of galaxies, but 
it also affects heavily the chemical composition of the interstellar medium. 
It is known to play a key role also on planet formation,  
as dust compounds made of metals and silicates build the bulk of the mass of terrestrial planets and asteroids,
and likely a significant fraction of giant planets~\cite{onyett2023}. 
The coagulation of heavy elements into dust plays a fundamental role during the early stages of star formation (SF) in molecular clouds (MCs),
shielding newly formed molecules against the  destructive effects of UV photons emitted by young stars, with their surfaces representing
a suited environment for new reactions \cite{tielens2022}.

In the coldest components of the interstellar medium several chemical elements, called refractory, 
can be easily captured by dust grains. 
In this process, refractory species are subtracted from the gas phase to become part of a solid phase,
thus hiding them from direct detectability in observational abundance studies. 
In this context, a common way to refer to this process is by stating that refractory elements are {\it locked} 
into dust grain, in the sense that they literally get abducted from their free circulation in the interstellar medium (ISM).
This is often called `depletion' of a chemical element, in which in the gas
it is observed as underabundant compared to a standard reference value \cite{whittet2010}, at a level  that 
is known to depend on the phase of the observed medium. 

The basic concept of the ``dust formation window'' serves as the foundation for any general dust formation scenario, in which, 
when a chemically enriched medium has a low enough temperature and a high enough density, 
effective dust condensation occurs \cite{elvis2002}. The time interval in which such conditions are favourable regulates
the amount of dust produced, the chemical composition, and the size of the grains. 
The current astrophysical knowledge of dust grains rests upon both observations and laboratory experiments that
attempt to recreate realistic physical conditions for the grain formation and their interactions with other chemical species \cite{nuth1982,rotundi2002}. 
In such current knowledge, dust can be produced in a variety of different environments, 
from the cold, expanding slow envelopes of post-asymptotic giant branch stars \cite{ferrarotti2006,ventura2020} and
rapidly cooling supernova (SN) ejecta \cite{temim2017},
to the coldest regions of the interstellar medium, where dust grains act as dominant cooling agents \cite{krumholz2010}. 
At the same time, dust grains can be destroyed in different ways, as violent as SN shocks,
where grains are shattered as they get accelerated to extreme velocities and eroded as the ISM is heated to 10$^6$--10$^7$ K,
and during the star formation itself, as refractory species disappear as they are incorporated in newly born stars, to end naturally what is often called the
`Life Cycle' of dust.  

This review is mainly focused on the very processes regulating the production and evolution of 
interstellar dust, with particular attention to models and observations of galaxies at high redshift.
The main aim here is to focus on some aspects which, in my view, are in particular need of comprehensive discussions.
These include how dust shapes the chemical abundances observed in various types of high redshift objects, i.e.,
Damped Lyman $\alpha$ systems, Gamma-ray burst host galaxies and Lyman Break galaxies. 
For different reasons, all these kinds of objects offer unique views on the chemical abundance pattern of high-redshift
systems. To correctly interpret the  measured abundances, one needs to account for the unknown amount of refractory elements
into solid grains. This task can be performed in different ways, mostly either by correcting the measured abundances via empirical
relations or  by interpreting the models with chemical evolution models. 
Therefore, a few open problems related to the formation and evolution of dust in high-redshift galaxies are presented and discussed.  
In parallel, I also describe the most basic ingredients of a chemical evolution model that can account for
the evolution of a galactic dust component. 
My hope is to provide the essential elements to build such a model, or to facilitate the implementation of
the described processes in more advanced ones, e.g., including a description of the dynamics of interstellar gas or in a cosmological
framework. 
The plan of the present review is as follows.
In Section \ref{sec_chem}, I provide a description of galactic chemical models and how they can incorporate 
the main processes regulating \linebreak dust evolution.

In Section \ref{sec_howdust}, I discuss how dust shapes the elemental abundances of high redshift galaxies. 
In particular, I show how 
the study of the abundances measured in Damped Lyman $\alpha$ (DLA) systems detected in quasar spectra 
advanced our knowledge of dust depletion at high redshift. This discussion is extended also to other samples,
such as Gamma-Ray Burst-DLAs and some distant starburst galaxies. 

In Section \ref{sec_ontheredshift}, I discuss a few relevant aspects concerning our knowledge of dust evolution as a function of redshift.   
I summarise how the mass of dust is commonly measured in galaxies and 
I focus on a few attempts to model how dust emerged in early galaxies.
I discuss the role of one fundamental process in particular, dust growth, in regulating the evolution of dust, and I 
conclude the section with an overview of the current knowledge of the cosmic dust budget evolution, through the study
of the comoving mass density. For this quantity, I also provide an up-to-date extensive compilation \linebreak of measurements. 

In Section \ref{sec_future}, I present a discussion of some outstanding problems and how they can be addressed by upcoming
instruments to advance our knowledge of interstellar \linebreak dust evolution.

Finally, in Section \ref{sec_summary}, I present a summary and draw some conclusions.

\section{Chemical Evolution Modelling of Interstellar Dust}
\label{sec_chem}
In this section, I will present the basic tools which are needed to model the chemical
evolution of interstellar dust, namely the set of differential equations that one needs to
solve to evaluate the behaviour of dust in the ISM. 
For a set of chemical elements of interest, it is often important to know the mass fraction of each species 
that is subtracted from the gas and included in the solid phase, i.e., in the
form of grains as a function of time and other fundamental quantities.
First, I will briefly describe the elements most commonly found in the dust phase and which will be particularly recurrent in this review.
To the purpose of this review, it is convenient to familiarise with a standard, simplified set of chemical evolution equations,
which is of general use in studies of the evolution of the interstellar composition in various environments. 
Finally, I will present a more detailed chemical evolution equation specifically aimed at studying dust evolution.

\subsection{Some Elements Like to Be in Dust, Some Others Don't}
Hydrogen, the most abundant element in the Universe, is only indirectly incorporated into interstellar dust,
playing a key role in surface chemistry and in the formation of molecular and icy mantles on dust grains.   
On the other hand, a large fraction of all the chemical elements heavier than H and He, commonly called {\it metals},
have been found as constituents of solid compounds in meteorites and chondrites (e.g., \cite{vankooten2022}). 
The capacity of an element to migrate from gas to solid phase is associated to
the concept of volatility. The volatility of a chemical element is useful to quantify its tendency to condense into dust grains,
and is commonly expressed by the condensation temperature. 
In geochemistry and cosmochemistry, this quantity is commonly defined as the 
temperature at which a specific element would condense from a gas of solar composition \cite{lodders2003,wood2019}. 

Volatile elements are defined as those that condense at relatively low temperature
(e.g., <1000 K) from a gas with solar composition whereas refractory,
or sometimes called involatile elements, tend to condense into solid grains at higher temperatures. 
%
However, there are a few elements presenting low $T_C$ values, which are frequently found in
solid compounds. The most common examples are the two most abundant elements after H and He, namely
C and O.  
As for C, in solar composition matter it is known to condense at low $T_C$ values, however it is commonly
regarded as a refractory elements, based on the empirical fact that 
Carbon solids are ubiquitous in the interstellar space \cite{ehrenfreund2010}.
However, the formation pathway of carbonaceous matter in astrophysical environments,
as well as in terrestrial gas-phase condensation reactions, is not yet understood. 

Together with C and O, the set of refractory elements most frequently discussed in astrophysical applications
include Mg, Si, Fe and together contribute to most of the dust mass in the ISM \cite{whittet2010}.
Many of the observed properties of interstellar dust are normally reproduced with models that include
silicates (inorganic dust particles having on [SiO$_4$]$^{4-}$ as building block, 
that can have a largely varying composition \cite{rimola2021}) 
and graphite or amorphous carbon dust (examples are CH$_4$ ices \cite{lodders2003}).
In Table \ref{table_tcond} I summarise the most frequent compounds in dust, along with their 
  equilibrium condensation temperature values for a Solar-System composition gas \cite{lodders2003} and
  the percentage estimates of their depletion fractions for the main refractory interstellar elements.
  The condensation temperatures reported in Table \ref{table_tcond} are defined as the temperatures at which 50\%
  of a given element is incorporated into solid phases under conditions of chemical equilibrium for a gas of solar composition at a total pressure of
  $10^{-4}$ bar. 
  
Compared to elements like Silicon or Iron, the fraction of the total cosmic hydrogen (H) reservoir contained in dust grains is extremely small.
However, H is a major chemical constituent within certain types of grains, particularly carbonaceous dust (such as
hydrogenated amorphous Carbon or polycyclic aromatic hydrocarbons) and ice mantles. 
For an order of magnitude estimate of the H fraction in carbonaceous grains,
in a solar composition medium, for every $10^6$ H atoms in the gas phase, there are about 250--300 atoms of Carbon (C) available in total,
half of which is locked in dust grains \cite{savage1996,jenkins2009} 
For an estimated typical ratio of H to C atoms H/C in solid amorphous carbon of $\sim$0.5 \cite{furton1999},
we will thus have 150 H atoms in this grain type, supporting a depletion fraction in number of 0.015 \%.
In dense, cold molecular clouds, hydrogen is primarily found also in the form of molecular ices that coat refractory silicate and carbon cores.
The dominant ice component is water ice (H$_2$O), with methane (CH$_4$), ammonia (NH$_3$) and 
methanol ($\mathrm{CH_3OH}$) as other frequent species. 
An estimate of the H depletion fraction in ices can be computed assuming
that the O abundance into ices in the highest density ISM is ~400 ppm \cite{whittet2010} and in the same medium, 60--70 \% of the ice is
in water \cite{jimenezescobar2024}; considering that every water molecule contains 1 O atom and 2 H atoms, 
the H depletion fraction is $D_H \sim 0.05\%$. 
H remains overwhelmingly in the gas phase, with only a minute fraction ($\sim$0.02--0.05\%),
incorporated into ices in grain mantles. 
This small percentage is nonetheless sufficient to drive critical surface processes, such as the catalysis of molecular hydrogen.
Besides the obvious importance of water for any living organism, the tiny fraction of H in H$_2$O ice 
plays a fundamental role in planet formation. 
In fact, beyond a certain radial distance from the central star (the so-called ``snow-'' or ``ice line''), the condensation of water ice leads to a significant
increase in the surface density of solid material in protoplanetary discs, facilitating rapid core growth and enhancing the likelihood of gas giant planet formation\cite{Kennedy2008}. 
This extra solid mass is what enables the formation of massive solid cores 
(roughly 10 $M_{\oplus}$, with\linebreak  1 $M_{\oplus}=5.97 \times 10^{27}$ g is the Earth mass) fast enough to capture gas from the disk before it dissipates \cite{Pollack1996}. 
Therefore, without the low percentage of Hydrogen locked in ice, gas giants like Jupiter likely would not exist.
Moreover, ice-covered grains are stickier than bare silicate grains, significantly increasing the collision and coagulation efficiency
 during the early stages of planet formation \cite{Gundlach2015}. 
 
Among elements that are usually present in grains, O is regarded as refractory despite its low condensation temperature, not too far from
the one of volatile elements like N, with T$_{cond}$ = 131 \cite{lodders2003}. 
Based on a collection of both indirect and direct evidence \cite{calura2009,jenkins2009,kimura2015}, 
the refractory nature of S is controversial. 
S is characterised by a condensation temperature higher than
the ones of other elements generally depleted, such as C and O, and for the fact that in 
cometary dust particles, S is present in iron sulfide (FeS) grains 
\cite{lodders2003,calura2009,jenkins2009} and, moreover, is known to be abundant in protoplanetary discs \cite{Kama2019}. 
Based on these results, if anything, S is expected to be depleted mostly in dense molecular gas \cite{Laas2019}. 
However, for reasons that will become clear later on, unless otherwise stated, in all the cases discussed here S will be regarded as a volatile element. 

Finally, the \% depletion fractions (last column of Table \ref{table_tcond}) are estimated mostly from ultraviolet 
absorption-line data in cool interstellar clouds toward $\zeta$ Oph \cite{savage1996} and from atomic column densities measured
  in 243 Galactic lines of sight \cite{jenkins2009}. 
Observations of diffuse interstellar clouds show that gas-phase depletions span a wide range, from mild values (40--60\%) 
for volatile elements such as C, O, to factors $>99\%$ for heavily refractory elements like Fe, Ti, and Ca.
This depletion pattern correlates strongly with condensation temperature, with higher-T$_{cond}$
elements being more heavily depleted \cite{savage1996}, reflecting their preferential incorporation into dust grains, although the exact pattern may be modified by grain growth and destruction processes in the ISM.

\begin{table}[H]

\caption{ List of the elements that are known to reside in interstellar dust grains, with estimated condensation temperatures and depletion fractions. }
\label{table_tcond}
\begin{adjustwidth}{-\extralength}{0cm}

\begin{tabularx}{\fulllength}{m{1cm}<{\raggedright}m{4cm}<{\centering}m{7cm}<{\centering}m{2cm}<{\raggedright}m{2cm}<{\raggedright}}
\toprule
\textbf{Element	} & \textbf{Atomic Number} &	\textbf{Main Dust Reservoir(s)} &	        \textbf{T$_{\textbf{\emph{cond}}}^ \textbf{1}$ }            & \textbf{Depletion} \\
         &               &                             &                \textbf{(K)  }                   &  \textbf{(\%)    } \\
\midrule

H   &	 1   &  	Organic mantles, ices	                        &   182         & 0.02--0.05    \\
C   &    6   &          Carbonaceous grains (amorphous C, PAHs)	        &   78          & $\sim$39  $^2$--61 $^3$       \\ 
O   &    8   &          Silicates (e.g., Mg–Fe–silicates), oxides, ices	&   182         & $\sim$42 $^2$--60 $^3$         \\
Mg  &    12  &          Silicates (olivine/pyroxene)	                &   1397        &  95 $^2$--97 $^ 3$         \\
Si  &    14  &          Silicates	                                &   1529        &  95 $^2$--96 $^3$       \\
Fe  &    26  &          Metallic Fe, Fe-bearing silicates/oxides	&   1357        &  99 $^{2,3}$         \\
Al  &    13  &          Oxides (e.g., Al$_2$O$_3$, Hibonite)	        &   1653        &  90--99 $^4$         \\ 
Ca  &    20  &          Carbonates, silicates	                        &   1659        &  >99 $^{3}$        \\
Ti  &    22  &          Ti-oxides, silicate trace	                &   1593        &  >99 $^{2,3}$        \\  
V   &	 23  &          Oxides, pyroxene                                &   1429        &  >99 $^{3}$      \\  
Cr  &    24  &          Fe alloy	                                &   1296        &  >99 $^{2,3}$         \\ 
Mn  &    25  &          Forsterite + enstatite	                        &   1158        &  96 $^{3}$--98 $ ^{2}$        \\ 
Co  &    27  &          Metallic Co inclusions	                        &   1352        &  >99 $^{3}$    \\
Ni  &    28  &          Metallic Ni                     	        &   1353        &  >99 $^{2,3}$   \\  

\bottomrule
\end{tabularx} 
\end{adjustwidth}
\end{table}

\begin{table}[H]\ContinuedFloat
\caption{\textit{Cont.}}

\begin{adjustwidth}{-\extralength}{0cm}

\begin{tabularx}{\fulllength}{m{1cm}<{\raggedright}m{4cm}<{\centering}m{7cm}<{\centering}m{2cm}<{\raggedright}m{2cm}<{\raggedright}}
\toprule
\textbf{Element	} & \textbf{Atomic Number} &	\textbf{Main Dust Reservoir(s)} &	        \textbf{T $_{\textbf{\emph{cond}}}^ \textbf{1}$ }             & \textbf{Depletion} \\
         &               &                             &                \textbf{(K)  }                   &  \textbf{(\%)    } \\
\midrule

Cu  &    29  &          Sulfides, metal inclusions	                &   1037        &  95 $^2$--96 $^3$  \\ 

P   &    15  &          Phosphides         	                        &   1248        &  68 $^3$--78 $^2$        \\ 
S   &    16  &          Sulfides (e.g., FeS), organosulfur compounds	&   704        &   -      \\
\bottomrule

\end{tabularx}
\end{adjustwidth}
\noindent{\footnotesize{$^1$ Data from Lodders \cite{lodders2003}; $^2$ From Jenkins \cite{jenkins2009}. We compute the depletion of an element X as (1$-$10$^{\rm [X_{\rm gas}/H]}$), with the interstellar ‘reduction’ of an element relative to the solar value due to dust [X$_{\rm gas}$/H] = log(X/H)$-$log(X/H)$_{\odot}$ calculated by means the fitting formula [X$_{\rm gas}$/H] = B$_{\rm X}$ + A$_{\rm X}$ (F$_*$$-$z$_{\rm X}$), considering the ‘heavy depletion’ case of  F$_* = 1$, corresponding to the densest ISM regions;  $^3$ From Savage \& Sembach \cite{savage1996}; $^4$ From Das \cite{Das2025}. }}

\end{table}

\subsection{General Chemical Evolution Equations in a Single-Phase Gas}
The basic assumptions which need to be set in order to solve a chemical
evolution equation are that (i) the modelled system is one-zone, i.e., it is an {\it open box}
and it does not contain any sub-regions, such as separate cells; 
(ii) that it is composed of single-phase gas, e.g., either neutral, or molecular as 
the star-forming gas typically is. A few attempts model the galactic ISM as not homogeneous and are aimed at
capturing its multiphase structure, dividing it into distinct components-typically cold, warm and hot phases
\cite{Ferrini1994,Molla2015,millanirigoyen2020}, 
with mass interchange between them  through photoionisation of atoms, recombination of electrons with ions and conversion of atomic
hydrogen into molecular hydrogen. However, they typically lack self-consistent gas dynamics, which are instead followed in full hydrodynamical simulations (See Section \ref{sec:hydro_dust}). (iii) at any time the system is instantaneously mixed, i.e., the typical timescale for
the diffusion and mixing of the heavy elements in the gas is negligible and the system presents always a homogeneous composition
and (iv) initially, the system is metal-free. 
The formalism is rather versatile and is aimed at describing the evolution of several different systems,
where with {\it system} we may refer to an entire galaxy (e.g., \cite{pipino2011}), a sub-galactic region, such as the solar neighbourhood
(e.g., \cite{chiappini2001})
or the galactic bulge \cite{ballero2008}  
or even an extended region of the Universe which may contain several galaxies surrounded by a diffuse gas distribution,
such as the intergalactic or intracluster medium \cite{pei1995}. 

The evolution of the interstellar fractional mass in the form of the chemical element $i$ can be computed
by solving the equation  

\begin{equation}
\label{eq_chem}
\dot{G_{i}}=-\psi(t)X_{i}(t) + R_{i}(t) + (\dot{G_{i}})_{inf} -(\dot{G_{i}})_{out}
\end{equation}
where $G_{i}(t)=M_{gas}(t)X_{i}(t)/M_{tot}$ is the gas mass in 
the form of an element $i$ normalized to the total mass of the system $M_{tot}$, whereas the quantity
$G(t)= M_{gas}(t)/M_{tot}$ is the fractional mass of gas present in the system at the time $t$. 

The quantity $X_{i}(t)=G_{i}(t)/G(t)$ represents the 
abundance by mass of an element $i$, with
the summation over all elements in the gas mixture being equal to unity. 
The quantity $\psi(t)$ is the star formation rate (SFR), i.e., the amount of gas turning into stars per unit time.

The SFR (expressed in $M_{\odot}$/yr) is calculated as:

\vspace{-6pt}
\begin{equation} 
\psi(t) \propto M_{gas}(t), 
\end{equation}
and it is assumed to be proportional to the gas mass, according to the Schmidt \cite{schmidt1959} law.
The proportionality constant is the star formation efficiency (SFE), largely unknown. In various models, different assumption can
be made regarding this constant. Moreover, in chemical evolution models this quantity may vary depending on the system that
one aims to model. 

In another, more recent formulation of the Schmidt \cite{schmidt1959} law, 
when dealing with surface densities such as the star formation rate per unit area $\Sigma_{SFR}$ and the gas surface density $\Sigma_{gas}$,
the SFR is calculated as 
\begin{equation} 
\Sigma_{SFR}  \propto \Sigma_{gas}^n,  
\end{equation}  
where $n=1.4$ \cite{kennicutt1998}.
This assumption relies upon the empirical results of Kennicutt \cite{kennicutt1998}, who studied the relation between
the SFR and gas density in a sizeable sample of local star-forming galaxies, including mostly spirals and dwarf galaxies, and
finding a tight correlation between these two quantities.

\textls[-15]{The quantity $R_{i}(t)$ is also known as the {\it Returned fraction}, namely the fractional amount of matter in the 
form of an element $i$ that the stars eject into the ISM through various evolutionary processes, such as, e.g.,  
stellar winds and SN explosions. 
In a complete treatment of chemical evolution, this term needs to include the contribution from (a) single low- and intermediate-mass stars, 
typically characterised by initial masses $m<8 M_{\odot}$, 
(b) type II SNe, originating from the explosion of massive stars with initial mass of 
$m>8 M_{\odot}$, and (c) from type Ia SNe, whose progenitors are still a matter of debate but which 
are generally assumed to originate in binary systems containing at least one degenerate star, specifically a white dwarf. }

In Equation~(\ref{eq_chem}), the two terms 
$(\dot{G_{i}})_{inf} = X_{i,inf} I(t) $ and  $(\dot{G_{i}})_{out} =  X_{i,out} O(t)$ account for 
the infall of external gas and for galactic winds, respectively.
The infall and outflow rates are the quantities $I (t)$ and $O (t)$, respectively,
whereas the mass fractions of the element $i$ in these components are $X_{i,inf}$ and $X_{i,out}$. 

Various assumptions can be made regarding the infall term.
It is often expressed by analytic forms, of which the most typical and frequently
used example is perhaps the exponential law \cite{matteucci2021} 
\begin{equation}
(\dot{G_{i}})_{inf}=X_{i,inf} A \exp{-t/\tau_{inf}},  
\end{equation}
where the e-folding time $\tau_{inf}$ can vary between different models. 
The constant $A$ is constrained by means of the total gas mass ($M_\mathrm{inf}$) through the following equation:  
\begin{equation}
\int_0^{t_\mathrm{G}} A~I(t)  dt=M_\mathrm{inf}, 
\label{inf}
\end{equation}
\noindent
which fixes the integral of the infall rate
over the entire lifetime $t_\mathrm{G}$ of the modelled system to a total infall mass $M_\mathrm{inf}$, which needs to be assumed a priori.

Various assumptions can be made also as far as the chemical composition of infalling gas is concerned. 
In chemical evolution studies of the Milky Way and its sub-regions, it is often convenient to assume that the infalling gas is of primordial
composition, i.e., that it is characterised by an H mass fraction $X \sim 0.75$, a He fraction $Y \sim 0.25$ \cite{valerdi2019} 
and it may present also some small traces of Li produced during primordial nucleosynthesis \cite{pinsonneault2002}.

In other cases, some authors have used chemical evolution models to study the effects of merging or interacting galaxies \cite{pipino2006}.
In such cases, the composition of the infalling gas is clearly different than primordial, as it is more realistic to
assume an infalling metal-rich gas accreted from a companion galaxy. 

As for the outflow or wind term $(\dot{G_{i}})_{out}$, one frequent assumption is
\begin{equation}
(\dot{G_{i}})_{out}= X_{i,out} \lambda \psi, 
\end{equation}
i.e., that the outflow rate is proportional to the
SFR, where the constant $\lambda$ is referred to as the wind parameter or mass loading factor \cite{spitoni2020}. 
Regarding the chemical composition of the outflowing gas, one can assume that it is the same as the ISM, in that case $X_{i,out} = X_{i}$. 
Other authors have assumed that the outflowing gas has different composition than the ISM.
One typical case of differential outflows is the assumption that in galactic outflows, heavy elements are lost more efficiently than H and He (e.g., \cite{recchi2008}).

The assumption of metal-enhanced outflows is supported by results from hydrodynamic simulations, showing that the fast, hot and 
metal-rich material ejected by massive stars is expelled from galaxies with particular efficiency, in some cases leaving the cold, diffuse
ISM relatively unperturbed \cite{dercole1999,romano2019}. This clearly depends on how much the sources of energetic and chemical {\it feedback}, namely
the OB stellar associations, permeate the ISM and are distributed across the galaxy. Previous works aimed at the study of metal ejection as a function of galactic mass concluded 
that metal ejection is much more efficient in lower mass galaxies \cite{maclow1999}. 

It is also important to add that, in order to have a closed system of equations,
\mbox{Equation (\ref{eq_chem})} needs to be complemented by two other equations 
for the conservation of the total mass $M_\mathrm{tot}$ and of the gas mass $M_\mathrm{gas}$, typically of the form

\begin{equation}
\begin{cases}
 \dot{M}_\mathrm{tot}(t) =   -\lambda \psi(t) + I(t) & \\[7pt] 
 \dot{M}_\mathrm{gas}(t) =  -\big(1 + \lambda \big) \psi(t) + R(t) + I(t)  
\end{cases} 
\label{system1}
\end{equation}
where $R(t)$ is the total ejection rate from stars of all masses and ages \cite{tinsley1980,spitoni2017}. 

In particular cases and with a set of particular assumptions, 
the above set of equations can be solved analytically and find expressions for the evolution of all the main physical quantities. 
These assumptions include the assumption that the stellar initial mass function, regulating the mass fraction in stars of various mass bins and
discussed later, is constant in time and space. 
Another fundamental assumption for an analytic solution is the instantaneous recycling approximation,
in which stars with mass above a certain value (that depends on the timescale of interest) die instantaneously after their formation,
whereas the others have infinite lifetimes \cite{matteucci2021}.
A detailed discussion of the derivation of analytic solutions for the system of Equations (\ref{system1}) is beyond the scope of the present paper \cite{tinsley1980,spitoni2017}.

\subsection{The Stellar Initial Mass Function}
\label{sec_imf}
One parameter which plays a very important role in chemical evolution models is the stellar IMF $\phi(m)$, a function 
describing the number of stars per unit mass interval. 
Traditionally, the most common and simplest assumption in chemical evolution models is a power-law of the Salpeter (1955) \cite{salpeter1955} form:
\begin{equation}
\phi_{S55}(m) \propto m^{-(1+\alpha)}, \quad \text{with } \alpha = 1.35
\end{equation}

for a mass range \( m_{\text{min}} \leq m \leq m_{\text{max}} \), typically in units of solar masses \( M_\odot \).
In chemical evolution models, the IMF $\phi$ is often normalized such that the integral over all stellar masses gives a fixed value, in general equal to 1. 
To normalize the IMF so that the total stellar mass integrates to 1, let us define:

\begin{equation}
A_{S55} = \frac{0.35}{m_{\text{min}}^{-0.35} - m_{\text{max}}^{-0.35}}
\end{equation}

This ensures:

\begin{equation}
 \int_{m_{\text{min}}}^{m_{\text{max}}} A_{S55} m\, \phi_{S55}(m) \, dm = 1.
 \label{eq_norm}
\end{equation}

The most widely used and observationally-grounded IMF is the piecewise-defined power-law of Kroupa (2001) \cite{kroupa2001}:
\begin{equation}
\phi_{K01}(m) \propto \begin{cases}
m^{-0.3} & \text{for } 0.01 \leq m/M_\odot < 0.08 \\
m^{-1.3} & \text{for } 0.08 \leq m/M_\odot < 0.5 \\
m^{-2.3} & \text{for } 0.5 \leq m/M_\odot \lesssim 100
\label{eq_imf_k01}
\end{cases}
\end{equation}

\noindent also known as the multi-component canonical IMF. 
In this case, the normalisation requires the calculation of three constants.
To have a closed system of equations, such constants are evaluated by imposing again the same as in Equation~(\ref{eq_norm}), 
plus two additional ones of 
the continuity of the IMF in the two intermediate points at $m=0.08$ M$_{\odot}$ and  at $m=0.5$ M$_{\odot}$. 

Another observationally-compliant IMF alternative to Kroupa \cite{kroupa2001} IMF is the \linebreak Chabrier \cite{chabrier2003}, obtained from a combination of a log-normal distribution
for low-mass stars and a power-law distribution for high-mass stars. 
It is defined as:

\[
\phi_{C03}(m) \propto \begin{cases}
\frac{1}{m} \, \exp\left[ -\frac{(\log_{10}(m) - \log_{10}(0.079\, M_{\odot}))^2}{2 \times 0.69^2} \right] & \text{for } m \leq 1 \, M_\odot \\
m^{-2.3} & \text{for } m > 1 \, M_\odot
\end{cases}
\]

At present, a complete theory to explain the origin 
of the stellar IMF and its relation with star formation is lacking. Another important unanswered question regarding the IMF
concerns its universality, i.e., whether its shape depends on particular properties of the ISM which cause
a variation as a function of galactic type or redshift \cite{larson1998,narayanan2013}. 
In this context, of significant value is the integrated galactic initial mass function (IGIMF) formalism initially developed by 
Kroupa \& Weidner \cite{kroupa2003}. 
This formalism is based on some relevant empirical properties of star formation in various local environments, which include 
(i) the clustering as primary mode of star formation, in particular the fact that stars are known to form in the dense cores of molecular clouds
in groups of at least a few stars \cite{lada2003,megeath2016};  in young star clusters, the IMF is observed to be universal and
well approximated by a canonical, multiple power-law form  of Equation~(\ref{eq_imf_k01}); 
(iii) young stellar clusters appear to follow 
a single-slope power law mass distribution \cite{lada2003} and (iv) within young clusters, the upper mass end of the IMF is known
to depend on the SFR of the host galaxy \cite{weidner2004}. 
The IGIMF is computed by weighting the
canonical IMF of Equation~(\ref{eq_imf_k01}) with the {\it embedded} young stellar cluster mass function (ECMF) \cite{kroupa2003}.
The IGIMF $\xi_{\rm IGIMF}(m, t)$ is a function of the stellar mass $m$ and the time $t$ and is expressed as 
   \begin{equation}
\label{e:IGIMF_def}
    \xi_{\rm IGIMF}(m, t) =\int_{M_{\rm ecl,min}}^{M_{\rm ecl,max}(\psi(t))} \phi_{K01}(m \le m_{\rm max}(M_{\rm ecl}))\,\xi_{ecl}(M_{ecl})\,dM_{\rm ecl},
   \end{equation}
where  $M_{ecl}$ is the cluster mass. 
Note that the dependence of time is due to the SFR $\psi(t)$ of the parent galaxy. 
For the ECMF, a single-slope power law is generally assumed:
    \begin{equation}
      \xi_{\rm ecl} (M_{\rm ecl}) \propto \bigg( \frac{M_{\rm ecl}}{M_{\rm ecl,max}} \bigg)^{-\beta},
      \label{e:ECMF}
    \end{equation}
    with a slope $\beta$ defined in the interval $0.5 \le \beta \le 2.35$.\\
    One fundamental parameter to define Equation (\ref{e:ECMF}) is the minimum cluster mass $M_{\rm ecl,min}$, 
    for which a value of $10^3~M_\odot$ is generally assumed \cite{weidner2011}. 

The adopted minimum cluster mass is $M_{\rm ecl,min}=10^3M_\odot$ \cite{weidner2011}.
The upper mass limit $M_{\rm ecl,max}$ is determined empirically from the 
correlation between SFR and most luminous cluster observed in local galaxies \cite{weidner2004}. 
This quantity is expressed as: 
\begin{equation}
M_{\rm ecl,max}=8.5 \cdot 10^4 \, \bigg(\frac{\psi(t)}{M_\odot\,yr^{-1}}\bigg)^{0.75} \, M_\odot,
\label{e:maxmass_cluster}
\end{equation}
 and with $10^7 M_\odot$ as the upper mass limit  $10^7 M_\odot$ \cite{weidner2004}.
 The most relevant result of this formalism is that the 
 'field' IMF calculated for disc galaxies (such as the Milky Way) is  steeper than the canonical IMF.  
 
A more generic form of Equation~(\ref{eq_imf_k01}) has a different high-mass slope $\alpha_3$ at $m>1 M_\odot$ that  
is parametrised as:
    \begin{equation}
        \alpha_3 (M_{\rm ecl}) = \left\{\begin{array}{ll}
        -1.67 \, \log_{10} \big( \frac{M_{\rm ecl}}{10^6 M_\odot} \big) + 1.05&\hspace{-0.25cm}  (M_{\rm ecl}\le 10^6 M_\odot),\\
       +1&\hspace{-0.25cm}  (M_{\rm ecl} > 10^6 M_\odot)\\
        \end{array} \right.
    \end{equation}
and is valid for clusters with masses $M_{\rm ecl} > 2 \cdot 10^5 M_\odot$ and in the most intensely star-forming galaxies. 
    
The upper stellar mass limit $m_{\rm max}$ is computed from $M_{\rm ecl}$ and, 
in any case, is always assumed $\le$150 $M_\odot$ \cite{weidner2004}. 
 
\begin{figure}[H]

\includegraphics[width=9 cm]{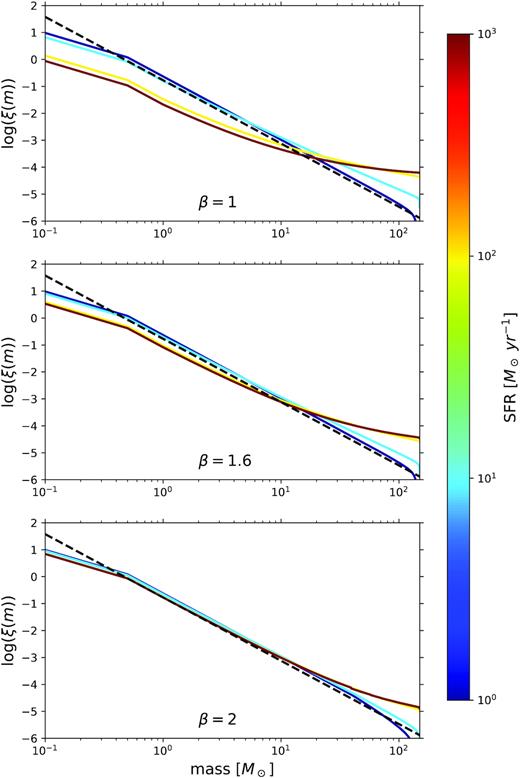}
\caption{The 
 coloured lines show the Integrated Galactic Initial Mass Function (IGIMF) as a function of stellar mass and star formation rate
  for different values of the slope parameter $\beta$ defined in Equation~(\ref{e:ECMF}), representing 
   the slope of the embedded young stellar cluster mass function. Upper
   panel: $\beta=1$; central panel: $\beta=1.6$; lower panel: $\beta=2$.
   In each panel, the black dashed lines indicate the Salpeter (1955) \cite{salpeter1955} IMF. Figure from Palla et al. \cite{palla2020b}.}
\label{f:IGIMF}
\end{figure}

In Figure \ref{f:IGIMF}, we show the IGIMF obtained with the above prescriptions and for different values of the SFR, in which 
three values of $\beta$ were selected: $\beta=1$, $\beta=1.6$ and $\beta=2$.
By adopting $\beta=1$ we obtain, for $\psi\gtrsim10 M_\odot yr^{-1}$, 
an IGIMF comparable to the single-slope form of \cite{Gibson1997}, generally regarded as a quite extreme top-heavy one
(characterised by an index $x=0.8$ and with $x=\alpha-1$, where $\alpha=2.35$ is the \cite{salpeter1955} IMF index over the whole stellar mass range).
On the other hand, the IGIMF obtained adopting $\beta=2$ is very similar to the \cite{salpeter1955} IMF over most of the stellar range,
except at the most extreme SFR values \linebreak ($>$10 $M_\odot yr^{-1}$). 
The IGIMF calculated at low SFR values ($1 M_\odot\, yr^{-1}$) shows a uniform decline with mass and a shape resembling a double-power law, with a knee at $0.5~M_{\odot}$. 
A cut-off is visible at masses larger than $\sim$100 $M_{\odot}$, where the decrease is steeper and where the behaviour is similar to the IGIMF shown in, e.g., \cite{recchi2009}. 
In general, the higher the SFR value, the flatter IGIMF, the higher the relative number of massive stars as due to increasing $M_{ecl,max}$ values with increasing SFR. 
Furthermore, the lower the $\beta$ value, the stronger the IGIMF dependence on the SFR \cite{palla2020b}.

In chemical evolution models, the adoption of the IGIMF leads to important results, in particular regarding its
capability to account for several observables 
in a variety of environments characterised by different star formation histories (SFHs), 
i.e., the solar neighbourhood \cite{calura2010}, dwarf 
galaxies \cite{vincenzo2015,lacchin2020} and 
local elliptical galaxies  \cite{recchi2009,demasi2018,jerabkova2018}.

\subsection{Chemical Evolution Equation for Dust Evolution}

\label{sec_dustchem}
In this subsection we present another, more detailed version of the equation that describes the chemical
evolution of interstellar dust. The version described here includes the expansion
of some of the terms present on the right side of Equation~(\ref{eq_chem}). The aim here is to describe each term 
in better detail, in a pedagogical way and to present a few fundamental quantities which need to be taken into account in chemical
evolution models to study a sizeable set of chemical elements, or of refractory species produced on different scales
and by means of various processes, of both stellar and non-stellar nature.

In a single-phase, instantaneously-mixed gas, 
the chemical evolution for a refractory element $i$ in dust  
can be computed by means of the following equation:

\begin{eqnarray}
 & & {d G_{dust,i} (t) \over d t}  =  -\psi(t)X_{dust,i}(t)\nonumber\\
& & + \int_{M_{L}}^{M_{B_m}}\psi(t-\tau_m) \delta^{SW}_{i}
Q_{mi}(t-\tau_m)\phi(m)dm \nonumber\\ 
& & + A\int_{M_{B_m}}^{M_{B_M}}
\phi(m) \cdot[\int_{\mu_{min}}
^{0.5}f(\mu)\psi(t-\tau_{m2}) \delta^{Ia}_{i}
Q_{mi}(t-\tau_{m2})d\mu]dm\nonumber \\
& & + (1-A)\int_{M_{B_m}}^
{8 M_{\odot}}\psi(t-\tau_{m})  \delta^{SW}_{i} Q_{mi}(t-\tau_m)\phi(m)dm\nonumber \\
& & + (1-A)\int_{8 M_{\odot}}^
{M_{B_M}}\psi(t-\tau_{m})  \delta^{II}_{i} Q_{mi}(t-\tau_m)\phi(m)dm\nonumber \\
& & + \int_{M_{B_M}}^{M_U}\psi(t-\tau_m)  \delta^{II}_{i} Q_{mi}(t-\tau_m) 
\phi(m)dm \nonumber\\ 
& & - \frac{G_{dust,i}}{\tau_{destr}} + \frac{G_{dust,i}}{\tau_{accr}} +({d G_{dust,i} (t) \over d t})_{\rm inf} -({d G_{dust,i} (t) \over d t})_{\rm out}  	
\label{eq_dust}
\end{eqnarray}

The  
 quantity  $G_{dust,i}(t)$ is defined as

\begin{equation}
  G_{dust, i}(t)=M_{g}(t)X_{dust, i}(t)/M_{tot}
\label{eq_mg}
\end{equation}
and represents the  mass in the form of an element $i$ in dust at the time $t$ 
normalized to a total fixed mass. 
Quantities such as the gas mass $M_{g}$ and the SFR $\psi$ 
can be defined also in terms of the surface densities, especially in spiral galaxies, where it is convenient to have all the masses
expressed in units of $M_{\odot}/pc^{2}$, for instance to study the evolution of the Solar Neighbourhood \cite{chiappini1997}. 
The first term on the right is sometimes referred to as the {\it astration} term, and accounts for the amount of dust which is 
subtracted from the ISM and destroyed as it is incorporated into stellar matter. 
$\phi (m)$ is the stellar IMF, for which one can assume one of the forms described in Section ~\ref{sec_imf}.

Its importance is that it regulates the amount of heavy elements produced by stars of different masses, and its shape
has generally strong consequences on the computed abundances. Later on, we will see how severely the IMF can affect also
the properties of interstellar dust, in particular its total amount and composition. 

The second term on the right side of Equation~(\ref{eq_dust}) is the rate at which each element is returned 
in the ISM in the form of dust by single stars with masses in the range between $M_L$ and $M_{B_m}$.
At a given time, 
$M_L$ represents the minimum mass contributing to the chemical enrichment of the ISM.
This quantity depends on the time of the run and on the adopted stellar lifetimes $\tau_m$, which in turn are a function of 
the initial stellar mass $m$. 
In general, in chemical evolution models it is assumed that the minimum mass contributing to chemical enrichment is $0.8~M_{\odot}$,
roughly corresponding to the turnoff mass at a time comparable to the present age of the Universe.
$\delta_i^{SW}$ is the dust condensation efficiency in stellar winds (SW), and represents the fractional amount of matter restored
into the ISM in the form of dust for the refractory element $i$, where we have omitted the dependency on the stellar mass,
which however holds true in the most general case. 
$M_{B_m}$ is the minimum mass  allowed for binary systems giving rise to type Ia SN (typically $3~M_{\odot}$ \cite{matteucci1986}). 
The quantity $Q_{mi}$ is computed from the so-called Talbot $\&$ Arnett \cite{talbot1971} matrix, and contains 
all the information about stellar nucleosynthesis for elements either produced or destroyed inside stars or both. 
Such a complex algorithm offers the advantage of correctly taking into account both the newly produced and unprocessed elements.  

In the literature, various studies have been devoted to the computation of the quantities $Q_{mi}$, or, more in general,
of the stellar yields, defined 
as the fraction of the initial mass newly produced and restored in the form of a given chemical element. 
In chemical evolution models, it is normally convenient to assume that the contribution of a star to the chemical enrichment 
takes place at the end of its life.  

The third term represents the enrichment due to binary stars exploding as type Ia SNe,
i.e., all the binary systems with total mass 
between the two extreme values $M_{B_m}$ and $M_{B_M}$. 
The physical details regarding the true progenitors of type Ia SNe are still poorly known.
Multiple possibilities have been proposed in the literature, most of which assume that the progenitors reside in binary systems.
These include binary systems with two degenerate stars (in general, two white dwarfs)
which, after losing angular momentum merge together, generating the explosion. 
(this is commonly known as the Double-degenerate scenario). 
One of the most popular assumptions is the single degenerate (SD) scenario,  
where a C-O white dwarf explodes by C-deflagration 
after having reached the Chandrasekhar mass ($1.44 M_{\odot}$), owing to progressive 
mass accretion from a non-degenerate companion \cite{whelan1973}.
In this case, the maximum mass for the binary systems hosting a Type Ia SN progenitor is $M_{B_M}=16 M_{\odot}$ \cite{matteucci1986},
resulting from the total mass of two $8 M_{\odot}$ progenitors, which is the highest-mass progenitor that, at the end of its life,
can give place to a white dwarf. 
The parameter $A$ represents the unknown fraction of binary stars
giving rise to type Ia SN, generally fixed by an observational constraint, e.g., the observed present-time SN Ia rate. 
In the third term, both quantities $\psi$ and $Q_{mi}$ refer to the time $t-t_{m_2}$, where $t_{m_2}$ indicates the lifetime of the 
secondary (i.e., the least massive) star of the binary system, determining the explosion timescale.
In the formalism introduced by \cite{greggio1983}, the quantity 
$\mu=M_{2}/M_{B}$ is the ratio between the mass of the secondary component $M_{2}$ 
and the total mass of the binary system $M_{B}$, whereas 
$f(\mu)$ is the distribution function of this ratio. 

Early studies indicated that mass 
ratios close to $1$ (corresponding to $\mu=0.5)$ are to be preferred (see \cite{tutukov1980}). In this case,  the formula:\\
\begin{equation}
f(\mu)=2^{1+\gamma}(1+\gamma)\mu^{\gamma}, 
\end{equation}
is commonly adopted, defined in the range $0 < \mu \le \frac{1}{2}$ and with $\gamma=2$ \cite{matteucci1986a,matteucci2006}. 

$\delta_i^{Ia}$ is the dust condensation efficiency as due to type Ia explosions. 
In a general form such as the one of Equation\ref{eq_dust}, this quantity is useful for exploring the potential contribution of these sources to
the production of dust. 
However, as we will see later in Section \ref{sec_dust_sne}, although their contribution to dust production has undergone some debate in the past, 
most current studies indicate that they are likely to produce a negligible amount of dust. 
The fourth and fifth terms are required to complement the contribution to the chemical enrichment of stars eligible for type Ia SN explosions. 
Taken together, these terms represent the enrichment due to stars in the mass range $M_{B_m}$$-$$M_{B_M}$ which are either single, or, 
if in binaries, do not produce a SN Ia event, which is the  reason why both terms include the factor (1$-$A). 
In this mass range, a common assumption is that all the single stars with initial mass $m > 8 M_{\odot}$
explode as type II SNe, generally assumed to originate from the core collapse (CC) of single massive stars.
Together with type Ib and Ic, this kind of SNe are often regarded as useful tracers of star formation in galaxies.  
The fact that a threshold mass value is believed to exist for stars exploding as CC SNe requires the 
fourth and fifth terms to be separated. When a single star is not massive enough to explode as CC SN, 
its contribution to dust production is expressed 
by the condensation fraction $\delta_i^{SW}$ and is assumed to occur by means of a stellar wind.
For the sake of simplicity, it is convenient to assume that each single star belonging to this mass range returns 
its entire amount of dust at the end of its life.
This assumption is useful for avoiding a time-consuming modelling of the instantaneous 
mass return in the form of an ejection rate, which in principle
could require the introduction of a new timescale and the adoption of a sub-Myr timestep, increasing considerably and, in most cases,
unnecessarily, the duration of the run. 
Such a small timestep is negligible compared to the typical timescale over which significant
abundance variations in galaxies occur, typically ranging from a few tens of Myr to Gyr and clearly depending on the nature of the element and of
its producers \cite{matteucci2021}.

The fourth term includes stars which release significant amounts
of dust during their asymptotic giant branch (AGB) phase, occurring through the ejection of their cold envelopes.
Together with CC SNe, AGB stars are regarded as one of the most important dust producers in galaxies.
The contribution of these sources will be compared and discussed more in detail later. 

In the fifth term, the fractional amount of dust ejected by stars with mass between $8 M_{\odot}$ and $M_{B_M}$ is represented
by the quantity $\delta_i^{II}$, which stands for the poorly known condensation efficiency of the \emph{i}-th element as due to type II SNe. 

This quantity is included also in the sixth term on the right, which accounts for the dust produced by all the stars
with mass between $8 M_{\odot}$ and $M_{U}$, all of which will explode as type II SNe.
The quantity $M_{U}$ is the uncertain maximum stellar mass contributing to the chemical enrichment of the ISM.
The existence of a maximum stellar mass is still an outstanding question and it has been the object of a large number of investigations. 
Theoretical stellar evolution studies generally indicate that a physical maximal mass is expected to exist and
its value to depend on various quantities of the parent gas cloud and of the external environment, including density, 
background temperature and metallicity. 
Several studies indicate larger maximum stellar masses at lower metallicity, mostly because of a less efficient
radiative cooling due to hydrogen, which implies a less efficient fragmentation due to a different cloud opacity
attainable in a gas with reduced content of metals, which in the end translate into larger stellar masses \cite{larson1998,Hosokawa2009}. 
The most extreme case is represented by Pop III stars, i.e., the first stellar objects ever formed, characterised by zero
metallicity and, according to stellar evolution theories, by very large typical masses, typically to $>$100~$M_{\odot}$  \cite{bromm2002,chantavat2023}
(although other indications suggest that they may also be characterised by a standard IMF and maximum stellar mass \cite{fraser2017}).

For a standard IMF such as the one of Salpeter \cite{salpeter1955}  or Kroupa \cite{kroupa2001}, 
the results of chemical 
evolution models are not much sensitive to the adopted value for $M_{U}$. This occurs because such IMFs are steep enough that
the contribution to the chemical enrichment from stars more massive than $\sim$a few $10~M_{\odot}$ is negligible.
As we will see later in Section \ref{sec_starburst}, the same is not true if a non-standard IMF is assumed, e.g., in particular the ones more 
skewed towards high mass values. 
Concerning the maximum stellar mass contributing to chemical enrichment, another important aspect to be mentioned
concerns the results of stellar nucleosynthesis models, which in some cases indicate that individual stars with mass larger
than a certain value fail to explode as SNe, but collapse directly in a black hole (BH), retaining all (or most of) the metals produced during their evolution.
Such maximum value for enrichment by type II SN explosions is known to depend on metallicity (e.g. \cite{maeder1992})
and can typically range between a few 10 and $\sim$100 solar masses \cite{woosley1995,portinari1998}. 
It is worth stressing out that the two terms describing dust production from massive stars (i.e., the fifth and sixth term of Equation (\ref{eq_dust})
do not include the amount of dust produced by massive stars during their pre-SN phase and through stellar winds, although they could
have a non-negligible role.  

The negative, seventh term accounts for the destruction of dust grains in the ISM as due to various violent processes, occurring mostly
in the warm/hot phase and due to their interaction with accelerated or high-temperature matter. The major responsibles for dust destruction are SN explosions, capable
of accelerating gas at velocities of $\sim$1000~\text{km/s}. 
In this term, dust destruction is assumed to depend on a typical destruction 
timescale $\tau_d$ which can depend on various physical quantities, such as the grain size, temperature and density.
In Section \ref{sec_destr} we will discuss how such quantity is calculated and its dependence on a few fundamental properties of the ISM. 

The eighth term accounts for another important mechanism for dust production, namely dust {\it accretion} or {\it growth}, which is expected to occur in the
densest, coldest regions of the ISM, namely in molecular clouds. 
In this process, refractory elements can condensate onto pre-existing grain cores. 
It is generally assumed that dust grains can undergo accretion on a typical timescale $\tau_{accr}$,
which may be a complex function of the interstellar metallicity, its density, temperature and of the grain size \cite{dwek1998}.  

The ninth term describes the infall of matter in the form of dust into the system, i.e., it is meant to  
account for the amount of dust accreted from an external component, which in principle might be
the intergalactic medium (IGM) or, in models in which gas flows are considered, a reservoir of cold gas present nearby the
system and ready to be accreted, such as, e.g., the ISM of a merging galaxy or a gas distribution leftover from a recent
merging event. 
The tenth and last term of  Equation~(\ref{eq_dust}) accounts for the possible ejection of dust into the IGM by means of galactic winds. 
Such term needs to be taken into account in the most general case in which outflows are to be included in the galactic type that one
aims to model, such as a dwarf, a starburst galaxy or a spheroid in the earliest phases of its evolution. 
In the specific case of the Milky Way, the chemical evolution used to describe the evolution of dust is typically a {\it closed box},
i.e., it does not include any term which accounts for an exchange of dust with the external environment. 
This occurs because in one-zone chemical evolution models for the solar neighbourhood, infall is assumed to contain negligible traces
of metals, hence also of refractory elements.

Given the variety of the different processes involved in dust production and, most of all, the fact that different sources act on different
timescales, Equation~(\ref{eq_dust}) (together with other equations required to close the system) is to be integrated numerically, therefore the derivation of its solution require computational resources.  
In general, an ordinary, non-high-performance computer is sufficient to compute the solution to Equation~(\ref{eq_dust}) for a significant set (order of 10--20) of chemical
elements of interest and for a significant fraction of cosmic time.

\subsection{Dust Production from Stars}
\label{sec_stardust}

In chemical evolution models, the prescriptions regarding the amount of dust produced by various 
stellar sources are represented by the dust condensation efficiencies $\delta^j_i$---
the fraction of heavy elements which are incorporated into dust for various elements  
(where $i$ is for the chemical element and $j$ stands for the source - generally AGB and SNe).

These are the most crucial quantities that define stardust production; however, 
the amount of dust produced by a given stellar type depends not only on this quantity but also 
on the chosen elemental yields.   
Few attempts have been made in the past to compute the condensation efficiencies by means of
physical, ab-initio approaches. 
In other cases, these quantities were assumed on an empirical basis.
In this Section we briefly review previous assumptions and estimates 
made in the past by various authors regarding the stellar contribution to stellar dust production in the form of
various elements, mostly as a function of fundamental parameters such as stellar mass and metallicity.

\subsubsection{Dust in Asymptotic Branch Stars}

The synthesis of dust in a suitable environment is known to proceed in two different steps: the nucleation and the condensation \cite{cherchneff2009,sarangi2013}. 
As historical note,  the  earliest approach to model dust creation in SN ejecta was based on the ‘Classical Nucleation Theory’ formalism, 
developed using the popular ‘liquid-drop model’, useful to study water droplets in the Earth’s atmosphere \cite{feder1966}. 

In low and intermediate mass stars (LIMS, i.e., stars with mass $0.8~M_{\odot} \le m \le 8~M_{\odot}$),
dust is produced  during  the  Asymptotic  Giant  Branch  (AGB) phase 
\cite{ferrarotti2006,dellagli2017}. 
Single stars with initial masses in this range evolve through the AGB after the core helium burning phase. 
The cold envelope of AGB stars is a good environment in which
nucleation and condensation of heavy elements in refractory {\it seeds}, or cores, \linebreak can occur. 

The total amount of dust produced in the previous evolutionary phase of LIMS, the Main Sequence, is
negligible, mostly because of the low amount of ejected material and  
because the physical conditions of their winds do not favor its formation. 
In the AGB phase, LIMS undergo a phase of thermal pulses after exhausting helium in their cores.
During this phase nuclear fusion of helium and hydrogen takes place in two distinct shells layered above a degenerate core.

Two primary processes can alter the surface chemical composition during this evolutionary stage:
hot bottom burning (HBB) and the third dredge-up (TDU) 
HBB occurs in more massive AGB stars and involves proton-capture nucleosynthesis at the base of the convective envelope \cite{dicriscienzo2013}. 
TDU, on the other hand, takes place when the convective envelope penetrates inward following each thermal pulse,
reaching deeper layers that have undergone $3\alpha$ reactions and are rich in carbon and oxygen. 
These two mechanisms modify surface abundances in distinct ways. TDU generally increases the levels of carbon—and to a lesser extent,
oxygen—at the stellar surface. In contrast, HBB reflects the outcomes of proton-capture reactions and its impact depends on the operating temperature. 
The combination of significant mass loss and low surface temperatures, typical of the extended, low-density envelopes of AGB stars, 
creates an environment conducive to dust formation, as gas-phase molecules condense into solid particles in the stellar winds. 

The species of dust grains formed during the AGB phase of LIMS 
depend strongly on the composition of the stellar surface \cite{ferrarotti2006}. 
The parameters which play a key role are stellar mass and the metallicity Z, as they determine the number of thermal pulses
and also shape the surface composition, and
therefore, favour the formation of particular dust species \cite{nanni2013,ventura2012}.  

In early works of chemical evolution of dust, it was  
assumed that the dust grains produced by AGB (and other sources) can be of two types:
carbon (C) dust and silicate (Si) dust \cite{dwek1998,calura2008}. 
The purpose of this choice was mostly to simplify the treatment of dust production, bearing in mind that 
dust composition  can  be  much more  complicated than that. 
For  instance,  Mathis \cite{mathis1996} presented  a  model  including  also  the  possibility  of  complex, composite  grains
containing  carbon,  silicates  and  oxides. In another work, \mbox{Li \& Greenberg \cite{li1997} }presented a  trimodal dust  model which included
silicate core-organic refractory mantle dust particles, small carbonaceous particles and polycyclic aromatic hydrocarbon particles.
These models were designed to describe more the spectral behaviour of dust grains that their chemical evolution, and were 
able to account for properties such as interstellar extinction and polarization. 
Furthermore, different, complex dust structure would also have an impact on dust destruction and
accretion \cite{greenberg1999,jones1996}.  
However, the inclusion of more complicated dust types was sometimes avoided in works focused on the composition of interstellar dust, 
since they clearly implied an increase of the already large number of free parameters involved in these studies. 

In the pioneering work of Dwek \cite{dwek1998} (D98), 
it was assumed that the amount of dust produced by AGB stars was depending on the composition of the stellar envelopes, and in particular
it was determined by the C/O ratio in the ejecta, which depends 
on the complex interplay between 
metallicity, mass loss and the capability of dredging up material in the form of heavy elements from
the innermost regions. 
The C/O ratio of stellar ejecta for various values of the stellar mass and metallicity 
can be easily computed from the adopted stellar yields. 
In this formalism, if $N_O$ and $N_C$ represent the  O and C abundances (in number) in the stellar envelopes,
respectively, it was assumed that stars with $N_O/N_C>1$  are  producers of  silicate  dust,
i.e.,  of dust particles composed by refractory elements such as O,  Mg,  Si,  S,  Ca,  Fe. 
On the other hand, C-rich stars, typically characterized by $N_O/N_C<1$, are producers of  carbonaceous solids,
i.e., carbon dust, ignoring the exact composition of the C dust, i.e., whether it was  graphite or amorphous carbon dust.

The masses of dust released in the form of various elements were then computed by multiplying the ejected masses in the form
of various elements which, as seen in\mbox{ Section ~\ref{sec_dustchem}}, can be computed from the Talbot \& Arnett {\cite{talbot1971} matrix. 

In D98, the quantity $M_{i, ej}(m)$, which represents 
the dust mass produced by the stars as functions of the initial mass $m$ 
and in form of the element $i$, was computed from the corresponding total ejected mass $M_{i, ej}(m)$ in the form
of the element $i$ as follows:

For stars with $N_{O}/N_{C}$ $<$ 1: 
\begin{equation}
M_{dust, C}(m) =  \delta^{SW}_{C} \cdot [M_{C, ej}(m)-0.75 M_{O, ej}(m)]
\end{equation}
with the condensation fraction in AGB ejecta for C typically $0.5 \le \delta^{SW}_{C} \le 1$ and
\begin{equation}
M_{dust, i}(m) =0, \\
\end{equation}
for all the other elements. For stars with $N_{O}/N_{C}$ $>$ 1 in the envelope, it was assumed instead that 
\begin{equation} 
M_{dust, C}(m) =0, 
\end{equation}
whereas for Mg, Si, S, Ca, Fe: 
\begin{equation}
M_{dust, i}(m) = \delta^{SW}_{i} M_{i, ej}(m)
\end{equation}
with $\delta^{SW}_{i}=1$ and finally 
\begin{math}
M_{dust, O}(m)=16 \sum_{i} \delta^{SW}_{i} M_{ej, i}(m)/\mu_{i}, 
\end{math}
where $\mu_{i}$ is the mass of the $i$-th element in atomic mass units.

In the following years, the production of dust from AGB stars has been
the object of various studies where model calculations  for dust condensation in stellar outflows have been combined
with synthetic models of stellar evolution (e.g., \cite{gail2009,dellagli2017}. 

A comprehensive comparison of the dust condensation efficiencies from AGB stars, computed considering various
stellar yields as inputs and for various chemical elements has been presented by Piovan \cite{piovan2011}.

Figure~\ref{fig:dustAGBy} shows the AGB stars condensation efficiencies  $\delta_i^{AGB}$ 
for a few refractory elements, i.e., carbon, silicon, magnesium, oxygen 
and iron (see \cite{gioannini2017a}).  
These quantities are calculated as a function of the stellar mass and 
metallicity, and take into account the dependence of the C/O ratio in stellar ejecta. 
At $Z\le 0.004$, the production of C dust dominates over the entire stellar mass range.
At higher metallicity, intermediate-mass stars with $M<4 M_{\odot}$ still play an important role
in C production, whereas higher-mass stars are sites where 
 the condensation of heavier elements is favoured, Fe- And Si-dust in particular. 
At low metallicity, most of the dust in intermediate-mass stars is in the form of C basically because
of its high condensation efficiency and since the progenitors of AGB stars are strong carbon producers
(see also \cite{dwek1998,valiante2009}), mostly through the CNO cycle \cite{abia2003,romano2022}. 
  
Elements heavier than carbon are not produced in large amounts by these stars, which causes their lower condensation
efficiency and the clear separation between carbon and silicate dust production, in which the latter is due mostly 
to pre-existing amounts of Si, O and Fe. 
On this regard, it is important to stress that LIMS can also produce a small amount of such dust species.
Also for these elements, during the AGB phase some fraction of the initial stellar mass that has not been processed in the inner regions
can be 'dredge up' and  expelled into the ISM.   
This returned mass composition reflects that of the star when it was formed, and could be chemically enriched by heavy elements.
Part of this material can condensate, forming dust silicates in a non-negligible amounts. 

A comprehensive comparison of the total (including both carbon and silicates) dust {\it yields} in AGB stars 
(i.e., the product of the condensation efficiency and the ejecta) 
was presented by \cite{rowlands2014} and is reported in the left panel of Figure~\ref{fig:row14}. 
Discrepancies of one (or more) order of magnitude characterise the dust yields from LIMS calculated
by various authors: the assumption that 100 \% of the available C and O end into dust \cite{dwek1998,calura2008}
is not supported by other studies \cite{ventura2012,ferrarotti2006}. 

The results from theoretical models are sometimes calibrated by comparing them with the observational properties of 
evolved stars of local galaxies, including the Magellanic Clouds 
and other systems \cite{dellagli2015,dellagli2019}, in order to constrain the production of
dust in AGB progenitors. 
The spectral energy distribution (SED) of post-AGB stars presents peculiar features that enable the determination of
fundamental properties, i.e., 
a characterization in terms of mass and formation epoch of the stellar progenitor and of the mineralogy 
of the dust responsible for this feature \cite{woods2011}. 
In particular, the SED of post-AGB stars exhibits a typical double-peak shape, referred to as a “shell-type” SED \cite{vanwinckel2003},
which allows one to disentangle the emission from the central star from the “IR excess”, typically in the\mbox{ 8--30 $\upmu$m} spectral region, 
due to the presence of dust in its surroundings \cite{garcialario2006,dellagli2023}. 

In a recent work, \cite{tosi2022} studied a sample of individual post-AGB sources in the Magellanic Clouds
and interpreted their SEDs and stellar parameters by comparing them with results from radiative transfer calculations and
from stellar evolution modelling of AGB and post-AGB stars (see also \cite{sarkar2022}). 
This allows for a detailed characterization of the individual sources, 
in terms of the initial mass, formation epoch of the progenitors, timescale of dust production and composition of the dust.
By exploiting the upcoming releases of astrometric data from {\it Gaia}, such a study can be extended to include also samples of Galactic stars,
fundamental to constrain the properties of dust production in various environments and to pinpoint the role of critical 
parameters, such as metallicity.

\begin{figure}[H]


\includegraphics[width=12 cm]{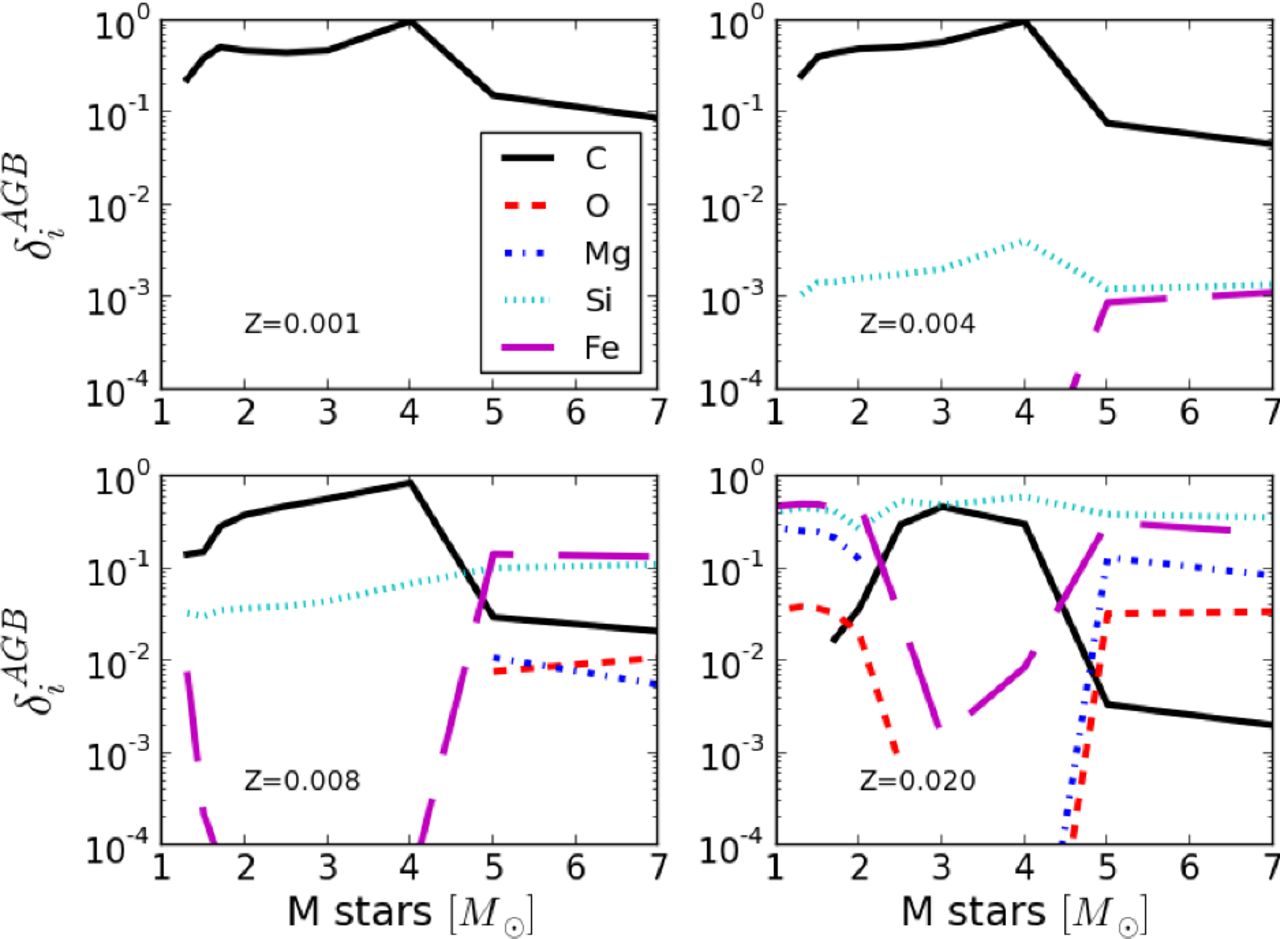}

\caption{Dust condensation efficiencies of C, O, Mg, Si and Fe for AGB stars as reported in Piovan et al. (2011) \cite{piovan2011}
  for various metallicities. Figure from Gioannini et al. \cite{gioannini2017a}.}
\label{fig:dustAGBy}
\end{figure}

\begin{figure}[H]
\begin{adjustwidth}{-\extralength}{0cm} 
\centering 
\includegraphics[width=19 cm]{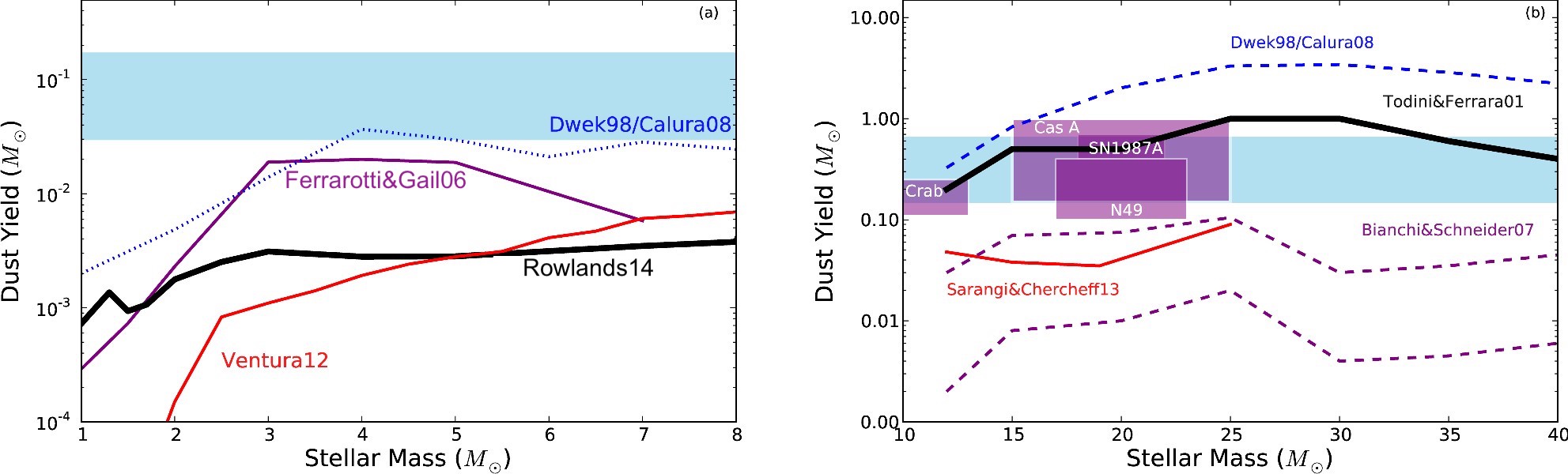}
\end{adjustwidth}

\caption{Theoretical and observational dust yields from low and intermediate mass stars and core-collapse SNe at solar
  metallicity ($Z=Z_{\odot}$).
  Panel (\textbf{a}): dust yields (expressed in $M_{\odot}$) from AGB stars computed by Rowlands et al. \cite{rowlands2014}  (black solid line) ,
  Dwek/Calura et al. \cite{dwek1998,calura2008} (blue dotted line), Ferrarotti \& Gail \cite{ferrarotti2006} (purple dashed line)
  and Ventura et al. \cite{ventura2012} (red solid line) 
  The shaded light-blue region shows the minimum average dust yield per AGB star required to explain observations
  of high-redshift submillimetre galaxies
  \cite{michalowski2010}.
  Panel (\textbf{b}): comparison of some theoretical dust yields from core-collapse SNe with observations.
  Black solid line: Todini \& Ferrara \cite{todini2001};  blue dotted line: Dwek/Calura et al. \cite{dwek1998,calura2008};
  red solid line: Sarangi \& Cherchneff \cite{sarangi2013}. The upper and lower purple dashed lines show the
  range of expected yields from the theoretical SN dust formation model of Bianchi \& Schneider \cite{bianchi2007}
  that survived the passage of the reverse shock and considering two different gas densities.
  The shaded light blue region shows the average dust yield per SN required to explain observations
  of high-redshift submillimetre galaxies. The observed dust masses from some Galactic and nearby young supernova remnants are indicated
  by the shaded purple regions (Rowlands et al. \cite{rowlands2014} and references therein).
  The boxes indicate the range of dust mass values derived from IR–submillimetre data as well as uncertainties
  in the mass of the progenitor stars \cite{michalowski2010}. 
Figure adapted from Rowlands et al. \cite{rowlands2014}.}
\label{fig:row14}
\end{figure}

\vspace{-13pt}
\begin{figure}[H]

\includegraphics[width=12 cm]{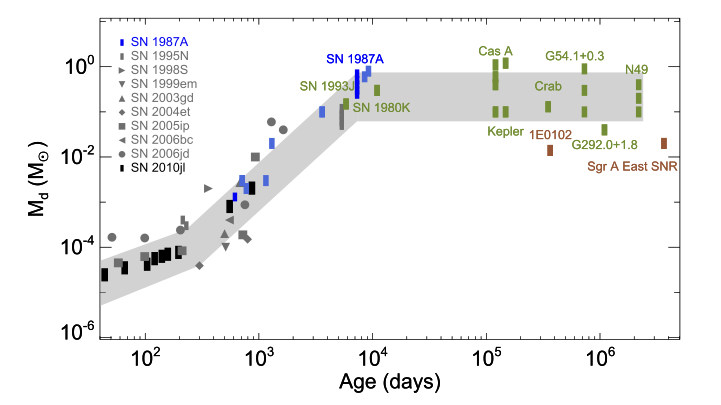}
\caption{Observed dust mass evolution in SN remnants as a function of the `Age' of the remnant, in many cases defined as the time after the explosion or the time past the maximum visual light. The collected data are for various SN remnants, including SN 1987A (blue symbols) Cas A and Crab (green squares; see Gall et al. \cite{gall2018} { for more details and references). The red bars indicate warm dust ($T>100$ K) measurements. The gray area illustrates a possibly increasing trend, followed by a flattening of the dust mass, characterised by significant scatter. Figure from Gall et al.} \cite{gall2018}.}
\label{fig:dust_SNE}
\end{figure}

\subsubsection{Dust Produced in SNe}
\label{sec_dust_sne}

Core-collapse SNe, the explosive deaths of massive stars (m $\ge$ 8--10 M$_\odot$), are other fundamental contributors to stellar dust production.
Dust formation in SNe is thought to begin soon after the explosion, 
as the ejected material, rich in refractory elements, expands and cools. 
Within the cooling ejecta, conditions become suitable for molecules to form and subsequently condense into dust grains in
dense, metal-rich clumps \cite{kozasa1989,todini2001}. These grains are composed of various species,
including silicates, carbonaceous compounds, alumina, and iron-bearing dust, depending on the local composition and temperature.

The progress in the study of dust production in SNe underwent a significant thrust in 1987, with the 
explosion of a core-collapse SN detected in the Large Magellanic Cloud, only 50 kpc away from us 
\cite{kunkel1987}. Since this discovery, a large amount of studies dedicated to SN 1987A and more SN remnants
led to a wealth of results that unveiled the production of dust in these environments.

The presence of dust in SN remnants is visible as early as a few hundred days after the explosion and 
is generally characterised by some common features \cite{sarangi2018}. 
First, an excess in the mid-IR region of the SED is visible \cite{roche1991,bouchet1993}, 
followed by a decline of the 
optical light curve \cite{danziger1991}, a blue-shift of the emission lines \cite{lucy1989} and diminishing line emission of elements and molecules
compared to the continuum \cite{kotak2009,smith2012a}. 
Some of these features are interpreted as the presence of blocking, newly-formed dust, but others can be also due to 
pre-existing dust in the circumstellar material, therefore it is generally assumed that the likely presence of SN-produced dust
can be ascertained when three/four of these phenomena are observed \cite{sarangi2018}.  

Together with SN 1987A, the detection of dust in the young supernova remnant (SNR) Cassiopeia A \cite{dwek1987}
was one milestone discovery which confirmed that dust can be formed in CC SNe. Sarangi et al.\cite{sarangi2018}
presented a list of known SN remnants with dust detection. 
All the remnants in which dust was observed by means of the above criteria were presumably produced
after the occurrence of a type II SN explosion.

{ Figure ~\ref{fig:dust_SNE} provides an overview of the current observational knowledge of dust production in CC SNe, showing 
a compilation of observed dust masses in SNe and SN remnants as a function of time since explosion \cite{gall2018}.  
One challenging aspect in this kind of measurements is to distinguish the amount of dust formed directly by the SN from interstellar dust and
foreground dust \cite{gall2014,delooze2017}. In fact, some amount can be formed from the interaction of the SN remnant
with the surrounding environment \cite{mattila2008} and even in the pre-SN phase, as massive SN progenitor stars in the 
luminous blue variable phase \cite{smith2012c} (see also \linebreak Sections ~\ref{sec_dust_WR} and~\ref{sec_growthrole}).  
Figure ~\ref{fig:dust_SNE} supports an overall increase of the inferred dust mass with age. 
This trend suggests that SNe can show significant amounts of dust, in a few cases $>$0.1 M$_{\odot}$, on timescales $>10^4$ yr.}  

The topic of dust production in SNe has gained considerable interest also due to a significant collection of observations of high-redshift galaxies. 
Through the years, a multitude of submm observations have evidenced copious reservoirs 
of dust in a large set of high-$z$ galaxies. 
Various observations of systems at redshift $z > 6$ have shown the presence of large dust 
masses in the hosts of quasi-stellar objects (QS0) \cite{wang2013} and in systems often characterised by vigorous starbursts \cite{bertoldi2003}. 
In many cases, the dust masses were comparable to the ones of the most dust-rich local galaxies, with $M_d$ 
values up to  $10^8$ M$_{\odot}$ or even more \cite{carilli2001,beelen2006,damato2020},
in systems often characterised by very high SFRs of 100--1000 M$_{\odot}$/yr as inferred from the
measured  sub-millimeter fluxes. 
In addition, some particular features in the extinction curves of various high-redshift objects have been linked to dust production from SNe.
As an example, the extinction curve derived for the QSO SDSS J1048+46 at redshift z = 6.2 exhibits a distinctive plateau around 1700--3000$\"A$.
This plateau is interpreted as resulting from amorphous carbon and magnetite dust from supernovae \cite{maiolino2004}.
Several more QSO spectra have been best-fitted using extinction curves indicative of SN-like \linebreak dust
\cite{gallerani2010,hirashita2008}. 
Another example is a similar feature reported for the afterglow of the Gamma Ray Burst (GRB) 071025 at z $\sim$ 5 \cite{perley2010}. 
Additionally, less prominent features in extinction curves, particularly flatter UV slopes compared to the Small Magellanic Cloud 
extinction curve, have been interpreted as evidence for dust originating from SNe. 
 Also the young infrared galaxy SST J1604+4304 located
at $z\sim 1$ has been suggested to fit best with a SN extinction curve \cite{kawara2011}.
The considerable amounts of dust in the highest redshift objects 
were found by some authors to be in contradiction with standard scenarios for interstellar
dust formation in which dust was expected to be produced by AGB stars,
whose lifetimes are typically longer than the age of the Universe at $z\sim 6$, i.e., 1 Gyr.

Moreover, despite their great popularity in the high redshift community,
observations of local SN remnants indicate that the production of dust grains in these environments is far from clear \cite{dwek2007a,kozasa2009}.
In fact, while sub millimetre observations of the Kepler and Cas A SN remnants indicate an amount of dust in the range 0.1--1~$M_{\odot}$ 
\cite{morgan2003,dunne2009,gomez2009}, infrared observation of other systems suggest $ \ll$0.1~$M_{\odot}$ \cite{sugerman2006,kotak2009}.
In principle, such discrepancy could be ascribed to a difference in the instrument sensitivity to the
different dust phases (`cold dust' in the case submillimetre, `warm' in the IR band). 

Theoretical calculations indicate between $\sim$0.1~$M_{\odot}$ \cite{cherchneff2009,slavin2020} and
$\sim$0.7--0.9~$M_{\odot}$ \cite{bocchio2016} of dust formed in supernovae, depending on the evolutive stage of the remnant. 
An important phenomenon to take into account is the reverse shock inside the SN remnant, 
that in principle can destroy a large fraction of this dust \cite{bianchi2007} but this amount is susceptible to troublesome details, such as 
the degree of asymmetry in the explosion \cite{nozawa2010}. 
Other theoretical studies that attempted to model the amount of dust produced in SNe \linebreak include
\cite{nozawa2003,schneider2004,kozasa2009,sarangi2022}. 

In the first attempt to account for SN dust production in a chemical evolution model, Dwek \cite{dwek1998} made the following assumptions regarding 
the dust mass produced by CC SNe (in all stars with initial mass $\ge 8~M_{\odot}$) as functions of the initial mass $m$ 
and in form of the element $i$: 

\begin{equation*}
M_{dust, C}(m) = \delta^{II}_{C}[M_{ej, C}(m)]
\end{equation*}

with  $\delta^{II}_{C}=0.5$; 

\begin{equation*}
M_{dust, i}(m) = \delta^{II}_{i}M_{ej, i}(m)
\end{equation*}
\indent with $\delta^{II}_{i}=0.8 $ for Mg, Si, S, Ca, Fe; \\
\begin{equation*}
M_{dust, O}(m)=16 \sum_{i} \delta^{II}_{i} M_{ej, i}(m)/\mu_{i} 
\end{equation*}
\indent This choice was calibrated to account for the elemental depletion pattern observed in the ISM \cite{savage1996},
i.e., the fractional abundance that is locked up in dust for a few \linebreak refractory elements. 

Nozawa et al.\cite{nozawa2003} presented a compilation of dust yields for an extended set of chemical elements,
but considering  Pop III (i.e., zero-metallicity) SNe.
They modelled the production of dust in stars with initial mass between 13 and 30 $M_{\odot}$ and
170 and 200 $M_{\odot}$, exploding as CC SNe and Pair-Instability SNe, respectively, based
on a non–steady state nucleation and grain growth model to account for the collisions of gaseous species in a cooling gas outflowing
from a star into the ISM. 
Starting from this comprehensive compilation of dust yields, 
Piovan et al.\cite{piovan2011} presented a detailed calculation of $\delta^{II}_{i}$ for various chemical elements,
under the assumption that 
dust formation in the ejecta is insensitive to the metallicity of the progenitor stars, as suggested by \cite{todini2001} and \cite{nozawa2007}. 
One crucial assumption is the degree of mixing of the ejecta. 
\cite{piovan2011} considered two extreme cases, i.e., an unmixed (i.e., reflecting the original onion-like structure of the pre-SN star)
and a uniformly mixed ejecta. 
The effects of the forward 
and reverse shocks were taken into account by multiplying the dust yields by a set of `destruction coefficients' \cite{nozawa2007}. 
The resulting condensation efficiencies are shown in Figure~\ref{fig:dustSNy} for the unmixed and mixed cases and as a function of the ISM density.
The efficiencies are here shown for C, O, Mg, Si, S and Fe in the unmixed case and O,
Mg, Si, Ca and Fe for the mixed one. 

An important role is played by the density of the environment surrounding the explosion of SN. 
In general, the higher the ISM density, the more resistance the shock will encounter and the more dust will
be destroyed \cite{nozawa2007}. On the other hand, in a 
lower-density environment, dust can easily resist to the passage of
the shock, causing a more efficient dust formation. 
Figure~\ref{fig:dustSNy} also highlights the importance of the assumption regarding the mixing. When it is considered, 
the predicted condensation efficiencies can vary by several orders of magnitudes, depending on the assumed density value. 

A study of various assumptions regarding the condensation efficiencies $\delta^{II}_{i}$ was performed later \cite{calura2008} 
in a model of the solar neighbourhood and considering the elemental dust depletion pattern observed in the Local
Interstellar Cloud \cite{kimura2003}, summarised in \mbox{Figure \ref{dep_pat}}. 
The study of Calura et al. \cite{calura2008} indicated that, if production of dust in stars only was considered,
the prescriptions of D98 allowed one to reproduce the observed depletion pattern for all the elements.
On the other hand, the depletion pattern computed in a more realistic model which includes also dust destruction and growth
is fairly insensitive to the assumed condensation efficiencies, as models including type II SN condensation efficiencies 
between  $\delta^{II}_{i}=0.1$ and  $\delta^{II}_{i} \sim 1$ were providing essentially the same results. 
It is worth stressing that the depletion pattern is affected also by the most crucial parameters that regulate dust destruction and accretion.
The most useful indication of this study is the significant degeneracy between the condensation efficiencies and other parameters, and 
that the depletion pattern is probably not the most suited quantity to constrain these quantities.  

\textls[-15]{A summary of the total theoretical and observed SN dust yields as a function of the progenitor mass is shown 
in the right panel of Figure~\ref{fig:row14}, in which results from various studies, both theoretical and from observations of SN remnants, are compared. 
This figure shows that the simple scaling relations adopted by D98 and Calura et al. \cite{calura2008} tend to overestimate
the observed dust yields at all masses. 
This is a remarkable result, in particular if one considers that some SN remnants could be very young for the reverse shock to have 
affected substantially the produced dust mass, that is therefore to be regarded as an upper limit to the 'final' \linebreak value \cite{gall2018}. 
At present, theoretical models suggest that the dust mass currently observed in SN remnants is only a fraction
of the initial dust mass formed in the explosion \cite{bocchio2016,marassi2019} (see also \cite{barlow2010} for a discussion
on possible ongoing dust destruction in Cas A).} 

As a final note, the dependence of dust yields on various crucial parameters, such as 
metallicity, explosion energy and rotation was presented by Marassi et al. \cite{marassi2019}. 
Their study shows that stellar rotation tends to favour more efficient dust production, particularly for more massive, low-metallicity stars.
Besides the intrinsic properties of explosion, in models where also rotation is considered, 
the stellar metallicity is found to have the largest effects on the dust mass,
with variations of several orders of magnitude in the metallicity range from [Fe/H] = $-$3 to solar metallicity ([Fe/H = 0]). 

Regarding type Ia SNe, even if they are expected to restore significant amounts of refractory elements, most of 
which in the form of Fe and other heavy elements, they are generally not expected to contribute significantly to the
interstellar dust mass budget. 
Theoretical studies have shown that after type Ia SNe explosions, no dust is finally produced
mostly because, even if favourable conditions are met at some stage, all the grains are destroyed in later phases \cite{nozawa2011}. 
These results show that, due to the high expansion velocity and low mass of
the ejecta, the resulting grains have very small size ($\lesssim$100$\AA$). 
Furthermore, it is less likely for the ejecta to be clumpy, therefore there is very small chance
for the grains to be shielded against eroding processes. 
As final result, all the 
dust that may have formed is expected to be destroyed by the SN reverse shock \cite{nozawa2011}. 
On the observational side, in most cases the searches for dust performed in the IR with {\it Spitzer} in Type Ia SNe remnants have given negative results. 
These include the Kepler remnant \cite{williams2012}, RCW 86 (where \cite{williams2011} find a non-zero, but very low dust mass), 
SN 1006 \cite{winkler2013},
and Tycho \cite{williams2013}. 
In some cases, these results were also confirmed by Herschel observations \cite{gomez2012}.
The only exception is SN2018evt, a very rare case of circumstellar medium---SN Ia interaction, where 
a conspicuous dust mass of $\sim$10$^{-2}$ M$_{\odot}$ was detected \cite{wang2024} a few years after the explosion. 
However, the conditions of this event are considered not very frequent in normal star-forming galaxies. 

\begin{figure}[H]

\begin{adjustwidth}{-\extralength}{0cm}
\centering 

\includegraphics[width=17 cm]{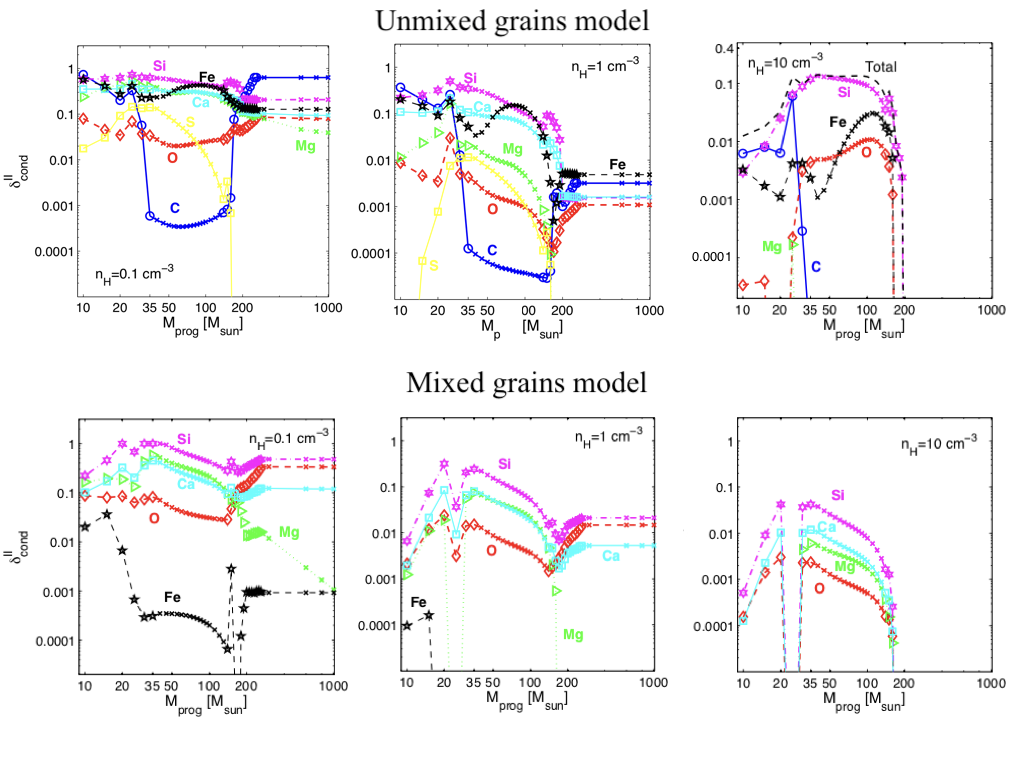}
\end{adjustwidth}
\caption{Dust condensation efficiencies of the elements C (empty circles and continuous lines), O (diamonds and dashed line),
  Mg (triangles and dotted line), Si (six-pointed
stars and dot-dashed lines), Ca (squares and solid lines), S (yellow squares and solid lines) and Fe (five-pointed stars
and dashed lines) in type II Supernovae as a function
of the progenitor mass, according to the unmixed (upper panels) and mixed (lower panels) grain models of Nozawa et al. \cite{nozawa2003,nozawa2007} and at various values of the hydrogen number density $n_H$ (left: $n_H=0.1~$cm$^{-3}$; centre: $n_H=1~$cm$^{-3}$; right: $n_H=10~$cm$^{-3}$).
The small crosses represent extrapolations of the dust yields to other mass ranges. Figure adapted from Piovan et al. \cite{piovan2011}. }
\label{fig:dustSNy}
\end{figure}

\subsubsection{Dust in Wolf-Rayet Stars}
\label{sec_dust_WR}
Other possible stellar sources of dust include Wolf-Rayet (WR) stars, i.e., massive stars (generally with m > 20--25~$M_{\odot}$)
in the pre-SN phase, experiencing significant mass loss though intense stellar winds \cite{crowther2007}.
Through such winds, high-mass stars can lose mass at a significant rate (up to a few $10^{-4} M_{\odot}/yr$, \cite{vanderhucht1992}), 
with appreciable effects on their structure and evolution \cite{maeder1991}. 
When they exceed a critical mass loss rate, WR stars will strip off their H-rich envelope, leaving their underlying core exposed.  
Such stripped cores will appear as WR stars of type N or C if the exposed material is N-rich or C-rich, respectively. 
However, while all WR stars experience rapid mass loss, only the coolest WC stars (WC8 and WC9) show an IR excess that can be attributed to the presence
of dust in their atmosphere \cite{cohen1975,williams1987}. 

The lack of features in the dust spectrum indicates carbon-type dust, which is compatible with carbon-enriched stellar ejecta
(whereas N is generally a non-refractory element). 

Peculiar structures like dust shells and dust rings have been observed in WR stars in several cases
\cite{vanderhucht1986,crowther2003,lau2020,lau2022}. 
One case is WR 112 \cite{lau2020}, a very hot and luminous star, characterized by powerful stellar winds that expel material at velocities of
several thousand kilometers per second. Under such extreme conditions, the intense radiation and turbulent winds 
would typically be expected to destroy dust grains. 
The observed presence of dust in this harsh environment therefore presents a puzzling and intriguing case, highlighting cases like WR 112 as  particularly unusual systems. 
Since they are mostly associated with particularly massive stars, for a normal stellar IMF, WRs will be considerably less common
than Type II SNe.   
Considering this, an interesting problem is to quantify their overall impact on the interstellar dust budget. 
In the Solar neighbourhood, D98 estimated for WR stars a present-day C dust production rate significantly lower (by a factor 75)
than type II SNe (see also Jones \& Tielens \cite{jones1994}). 

A rough estimate for the contribution of WR stars to dust production in a simple stellar population can
be derived by assuming a specific WR number of $1 \times 10^{-3}$, obtained by integrating between $20~M_{\odot}$ and $40~M_{\odot}$
a Kroupa (2001) IMF \cite{kroupa2001}  
(for comparison, the specific number of massive stars, i.e., in the range $8 \le m /M_{\odot} \le 100 $ is $\sim$0.01).
If these stars lose mass with an average rate of $\dot{M} \sim 10^{-5}~M_{\odot}/yr$ \cite{nugis2000},  
assuming a solar metallicity and a dust-to-metals fractions in the ejecta of $ \zeta_d \sim 0.5$,
in a time span of 5 Myr these stars can produce a specific amount of dust
$m_{d,WR} = \dot{M} Z_{\odot} \zeta_d \times 10^{-3} \cdot 5 Myr \sim 0.0004~M_{\odot}$, 
little less than half the specific amount of dust produced by SNe, $m_{d,SNe}= 0.01 \cdot 0.1 = 0.001~M_{\odot}$, computed  assuming $0.1~M_\odot$ as
average dust mass produced per supernova \cite{cherchneff2009}.
This simple calculation indicates that the early assumption of neglecting WR stars as dust factories was probably an oversimplification.
WR as possible dust factories will be considered again later on, when discussing observations of dust-embedded young star clusters (Section ~\ref{sec_growthrole}).  

\subsubsection{Dust in Novae}

A “nova” (Latin for “new”) is a transient astronomical phenomenon that results in the rapid appearance of a bright, seemingly “new” star
that gradually fades over the course of weeks or months. 
Classical novae (CNe) are the most common types of novae, with 
recurrent novae and dwarf novae as other primary sub-classes. 
White dwarfs in close binary systems are present in all the observed novae.
In a nova system the white dwarf is close enough to its companion star to begin dragging accreted matter onto its surface,
creating a dense, shallow atmosphere, when the orbital period falls in the range of several days to one day.
The white dwarf thermally heats up this atmosphere, primarily made of hydrogen, until it reaches a threshold temperature that ignites rapid runaway fusion.
For this, CNe can be regarded as explosive contributors to 
the chemical enrichment of the ISM, although with an impact significantly less remarkable than SNe. 
The possibility of some sort of dust grain production in CNe was first suggested in the past,  
in connection with the earliest observations and interpretation of photometric studies of systems such as DQ Her 1934 
\cite{Payne-Gaposchkin1964,shore2018,bode1983}. 
Many novae eruptions show a remarkable evolution of their light curve, that develops roughly
three to four months after the outburst and is accompanied by a sharp decline at UV and visible wavelengths
and a quick rise at infrared wavelengths \cite{gehrz1988}. 
This light curve behavior, discovered in the '70 s, was interpreted as the proof that dust had formed in the nova ejecta \cite{geisel1970,clayton1976}. 
Bode \& Evans\cite{bode1981} argued that the rise in the IR light curve is the result
of the delayed re-radiation of the UV-optical light by pre-existing circumstellar dust (an infrared echo),
refuting the interpretation that it was due to dust newly produced by the nova.  
The dust production scenario is currently the most widely accepted explanation for the IR development in novae,
as the echo model fails to adequately explain a number of characteristics of nova evolution (\cite{gehrz1988} and references therein). 
Over the years, a number of speculative theories regarding the formation and development of the grains have been put forth.
Each one of them is dependent on a particular scenario, such as chemistry \cite{rawlings1988}, 
kinetic agglomeration and photoionisation processing \cite{shore2004}, and shock-driven chemistry \cite{derdzinski2017}.
All agree that for grains to grow, the matter must be sufficiently kinetically cold, but when this happens is up for debate,
as is the final fate of the produced dust. 
In the overall consensus, dust in novae is expected to form in the dense cool shell generated by radiative shocks \cite{chomiuk2021}. 
The optical indication of dust formation and/or growth in the post-maximum stage has been represented by DQ Her-like events, 
but its true fingerprint was traced by IR features such as continuum and polycyclic aromatic hydrocarbon emission
and the broad silicate-related emission at 10--20 $\upmu$m
(e.g., \cite{evans2012,shore2018}). 
When the ejecta become sufficiently transparent to allow UV and X-ray illumination by the 
hot white dwarf remnant to reach the part of the ejecta harboring the newly formed dust,
the decrease in the IR luminosity is typically interpreted as grain destruction. 

Dust masses produced in novae ejecta can range from
a few times $10^{-9}~M_{\odot}$ to $\sim$ a few  $10^{-7}~M_{\odot}$ \cite{gehrz1988,dwek1998,shore2018}. 
From a detailed estimate of the local nova rate and assuming that each nova produces  2$\times$10$^{-7}~M_{\odot}$ of dust,
whereas \cite{dwek1998} estimated a nova dust production rate of 3$\times$10$^{-6}~M_{\odot}~pc^{-2}~Gyr^{-1}$, i.e.,
lower than the SN production rate by several orders \linebreak of magnitude. 

In Table \ref{table_prod}, I present a list of the main stellar and non-stellar dust production sources,
  highlighting the distinct evolutionary stage, physical conditions, chemical products, and production timescales. 

  \captionsetup{width=\linewidth}
\begin{table}[H]

\caption{\textls[-15]{Summary of stellar and non-stellar dust-producing sources and their main features. 
  References: $^1$: Ferrarotti \& Gail (2006) \cite{ferrarotti2006}; $^2$: Ventura et al. (2012) \cite{ventura2012};
  $^3$: Sarangi et al. (2022) \cite{sarangi2022}; $^4$: Gall et al. (2018) \cite{gall2018}; $^5$: Gehrz et al. (1998) \cite{gehrz1998}; $^6$: Chong et al. (2005) \cite{chong2025};
  $^7$: Lau et al. (2020) \cite{lau2020a}; 
  \linebreak $^8$: Jim{\'e}nez-Hern{\'a}ndez et al. (2020) \cite{jimenezhernandez2020};  $^9$: Elvis et al. (2022) \cite{elvis2002}; $^{10}$: Asano et al. (2013) \cite{asano2013}.}}
   \tablesize{\footnotesize}
\centering
\renewcommand{\arraystretch}{1.4}
\tiny

\begin{adjustwidth}{-\extralength}{0cm}
\centering 

\

\begin{tabularx}{\fulllength}{m{2cm}<{\raggedright}m{2.5cm}<{\raggedright}m{2cm}<{\raggedright}m{3cm}<{\raggedright}m{2cm}<{\raggedright}m{2cm}<{\raggedright}m{2cm}<{\raggedright}}
  \toprule
  \textbf{Stellar Sources} &  &  &  & & \textbf{Non-Stellar Sources} \\
  \midrule
\textbf{Property} & \textbf{AGB Stars $^{\textbf{1,2}}$} & \textbf{Core-Collapse Supernovae $^{\textbf{3,4}}$} & \textbf{Novae $^{\textbf{5,6}}$} & \textbf{Wolf-Rayet Stars $^{\textbf{7,8}}$} & \textbf{QSOs $^{\textbf{9}}$} & \textbf{Accretion in MCs $^{\textbf{10}}$} \\
\midrule
Evolutionary Stage & Late stage of low/intermediate-mass stars (1–8 $M_\odot$) & Massive stars ($>$8 $M_\odot$) after core-collapse & Thermonuclear runaway on accreting white dwarf in binary system & Late He-burning phase of massive stars (especially WC subtype) & During the peak of the QSO luminosity & After Molecular cloud formation\\
\midrule
Main Dust Types & Amorphous carbon (C-rich), silicates (O-rich), SiC & Silicates, alumina, carbonaceous grains, Fe oxides & Amorphous carbon, silicates, alumina & Amorphous carbon (WC), possibly silicates & Carbonaceous dust & Ice Mantles\\
\midrule
Dust Production Efficiency & High & Moderate to High (uncertain due to reverse shocks) & Low to Moderate & Low to Moderate & Low (relative to the cumulative stellar production) & High\\
\midrule
Dust Mass per Event & $10^{-4}$ to $10^{-2}$ $M_\odot$ & $10^{-3}$ to 1 $M_\odot$ (pre-shock); \linebreak $\leq$10$^{-2}$ $M_\odot$ (post-shock) & $10^{-8}$ to $10^{-6}$ $M_\odot$ & $10^{-6}$ to $10^{-1}$ $M_\odot$ (per event) & Up to $10^7$ $M_\odot$ & N/A-(Of secondary origin)\\
\midrule
Environment/ Conditions & Cool, dense winds (1000–1500 K) & Expanding ejecta with rapid cooling; subject to reverse shocks & Transient, short-lived mass ejection & Hot, dense winds; dust often forms in WC+O binaries &  Adiabatically expanding Broad Emission
Line regions & Cold ($T<50$ K) and dense $n$>10$^3$ cm$^{-3}$ gas\\
\midrule
Formation Timescale & $10^3$–$10^5$ years & Weeks to years & Weeks to months & Episodic; months to years &  A few years (cloud expansion timescale) & Molecular cloud lifetime \mbox{($<10^7$ yr)} \\
\bottomrule
\end{tabularx}
\end{adjustwidth}
\label{table_prod}

\end{table}

{\subsection{Dust Shattering, Coagulation and Grain-Size Evolution}
\label{sec_shat} 
Grain-size evolution plays a fundamental role in regulating the lifecycle of interstellar dust, 
as many key physical processes depend explicitly on grain radius and surface area. 
In particular, the susceptibility of grains to destruction in shocks, and their optical properties are all strongly size dependent. 
A central quantity underlying all grain-size–dependent studies is the grain size distribution (GSD), 
usually defined as the number of grains per unit size interval $dn/da$, where 
$a$ is the grain radius. In the diffuse interstellar medium of the Milky Way, the GSD is commonly approximated by a power-law distribution,

\begin{equation}
  \frac{dn}{da}\propto a^{-3.5}
\label{eq_mrn}  
\end{equation}
over the range $a \sim$ 0.005--0.25  $\upmu$m, also known as the Mathis–Rumpl–Nordsieck (MRN) distribution \cite{Mathis1977}.
This functional form successfully reproduces Galactic extinction curves and has therefore become the standard reference model in
both analytical studies and numerical simulations. More recent models introduce refinements to the MRN prescription,
including broken power laws, log-normal components at small sizes, or composition-dependent distributions,
in order to match observations of infrared emission and extinction in different environments (e.g., Weingartner 2001 \cite{weingartner2001a}). 
In the following, we discuss in more detail how the various physical processes acting in different ISM phases 
contribute to shaping the grain size distribution, and which of them dominate its evolution under typical galactic conditions. 

Stellar sources predominantly inject large grains into the ISM, with typical sizes \mbox{$\gtrsim$0.05 $\upmu$m}. 
This is supported by dust condensation models in AGB winds \cite{ferrarotti2006,nanni2013} 
and by the preferential survival of large grains in supernova ejecta \cite{nozawa2007,hirashita2015}. 
The abundant small grains observed in galaxies are thought to arise mainly from later processes occurring in the ISM.
One of the dominant mechanisms shaping the grain-size distribution in the diffuse ISM is grain shattering,
driven by high-velocity (typically $v \gtrsim 10$ km/s, \cite{hirashita2015}) grain–grain collisions induced by turbulence and supernova shocks. 
Shattering leads to the fragmentation of grains, transferring mass from large grains to smaller fragments,
increasing the total grain surface area and thereby enhancing the efficiency of gas-phase metal accretion and molecular hydrogen formation.
Conversely, in dense and cold environments, the collisions between grains occur at lower speeds, promoting the sticking of grains into large particles \cite{hirashita2009}.  
This process, commonly referred to as grain coagulation, 
reduces the abundance of small grains modifying other galactic properties, such as dust extinction and emission.  
These competing processes naturally produce spatial and temporal variations in the grain-size distribution across different ISM phases. 
Therefore, the models including a basic treatment of the grain size evolution require the inclusion of 
 shattering in combination with coagulation. 

To account for these effects while remaining computationally feasible, several studies have introduced grain-size–resolved dust models,
either by discretising the grain-size distribution into multiple bins or by adopting simplified schemes such as the two-size approximation,
where dust is separated into representative small and large grain \linebreak populations \cite{hirashita2015,aoyama2017,gjergo2018}.
One motivation for such an approach is the general knowledge that in some models, the evolved full-size distribution presents
two separate peaks, with a boundary located approximately at $a \sim$ 0.02 $\upmu$m \cite{asano2013b}. 
This approach allows simulations and chemical evolution models to follow size-dependent processes such as shattering, coagulation,
and sputtering, and to link grain evolution more directly to observable quantities like extinction curves and infrared emissivities. 

The simplest models adopt a single shattering timescale $\tau_{sh}$ inversely proportional to the large grains dust-to-gas ratio $D_{l}$
(see, e.g., \cite{hirashita2015}) 
\begin{equation} 
\tau_{sh} = \tau_{sh,0} \frac{D_{MW,l}}{D_{l}}, 
\end{equation}
where $\tau_{sh,0}$ is a normalization value and $D_{MW,l}$ is the large-grain dust-to-gas ratio in the Milky Way,
which can be computed from a GSD such as the MRN of Equation~(\ref{eq_mrn}), 
considering a given size range and a transition value between the two regimes.
As for the value of $\tau_{sh,0}$, a fiducial value of $10^8$ yr is sometimes assumed \cite{hirashita2015};
however, large variations of this parameter are expected depending on the ISM density and 
the ISM phase filling factor and mass fraction, 
which makes it recommendable to consider a reasonable range, with an order of magnitude variation or similar 
\cite{hirashita2015,aoyama2017}. 

In a similar way, it is possible to define a single coagulation timescale of the small grains into large grains:
\begin{equation} 
\tau_{co} = \tau_{co,0} \frac{D_{MW,s}}{D_{s}}, 
\end{equation}
which includes the normalization $\tau_{co,0}$, as uncertain as $\tau_{sh,0}$ and assumed to vary considerably
in different models, from $10^5$ yr  \cite{aoyama2017,gjergo2018} to $10^7$ yr \cite{hirashita2015}.
In this case, $D_{s}$ is the small grains  dust-to-gas ratio, with the total dust-to-gas ratio $D$ defined as
$D= D_{l}+D_{s}$. 
$D_{MW,s}$ is the small-grain dust-to-gas ratio in the Milky Way, defined in a similar way as above, computed
from the MRN distribution and with a total Milky Way dust-to-gas ratio defined as $D_{MW} = D_{MW,s}+D_{MW,l} = 0.01$. 

Other treatments prefer a density-dependent parametrisation of the shattering timescale as follows 


\begin{equation}
  \tau_{sh}=\begin{cases}
  \tau_{sh,0} \left( \frac{D_{MW,l}}{D_{l}} \right) \left( \frac{1~cm^{-3}}{n_{gas}} \right),~{\rm if}~n_{gas} < n_{thr} \\
   \infty ~~~~{\rm if}~n_{gas} \ge  n_{thr}, 
  \end{cases}
\end{equation}
with the adoption of a suitable density threshold $n_{thr}$ below which shattering can \linebreak occur \cite{aoyama2017}
and for which a value of 1~cm$^{-3}$ is a good choice, considering that shattering tends to occur preferentially in
the diffuse medium \cite{hirashita2015,gjergo2018}.

It is also possible to capture the coagulation of small grains in dense clouds considering a crude dynamical turbulence model, expressed as  
\begin{equation}
  \tau_{co}=\tau_{co,0} \left( \frac{D_{MW,s}}{D_{s}} \right) \left( \frac{0.1~\text{km~s}^{-1}}{v_{co}} \right) \frac{1}{f_{dense}}, 
\end{equation}
see \cite{aoyama2017,granato2021}. In this case, the relevant parameters are the velocity dispersion of small grains $v_{co}$
(typically assumed of the order of 0.1~km~s$^{-1}$) and the dense gas fraction $f_{dense}$, for which a fiducial value of 0.5 can be assumed
in multiphase models (e.g, \cite{aoyama2017}), in which most of the times the cold gas is unresolved \cite{granato2021}.

A different process that can lead to significant variation of grain size and mass is the dust accretion, occurring  in dense clouds  
and discussed later in Section ~\ref{sec_accr}.   
Numerical simulations can account for the multiphase gas components and therefore they represent ideal tools to study grain size
evolution (Section \ref{sec_simulations}}). 
}

\subsection{Dust Accretion in the Cold ISM}
\label{sec_accr}
In the process of dust accretion or growth, refractory elements can condensate onto pre-existing grain cores, originating a volatile  
part called mantle \cite{dwek1998,inoue2003,calura2008,gall2011}. The sites where this process occur are dense molecular clouds, 
where it is favoured by the large particle density and the significant presence of molecules which are main constituents of the mantle 
\cite{ceccarelli2018}. 
The most direct evidence of dust accretion in molecular clouds comes from the  
large variations of the depletion levels as a function of column density \cite{savage1996}. 
Indirect evidence for accretion can be obtained from the average estimate of the grain lifetimes,
too small to justify the substantial presence of interstellar dust if one considers stardust production and destruction only 
\cite{mckee1989,draine1979,dwek1998,calura2008}. 
In the current knowledge, an important physical and chemical distinction needs to be done between dust cores and mantles. 
The core is more immune to various grain destruction processes operating in the ISM, whereas the mantle, 
that represents most of the dust mass, consists of more loosely bound material, and is therefore more prone to destruction. 

It is worth noting that the processes of shattering and growth are interconnected,
  as the former tends to increase the abundance of small grains; since growth is more effective on small-sized particles,
  the overall effect is to boost up the grain growth process \cite{parente2025}.

D98 \cite{dwek1998} wrote the rate at which the dust grains grow by accretion of metals for the element $A$ as: 
\begin{equation}
\frac{dN_{\rm A}}{dt}= \alpha \pi a^2 n (A) n_{\rm gr} \overline{\rm v},
\label{eq_rate}
\end{equation}
where  $\alpha$ is the sticking coefficient of the element $A$ to the grain,  $n$ and
$n_{\rm gr}$ are the number density of gas particles and dust grains, respectively, and 
$\overline{\rm v} = \frac{8 k T}{\pi m_{\rm A}}$ is the mean thermal speed of atoms with mass $m_{\rm A}$ in a gas 
with temperature $T$. 

Starting from Equation (\ref{eq_rate}), D98 showed that the accretion rate can be written (using the notations of Equation (\ref{eq_dust})) as 
\begin{equation}
\tau_{accr}=\tau_{0,i}/(1 - f_i) 
\label{accr_t}
\end{equation} 
where 
\begin{equation}
f_i=\frac{G_{dust,i}}{G_{i}}
\end{equation} 

In Equation~(\ref{accr_t}), the accretion timescale is an increasing function of the dust mass. 
Typical values for the timescale $\tau_{0,i}$ are of the order of 
the lifetime of a typical molecular cloud \linebreak ($\simeq$10$^{7}$ yr). 
An important test for dust evolution models is the local elemental
depletion pattern, expressed for each element as the ratio between the gaseous abundance measured in the ISM and the `cosmic', reference value,
namely the solar abundance or the one measured in local B stars, assumed to represent the intrinsic, undepleted  abundances \cite{kimura2003}.

\begin{figure}[H]
\centering

\includegraphics[width=14 cm]{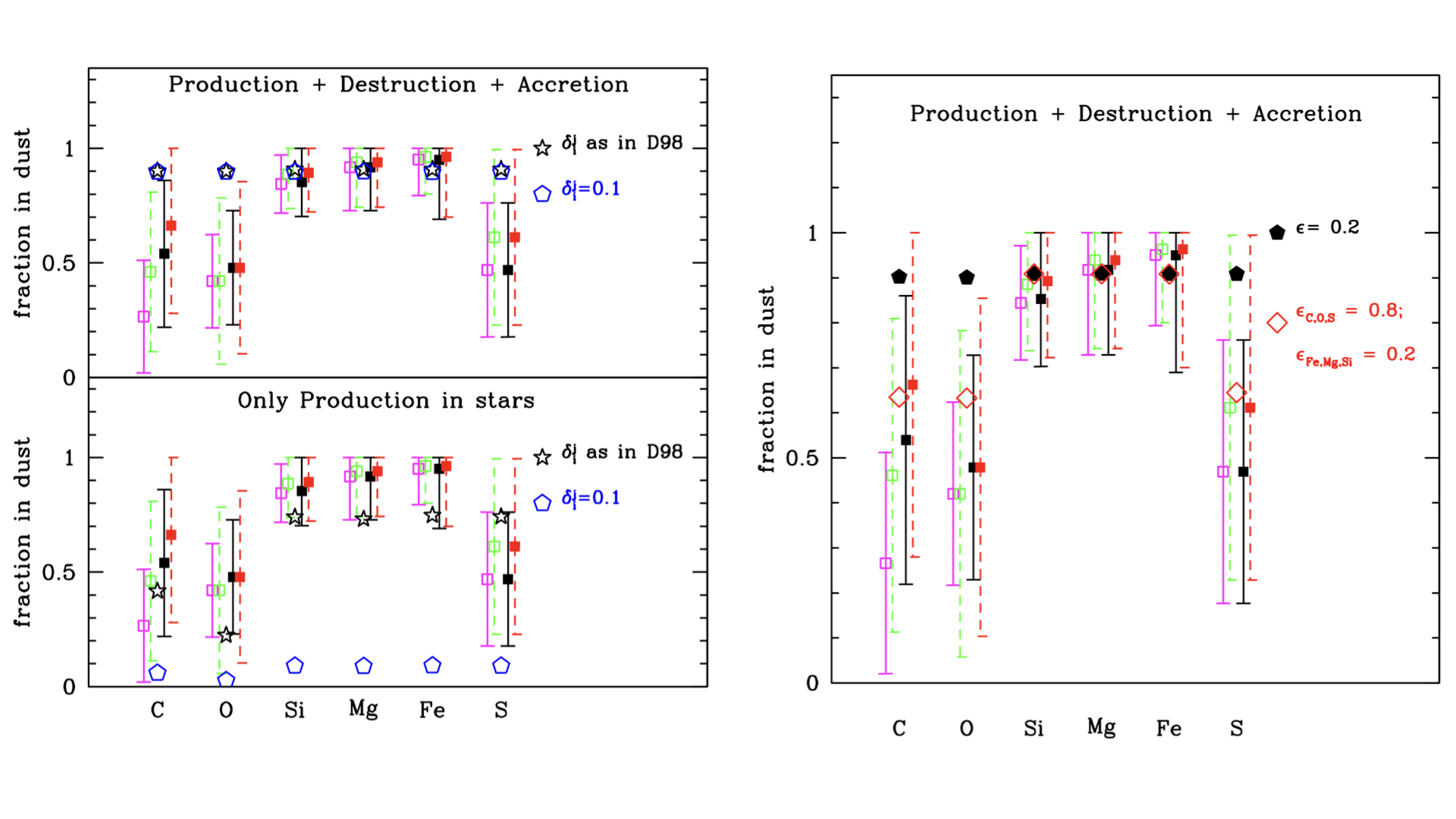}
\caption{Fractions in dust for various elements. In the top-left panel,
the solid and open squares are the fractions observed by Kimura et al. (2003) \cite{kimura2003} in the Local
Interstellar Cloud using the set of cosmic abundances specified in their Tables 2 and 3, respectively and assuming different values for the  H ionisation fraction (for more details of the adopted prescriptions, we refer the reader to \citet{kimura2003} and references therein). The squares with solid and dashed error
bars have been calculated assuming a H ionization fraction of 0.25 and 0.45, respectively. 
The open stars and pentagons are the predicted present-day fractions of Calura et al. (2008) \cite{calura2008} calculated by adopting the prescriptions suggested
by D98 for the dust condensation efficiencies and assuming a constant value
of 0.1 for the dust condensation efficiencies, respectively. 
The predicted values are calculated considering dust production in stars, dust destruction by SNe, and dust accretion in the ISM.
In the bottom-left panel, the predicted values are 
calculated considering only dust production in stars, while the other symbols are as in the top-left panel. 
In the right panel, the solid pentagons are 
the predicted present-day fractions calculated by assuming a grain destruction efficiency $\epsilon$ = 0.2 for all elements.
The open diamonds are the present-day fractions calculated assuming a grain destruction efficiency $\epsilon$ = 0.8
for C, O, and S and $\epsilon$ = 0.2 for Fe, Si, and Mg, while the other symbols are as in the top-left panel. 
Figure adapted from Calura et al. (2008) \cite{calura2008}.}
\label{dep_pat}
\end{figure}

Chemical evolution models have been used to reproduce the local depletion pattern in the solar neighborhood \cite{dwek1998}. 
Using a chemical evolution model that included stardust production, dust destruction, and accretion, Calura \cite{calura2008} 
aimed to replicate the observed dust depletion patterns for various refractory elements. 
The elemental depletion pattern for various elements, expressed as the mass fractions in dust with respect to
the total (i.e., in both gas and dust) abundance, for  various assumptions about dust production and destruction in the solar neighbourhood is shown in Figure ~\ref{dep_pat}.

With the standard prescriptions of D98, in a complete model which includes stardust production, destruction and accretion, 
the observed pattern depends mainly on the balance between growth and destruction, with very little dependence on the 
stellar condensation fractions. 
In fact, in such conditions, D98 obtained a dust abundance pattern which does not depend on the considered elements,
characterised by a $\sim$ 0.5 fraction in dust for all elements (top-left panel of Figure ~\ref{dep_pat}). This result 
is at variance with respect to the observations, which show 
depletion values $>0.9$ for the most refractory species (e.g. Si, Fe) and
fractions between 0.36 and 0.6 for more volatile elements, such as C, O.
In order to account for the observed pattern, D98 noticed that it was necessary to have
a dust depletion pattern shaped by stardust production or, equivalently, grain destruction had to be
exactly balanced by growth in molecular clouds. 
Clearly, a scenario which includes destruction but in which the effects of growth are negligible
is disfavoured, as it gives place to too low depletion fractions ($\lesssim$0.1) for all elements (D98).
The observed dust fractions can be reproduced by assuming either a different dust accretion timescale or destruction efficiency for different 
chemical elements. A physical justification can be ascribed to the different condensation temperatures of these elements \cite{lodders2003}.

Asano et al. \cite{asano2013} investigated the role of interstellar dust growth using a closed-box chemical evolution model.
Besides accretion, this model included factors such as stardust production and destruction by SN shocks.
In their study, they emphasized the significance of interstellar metallicity in determining the dust accretion timescale.
Specifically, they identified a critical metallicity value that decreases with decreasing star formation efficiency.
Above this critical value, accretion becomes the primary process regulating the galactic dust budget,
which also leads to a significant depletion of heavy elements from the \linebreak gas phase.

Starting again from Equation (\ref{eq_rate}), Asano et al. \cite{asano2013} assumed that
the dust mass accretion rate is proportional to the gas mass in the form of metals $\rho_Z^{gas}$ that
are not already contained in dust grains. 
This quantity is directly connected to the total ISM density $\rho_{ISM}$ and the metallicity $Z$ via
the formula 
\begin{equation}
  \rho_Z^{gas} =   \rho_{ISM}~Z~(1-\delta),
\end{equation}
  where $\delta$ is the dust mass fraction in the form of metals.  

The characteristic timescale $\tau_g$ can be expressed as a function
of the grain size $a$, the sticking coefficient of refractory elements $\alpha$, and the mean velocity of metals in the gas,
denoted as $\langle v \rangle$ as follows: 
\begin{equation}
\tau_g = \frac{4 \langle a\rangle^3 \rho_{\rm d}}{3 \langle a\rangle^2 \alpha \rho_{\rm ISM} Z \langle v\rangle}.
\label{eq_tg} 
\end{equation}
\indent In Equation (\ref{eq_tg}), $\langle a \rangle ^2$ and $\langle a \rangle^3$ are the second and third moment of the grain size distribution, respectively, 
whereas $\rho_{\rm d}$ and  $\rho_{\rm ISM}$ are the average dust and ISM mass densities. 
Assuming for the typical molecular cloud temperature a value of 50 K, 
100 cm$^{-3}$ for the cloud ambient density and an average value of 0.1 $\upmu$m for the grain
size, ref. \cite{asano2013} \mbox{expressed $\tau_g$ as:}
\begin{equation}
\tau_g =  2.0\cdot 10^7 \times \bigg( \frac{Z}{0.02} \bigg)^{-1} \, [{\rm yr}],
\label{eq_asa_tg} 
\end{equation}

\textls[-15]{Several works discussed accretion in dust evolution models,
highlighting its fundamental role in regulating the dust budget in galaxies \cite{dwek1998,calura2008,gall2011,valiante2011,mattsson2014}.
The different ways to include accretion in models express the considerable complexity and our basically poor knowledge of
this process. 
The results of several models outline in particular the significant degeneracy of the prescriptions that,
despite the different formalism, can in several cases provide quite similar results \cite{calura2023}.
The model results often converge towards the indication that the accretion term, if present,
has to be roughly balanced with the one of destruction \cite{gall2018}.
Essentially, this means that they can be both present and they can either dominate over the
stardust term, or they can be both negligible.
At the present time, it is very difficult to determine which one of these two possibilities is
the most realistic \cite{calura2023}. 
A collection of further arguments supporting or refuting accretion will be presented later in Section~\ref{sec_growthrole} ,
along with a more detailed and extended discussion of its role in shaping the dust mass in galaxies.}

\subsection{Dust Destruction in SN Shocks} 
\label{sec_destr}

Supernova shock waves propagating in the heated and ionised interstellar medium are the main cause for dust destruction
\cite{mckee1989,jones1994}. 
Dust destruction can occur in the shocked warm medium with T $\sim$ 10$^4$ K and density $\sim$0.1 cm$^{-3}$ driven by SN explosions, predominantly 
through collisions with high-velocity ions ($\ge$100 km/s). 
Also gas-grain collisions may occur, sometimes referred to as “sputtering”, that may return part of the refractory material to the gas phase \cite{tielens1994,jurac1998}. 
In the hot gas of supernova remnants, an important distinction is made between thermal sputtering, in which grain erosion depends
  on the gas temperature, and non-thermal (inertial) sputtering, where the erosion rate depends on the relative velocity between the gas and the dust grains \cite{hu2019}. 

Early observations of high-velocity clouds that demonstrate an inverse relation
between depletion levels and cloud velocities \cite{shull1978,mckee1989} 
provided weight to this concept and revealed that grain destruction occurs primarily in supernova shocks. 

In line with the suggestion of McKee\cite{mckee1989} and D98, 
the destruction timescale for a given element, represented by $\tau_{destr, i}$ in Equation~(\ref{eq_dust}) can be expressed in chemical evolution models as: 
\begin{equation} 
  \tau_{destr, i}=(\epsilon M_{SNR})^{-1} \cdot \frac{\sigma_{gas}}{R_{SN}},
\label{eq_taud}
\end{equation}
where $M_{SNR}$ is the mass of interstellar gas that is swept up by a SNR,
$\sigma_{gas}$ is the ISM surface density and $R_{SN}$ is the total SN rate, i.e., that take into account Type Ia and CC SNe.  
One crucial assumption here concerns $M_{SNR}$. 
Several works were devoted to the assessment of this quantity, both theoretical and observational.
Based on theoretical arguments, in idealised conditions 
the amount of ISM mass swept up  by an isolated SNR is expected to be weakly dependent on the 
ambient density, as a consequence of momentum conservation during its evolution \cite{cioffi1991,dwek1998}.  
Numerical simulations of the evolution of SNRs showed that this quantity should be of the order of several $10^3~M_{\odot}$ \cite{thornton1998}.
On the observational side, \linebreak  Temim et al. \cite{temim2015} analysed more than 80 SNRs in the Magellanic Clouds, 
in which the destroyed dust masses were estimated by means of multi-wavelength observations. 
In particular, X-ray measurements were used to calculate the diameters of the SNR, while IR spectra obtained with Herschel 
allowed for an estimation of the dust mass, using a two-component SED fitting that takes into account carbon and silicate dust. 
The results of Temim et al. \cite{temim2015} support a weak dependence on ISM density for the overall effective swept-up mass of dust in SNRs.
For both dust types, the derived swept-up mass values are typically of the order of $\sim$2000~$M_{\odot}$.

Motivated by extant studies, various assumptions have been made for $M_{SNR}$ in models of dust evolution.
For this quantity, in some models a constant value of the order of a few $\sim$1000--1300~$M_{\odot}$ has been assumed \cite{dwek1998,calura2008},
consistent with the suggestion of \cite{mckee1989}, who computed a total mass of shocked material of $ 6800~M_{\odot}$ in a 
multiphase ISM. 

In Equation~(\ref{eq_taud}), the ISM mass cleared out of dust by a SNR is simplified and is given by the product between the grain destruction efficiency $\epsilon$ and
the swept-up mass $M_{SNR}$, both assumed as constants. 
However, the calculation of the effective swept-up dust mass can be more complex, e.g., in the case where $M_{SNR}$ and $\epsilon$ are functions
of other parameters, such as the shock velocity \cite{temim2015} or density.
For shock velocities of the order of \linebreak $\sim$1000 \text{km s}$^{-1}$, $\epsilon$ can be assumed of the order of 1 \cite{temim2015}. 
In a two- or three-phase ISM, McKee \cite{mckee1989} estimate for the dust destruction efficiency $\epsilon=0.2$, so that $\epsilon M_{SNR}\sim 1000~M_{\odot}$. 

Other authors have assumed a more complex behaviour for the effective ISM swept-up mass of dust  $\epsilon M_{SNR}$. 
In the formalism of Asano et al. \cite{asano2013}, this quantity depends on the interstellar metallicity $Z$ and  ISM density  $n_{\rm ISM}$ as:

\begin{equation}
   \epsilon M_{SNR} = 1535 n_{\rm ISM}^{-0.2} \cdot (Z/Z_\odot + 0.039) ^{-0.289}\,[{\rm M}_\odot],
\label{eq_mclear_asa13}   
\end{equation} 

The notion behind a metal-dependent swept-up mass is that metal-poor gas cools less effectively and stays at high temperatures for longer.
In principle, this increases the amount of dust destroyed by thermal sputtering in shocked gas with low metal content \cite{yamasawa2011}. 
Note that in this case, for typical ISM density values in the range $0.1 \le n_{ISM}/(\text{cm}^{-3}) \le 1$ and metallicity <$Z_{\odot}$, 
the swept mass is always above the constant value of $1300~M_{\odot}$, as assumed in \cite{calura2008}.

Priestley et al. \cite{priestley2022} envisage a stronger dependence on the metallicity, expressing 
the effective mass cleared of dust as:
\begin{equation}
    \epsilon M_{SNR} =\frac{8143}{1 + Z/(0.14 Z_\odot)} [{\rm M}_\odot], 
    \label{eq:Priestley}
\end{equation}

Equation~(\ref{eq:Priestley}) might overestimate the metallicity dependence of $\epsilon M_{SNR}$ since it neglects kinetic sputtering.
Other studies \cite{yamasawa2011} that include this process find a weaker dependence, as in Equation~(\ref{eq_mclear_asa13}). 
However, Priestley et al. \cite{priestley2022} argue that a complete treatment of destruction that includes a multitude of processes should obtain
a scaling similar to the one of Equation (\ref{eq:Priestley}). 

Besides the dependence on local properties of $\epsilon M_{SNR}$ postulated by some authors, the dust destruction timescale
of Equation~(\ref{eq_taud}) is expected to be a function of the environment because of its dependence on the total SN rate, which will
depend on the star formation history. 
In the solar neighbourhood, the global destruction timescale $\tau_{destr}$ is expected to vary between $\sim $0.5 Gyr and 100 Gyr across
its evolutionary history, whereas it varies in a narrower range in the Galactic center, i.e.,
between $\sim $0.2 Gyr and 1 Gyr \cite{dwek1998}. 

Several studies agree on the fact that the calculated dust destruction rates are at least one orders of magnitude higher
than the dust injection rates from AGB stars and SNe, 
suggesting that other process may be significant in regulating the dust budget, as indicated also by the current IR emission in
the Magellanic Clouds and other local\linebreak  galaxies \cite{dwek1998,calura2008,gall2011,temim2015}, such as dust accretion (Section \ref{sec_accr}). 

\subsection{Dust Destruction or Removal by Other Processes (Astration and Winds)} 

Other processes can contribute to the grain destruction.
In the astration process, refractory material is removed from the ISM to be incorporated into newborn stars.
Because of its nature and the complicated interplay between dust condensation and removal in star-forming regions,
which, in the local Universe, tend to be dust-rich environments, the role of astration cannot be assessed by means of observations.
On the other hand, this topic has been addressed by several theoretical studies, in which it is 
possible to perform controlled numerical tests where each process can be directly quantified. 

D98 \cite{dwek1998} compared the evolution of dust in two separate regions of the MW disc, characterised 
by different SFHs, i.e., the Solar Circle and the Galactic Centre.  
The latter presents a surface gas density and, therefore, a SF activity significantly higher (by a factor $\sim$10) than the latter.
Their study showed that the relative roles of astration and destruction by SN shocks is a function of 
time and of the star formation history (SFH), 
with the former dominating the global destruction at the earliest epochs, soon after the beginning of SF. 
As time progresses and the SF activity decreases, the role of astration becomes comparable to
the one of destruction by SN shocks due both to type Ia and CC SNe.  

These results were confirmed by Calura \cite{calura2017}, who  studied the relative roles of such processes
in models for proto-spheroids. Also in such systems, at the earliest times, when SF is particularly intense, astration dominates the
global destruction rate. 

The role of different processes in the evolution of interstellar dust has been studied 
by means of hydrodynamical simulations. With these tools, dust evolution can be modelled physically and
in a self-consistent fashion and within a cosmological framework \cite{Mckinnon2017,aoyama2018}.
With these tools, basic mechanisms such as star formation, stellar feedback and dust production are generally modelled
via sub-grid recipes that often depend on a set of parameters.
Often, this approach is useful to perform controlled numerical experiments, sometimes aimed at describing isolated
systems, where the role of single processes can be singled out by means of a switch, i.e., by turning it on/off in separated
runs. 
By exploiting these methods, a few of them have been devoted to study in detail the role of astration. 

At variance with other computational studies on dust evolution in galaxies, in the study of McKinnon 
\cite{mckinnon2018} dust is not produced with the assumption 
that it is fully dynamically connected to gas. Their galaxy simulations, however, demonstrate that the `drag' force term
between the dust and gas components may prove to be strong enough to enable their effective coupling.  
Within their detailed, moving-mesh simulation set, \mbox{McKinnon et al. \cite{mckinnon2018}} suggested that the global dust destruction rate is 
dominated by SN shocks and astration plays a negligible role. 

\textls[-15]{With a different approach based on smooth particle hydrodynamics, Granato et al. \cite{granato2021}
run tests cases in which their particles retained their full dust content
when spawning new stars, i.e., ignoring the role of astration, and others in which SN destruction was suppressed.
Their tests indicated that both astration 
and destruction by SNe are equally important and that their roles are comparable, showing that 
ignoring either astration or SN destruction results in about a  50\% higher dust mass at the present time.
This may have consequences on the predicted Si/C ratio in the dust as a consequence of the fact that 
when metals are destroyed by SNe they are entirely returned into the ISM and ready to be incorporated into new dust by accretion,}
but the same does not hold when refractory elements are astrated. 
It is also important to note that other ingredients may play a fundamental role in the different conclusions of  \cite{granato2021} and 
\cite{mckinnon2018}, such as a more or less top-heavy stellar IMF or slight differences in the implementation of
CC SN explosion, such as the lower mass limit for their progenitors.  
Other results indicate that SN destruction is not enough to explain the decline of the comoving dust density
observed at redshift $z<1$ \cite{aoyama2018}. 

Dust can be removed from the ISM also through the development of galactic winds (see Equation~(\ref{eq_chem})).
Galactic winds play an important role in galaxy evolution in various phases of their history. 
Intense stellar feedback resulting from strong star formation activity or active galactic nuclei (AGN) feedback 
can produce massive galactic outflows  
in which the metal-enriched star-forming gas can be ejected into the circumgalactic or inter-galactic medium. 
Infrared observations of local systems showed that galactic outflows contain significant amounts of dust \cite{veilleux2005}.
Filamentary cold dust emission at high galactic latitude is observed in some local galaxies experiencing a galactic wind
(NGC 1808, \cite{phillips1993}; \linebreak M82 \cite{ichikawa1994} and other more recent cases observed with Herschel \cite{melendez2015}). 
Some of these studies are relevant also to investigate the presence of circumgalactic dust \cite{farber2022}, 
observed in a few cases around nearby dwarf galaxies and of unclear origin \cite{mccormick2018}. 

The ejection of dust caused by the galactic wind is the most obvious mechanism to explain the presence
of dust at large distances from the galaxy and/or in galactic halos.  
These environments, and the IGM in general, do not offer ideal conditions for the 
growth and survival of dust because of their typically low densities and high\linebreak  temperatures \cite{bianchi2005,farber2022}. 
Galactic winds are particularly efficient in galaxies with active star formation and/or active galactic nuclei \cite{kaneda2010,mccormick2013} but, 
in some cases, extra-planar or high-latitude dust is also observed in galaxies with low star formation \cite{Irwin2006}. 
Considering also that the spatial distributions of dust and gas in galactic halos may differ, 
the wind may only be one of the processes that sweeps dust out of a galaxy and stellar radiation pressure 
may represent an alternative mechanism \cite{Sivkova2021}. 

In chemical evolution models of spiral galaxies, galactic winds are generally neglected~ \cite{chiappini2001,calura2010,palla2020,calura2023}. 
This is based mostly on the knowledge that in massive discs the gas ejected by OB associations is likely to remain within the dark matter (DM) halo
gravitational potential well \cite{lapi2020} and tends to fall back in the form of galactic \linebreak fountains \cite{bregman1980,spitoni2009}. 

By means of chemical evolution models for galaxies of different morphological types, Calura et al. \cite{calura2008} investigated the amount of dust removed via galactic winds in elliptical and dwarf irregulars.  
In chemical evolution models for elliptical galaxies, an intense starburst triggers a strong galactic wind as soon as 
the energy of the feedback-driven ISM exceeds its binding energy \cite{matteucci1994}.
In such models, the mass of dust lost in the wind is estimated through\linebreak  the formula
\begin{equation}
{\Delta M \over M_{\rm ISM}}= {\Delta M_{dust} + \Delta M_{gas} \over M_{\rm dust}+M_{\rm gas}} =
{\Delta E \over E_{\rm th}}, 
\label{eq_dust_esc}
\end{equation}
in which $\Delta \, M$ is the total mass ejected through the wind, $M_{\rm ISM}$ is the total mass of the ISM,
$M_{dust}$  and $M_{gas}$ are the total dust and gas mass in the ISM, respectively,
whereas $\Delta E $ is the difference between the thermal energy $E_{\rm th}$ and the gas binding energy.

The fiducial model of Calura \cite{calura2008} describes and elliptical galaxy of
$10^{11}~M_{\odot}$ stellar mass and can account for the dust mass observed in local early types,
typically of \linebreak $\lesssim$10$^6~M_{\odot}$ \cite{kaneda2007,panuzzo2007,smith2012b}.
In this model, a cumulative ejected dust mass of $5 \times 10^8~M_{\odot}$ was obtained, comparable
to the characteristic dust mass at the break of the dust mass function obtained at high redshift
in SCUBA galaxies \cite{dunne2003}. 

In a similar way, Calura et al. \cite{calura2008} estimated the dust ejected in galactic winds in irregular galaxies, showing that 
this quantity is sensitive to the assumed SFH. They considered a dwarf 
irregular with two significantly different SFHs, i.e., a low-level, continuous SF and a starburst, both suited to account for 
the properties of local, low-mass galaxies \cite{recchi2008,calura2008a}. 
They found that large differences are expected for the features of the winds 
in these two systems, both regarding the epochs at which the wind develops and the total amount of ejected material. 
In the case of a continuous SF and a starburst, the ejected dust mass is $2 \times 10^5~M_{\odot}$ and  $10^4~M_{\odot}$, respectively. 

By means of hydrodynamical simulations, Kannan et al. \cite{kannan2021} outlined the importance of accounting for the multi-phase
nature of the ISM to study realistically the entrainment of dust in galactic winds. 
Their results suggest that the amount of dust entrained in the wind shows a strong dependence on its temperature.
Moreover, the amount of ejected dust depends on the location from which the winds are launched. 
Compared to the hot and tenuous material, characterised by a large filling factor,
the outflowing cold and dense gas is substantially more dust-enriched. 

{
\subsection{Grain Charge Effects}
\label{sec_charge}
Interstellar dust grains can be electrically charged, as a consequence of their interaction
with the surrounding plasma and radiation field. Grain charging arises primarily
through photoelectric emission \cite{weingartner2001b},
induced by ultraviolet and optical photons, and through collisional charging due to impacts with electrons and ions \cite{draine1987}. 
In particular, ``sticking'' collisions with ions and electrons charge the grains positively or negatively, respectively. 
In general, given their lighter weight, 
collisions with electrons are more frequent, resulting in negatively charged grains \cite{ibanezmejia2019}. 
Photoelectrons ejected from dust grains may also be one important heating mechanism in both the warm and
cold phases of ISM \cite{weingartner2001b}. 
These processes are heavily dependent on the local properties of the ISM, such as temperature and density, 
affected in turn by the stellar radiation field \cite{draine1979} and the cosmic ray ionisation rate \cite{padovani2018}.

Grain charge has important implications for dust dynamics.
Dust grains moving through the interstellar medium are always subject to an aerodynamic drag force,
which depends on the 
relative velocity between dust and gas, on grain size, gas density and temperature, dominating dust dynamics in dense, weakly ionised environments.
When grains are electrically charged, however, additional forces come into play that can substantially modify their behaviour. 
In magnetised plasmas, charged grains experience the Lorentz force, coupling their motion to magnetic field lines \cite{lee2017}. 
Electric fields generate a Coulomb force that can alter grain trajectories in shocks and ionised regions where charge separation occurs.
Grain charge also affects dust–radiation interactions indirectly. 
The photoelectric emission is influenced by the charge, as the {\it photoelectric yield}  
(i.e., the probability that an absorbed photon ejects an electron) depends on the grain charge potential.
It is worth noting that positively charged grains have a higher potential barrier which tends to suppress electron escape; 
conversely, negatively charged grains have enhanced yields.   
Furthermore, electrostatic interactions between charged grains and impinging ions modify sputtering efficiencies through Coulomb
focusing or shielding, 
leading to charge-dependent destruction rates in hot and shocked gas \cite{hu2019}. 
Altogether, these effects imply that grain charging can significantly influence the motion of dust grains, their 
survival, and coupling to feedback processes in ionised environments. 
In these environments, the equilibrium charge depends on various parameters that include grain size,
composition, local radiation intensity, gas temperature, and ionisation state \cite{ibanezmejia2019}. 
The charge of the grains has a important role in determining the conditions in which the dust is spatially coupled 
with the gas. 
In fact, when dust grains are charged, they gyrate around the interstellar magnetic field 
with a Larmor radius that depends on the grain mass $m_{gr}$, on the relative velocity between gas and dust $v_{rel}$, the grain charge $z_{gr}$
and the magnetic field strength $B$ as 
\begin{equation}
r_L = \frac{m_{gr}~v_{rel}}{z_{gr}~e~B}, 
\end{equation}
where $e$ is the electron charge, $m_{gr} \simeq 4/3 \pi a^3 \rho_{gr}$ and with $\rho_{gr} \sim 3$ g cm$^{-3}$ \cite{hu2019}
The corresponding Larmor (gyro) timescale can be defined as $t_{\rm L} = 2\pi/\omega_{\rm L}$,
where \linebreak $\omega_{\rm L} = |z_{gr}| e B / m_{gr}$ is the grain gyrofrequency \cite{yan2004}. 
The drag (or stopping) time of a grain moving through a particle distribution and due to collisions with atoms is 
\begin{equation}
t_{\rm drag}=\frac{a\,\rho_{\rm gr}}{n_n}
\left(\frac{\pi}{8\,m_n k_{\rm B}T}\right)^{1/2},
\end{equation}
where $a$ is the grain radius, $n_n$ the particle number density, $m_n$ the mass of the gas species,  
and $T$ the gas temperature \citep{yan2004}.
Magnetic forces dominate the grain dynamics when the Larmor timescale is shorter than the drag timescale, 
$t_{\rm L} \ll t_{\rm drag}$, in which case the grain is effectively tied to magnetic field lines.

If $r_L \ll L$ (where $L$ is the characteristic length scale of the flow or magnetised global structure),
the grain is magnetically tied to field lines and therefore the dust is spatially
coupled with the magnetic fields and the gas. The degree of coupling depends on both local gas conditions and on the grain size. 
In SN shocks,  small grains (with size $a\ll$ 1 $\upmu$m) are well coupled with the gas on scales much smaller of the shock itself ($\sim$a few pc),
but the same may not be true for large grains (\mbox{a $\gtrsim$ 1 $\upmu$m}), for which the Larmor radius can be larger than
this scale.  
In this case, they can escape the magnetic fields and be decoupled from the gas \cite{hu2019}. 

The grain charge is described by the integer charge number $z_{gr}$, 
defined through a discrete probability distribution $f(z_{gr})$ over the interval $z_{gr,\min} \le z_{gr} \le z_{gr,\max}$,
where the bounds depend on the grain size and composition as well as on the local physical conditions
(e.g., radiation field intensity, gas temperature, and electron density).  
The physical grain charge is $q = z_{gr} e$, where $e$ is the elementary charge.
The underlying assumption for the definition of  $f(z_{gr})$ is that  
the grain charges have sufficient time to relax to an equilibrium state \cite{ibanezmejia2019}.  
The two extremes of the distribution $z_{gr,\min}$ and $z_{gr,\max}$ are sensitive to the grain
composition (e.g., carbon or silicate dust), size and ISM phase.  

Grain charge becomes more positive and more broadly distributed as the ISM becomes more diffuse, warmer,
and more strongly irradiated, whereas it becomes smaller (and can turn negative) in colder, denser, and more shielded environments. 
The primary interstellar regime where grain charge matters is the warm/diffuse ISM
(with average temperature $T\sim10^4$ K and density $n\sim$ 0.1--1 cm$^{-3}$ \cite{hirashita2009}),
where the gas is maintained ionised by the UV interstellar radiation field, allowing free electrons to be abundant.
In this ISM phase, the grain charge is regulated primarily by the balance between photoelectric 
emission and collisional charging by thermal electrons. 
Photoelectric emission driven by the interstellar ultraviolet radiation field constitutes the dominant positive charging mechanism,
while collisions with electrons provide the main negative charging channel; 
collisions with ions typically play a secondary role due to their lower thermal velocities. 
Dust grains in the warm neutral and warm ionised media attain a statistical equilibrium charge distribution $f(z_{gr})$ that is
typically skewed toward positive values,  
with both the mean charge value and the width of the distribution increasing with grain size and radiation field strength, and decreasing
with electron density \cite{weingartner2001b,ibanezmejia2019}. 
The grain charge distribution $f(z_{gr})$ depends not only on the local physical conditions but also on the grain composition. 
Carbonaceous grains typically attain higher mean positive charges 
and broader charge distributions than silicate grains of the same size, owing to their lower work function
({the work function can be figured out as the minimum energy required to eject an electron from a grain surface;
  typical values are $W \sim 4$--$5\,\mathrm{eV}$ for carbonaceous grains and $W \sim 8\,\mathrm{eV}$ for silicates \cite{weingartner2001b,ibanezmejia2019}).
and higher \mbox{photoelectric yields.} 

As photoelectric emission dominates grain charging in more diffuse, irradiated environments,
these compositional differences directly translate into systematic shifts in the centroid $f(z_{gr})$ \cite{ibanezmejia2019}, 
with a strong dependence on grain size. 
Small grains with size $a \lesssim 10\,\text{\AA}$ generally have modest mean charges, $\langle z_{gr} \rangle \sim 0$--$1$,
while intermediate-size grains ($a \sim 100\,\text{\AA}$) reach centroids of order a few tens. 
For large grains ($a \sim 0.1\,\upmu\mathrm{m}$), the mean charge can rise to $\langle z_{gr} \rangle \sim 10^2$,
and in strongly irradiated warm environments may reach several hundreds, particularly for carbonaceous grains \cite{cervetto1995}. 

In the cold neutral medium ($T\sim 10^2$ K and $n\sim1$-to a few 10 cm$^{-3}$)  $f(z_{gr})$ differs markedly from that in the warm ISM. 
The reduced ultraviolet radiation field and low gas temperature suppress photoelectric emission, 
while collisional charging by electrons becomes more important.  
As a result, $f(z_{gr})$ is typically narrower and shifted toward lower charge values, with small grains often neutral or weakly negative and larger grains only weakly positively charged.  
Typical centroid values range from $\langle z_{gr} \rangle \sim 0$--$1$ for small grains to at most $\sim$10 for $a \le 100\,\text{\AA}$,
and up to a few $\sim$10 for larger grains, depending on composition and local conditions\cite{ibanezmejia2019}. 
In this regime, the dependence of the charge distribution $f(z_{gr})$ on grain composition is also significantly reduced. 
As photoelectric emission is suppressed by the weaker and attenuated ultraviolet radiation field,
grain charging is controlled primarily by collisional processes, which depend only weakly on composition \cite{yan2004}. 
As a result, carbonaceous and silicate grains exhibit similar charge centroids and narrower charge distributions in
the cold neutral medium 
compared to the warm ISM \cite{ibanezmejia2019}. 

In cold molecular gas ($T\lesssim 10$ K and density $n> 100$ cm$^{-3}$), grain charging is fundamentally different from that in diffuse and warm environments. 
The strong attenuation of ultraviolet radiation suppresses photoelectric emission, 
such that grain charge is maintained primarily by collisional processes with electrons and molecular ions,
with the ionisation fraction maintained by cosmic rays.
In fact, cosmic rays produce suprathermal electrons and ions that can collisionally ionise the grains.
They can also induce an ultraviolet radiation field due to H$_2$ fluorescence, leading to photoelectric charging and  
influencing the charge distribution \cite{ivlev2015}. 
Under these conditions, grains are predominantly neutral or weakly charged, with narrow charge distributions centered near
$z_{gr} \sim 0$ or a few \cite{yan2004}, with a slightly increasing trend with size 
and only a weak dependence on grain \mbox{composition \cite{ibanezmejia2019}. }

We will discuss further on the role of grain charge on dust evolution later on, in relation to the evolution of 
intergalactic dust (Section \ref{sec_intergalactic}) and of dust growth \linebreak (Section \ref{sec_growthrole}). }

\subsection{Dust Produced by QSOs} 
Some years ago, the result that QSOs contain significant amounts of dust
even at high redshift ($z>6$ \cite{bertoldi2003}), when the Universe is still $<$1 Gyr old was considered
surprising by a significant part of the community. 
This is because it is widely believed that the quasar environment is hostile to dust \cite{elvis2002}. 
However, the outer radius of the region producing strong broad emission lines (BELs), which typically
coincides with the dust sublimation radius, suggests that such dust can have an intrinsic origin \cite{clavel1989}.
In principle, the free expansion of the clouds that cause 
the BELs is expected to yield favourable conditions to the formation of dust \cite{elvis2002}.
In fact, as already pointed out, in environments that appear to be similarly unfavorable,
such as the UV-bright WR stars (and R Corona Borealis stars as well \cite{jeffers2012}), carbon-rich dust can form. 
It is worth noting that the effective photospheric temperatures and wind densities of these hot giant stars (and AGB stars) 
are remarkably similar to the initial BELs cloud conditions \cite{elvis2002,sarangi2019}.  
In addition to their potential role as dust sources at high redshifts, 
local AGN are known to contain significant amounts of dust. Structures known as ``dusty tori''
are thought to be ubiquitous features of AGN and are 
related to the unification of Seyfert 1,2 \cite{antonucci1993} 
and Broad and Narrow Line Radio Galaxies \cite{vanderWolk2010}. 
In the grand AGN unification scheme (e.g., \cite{meier2002,fanidakis2011}),
these several classes all represent the same system, 
differing only in terms of how they are angled in relation to the viewer's line of sight.
For observers at sufficiently high inclination sites, dust is likely present in toroidal formations preventing direct observation of the AGN central regions \cite{mason2015}.
However, even for the nearest AGN, the diameters of the dusty tori, typically of the order of $\sim$1 $pc$,
are too tiny to be  angularly resolved. 
Over the past few decades, spectroscopic evidence has established the existence of dust in galaxies
harboring AGNs, with the indication that
a sizeable fraction of the continuum source brightness is subtended by the tori \cite{brightman2012,sarangi2019}. 

As pointed out by \cite{elvis2002}, the BELs clouds' initial conditions are quite different from those
of cool stellar atmospheres. 
These clouds are typically much warmer ($T_{BELs} \sim 10^4$ K, e.g., \cite{osterbrock1989}),
and temperatures this high are unfavorable to the formation of dust. 
However, the density of BEL clouds is by a few orders of magnitude higher than the region of cool star atmospheres where dust condenses 
\cite{osterbrock1989,sarangi2019}. 
They might therefore meet the conditions for the dust formation ``window'' if the BEL cloud expands and cools as the result of an outflowing wind. 
This wind can host both the formation of 'new' dust, i.e., 
the creation of molecules, dust grains or grain seeds from an initial fully or partly
dust-free material, containing refractory elements synthesized elsewhere, and
dust from pre-existing seeds born elsewhere \cite{sarangi2019}.
In fact, as the crystallinity degree of the AGN dust grains is higher than upper limits derived in the local ISM,
this supports the idea that new dust grains could primarily form in these systems \cite{sturm2005,srinivasan2017}.

AGN are known as copious reservoirs of refractory elements even at high redshift, as indicated by
the lack of evolution of the metallicity measured in BLRs, found to be oversolar in a wide redshift interval
(2 $\le$ z $\le$ 4.5 \cite{maiolino2006}). 
The AGN torus is composed primarily of silicates \cite{stenholm1994,srinivasan2017},
as indicated by the strong emission features detected by Spitzer in\linebreak  QSOs
\cite{hao2005,sturm2005}, likely characterised by a clumpy distribution \cite{esparzaarredondo2021}. 

Elvis \cite{elvis2002} proposed a model in which the BEL region is composed of clouds initially in pressure equilibrium with a surrounding warmer medium.
The onset of a QSO wind takes the system out pressure balance and the clouds can expand and cool to suitable temperatures
for the formation of grains.  

In an attempt to account for the dust mass observed in the distant QSO SDSS J11481, located at z = 6.4,
Pipino et al. \cite{pipino2011} used a chemical evolution model including dust production in the QSO.
For the growth of the supermassive black hole, they assumed an empirical law able to roughly account for the local
BH mass---$M_{*}$ correlation, a QSO dust production rate that scales as the mass flow rate in the QSO wind and
condensation efficiencies similar to AGB stars. They found that the contribution of the QSO to the global
dust budget is negligible compared to the cumulative contribution from stellar sources. 

In an attempt to model the circumnuclear dusty torus, Sarangi et al. \cite{sarangi2019} invoked a magneto-hydrodynamical,
geometrical model of the AGN wind, that consists of a Keplerian, thin disk around a supermassive BH.
The model includes detailed  molecule nucleation, condensation and grain coagulation to follow grain growth in the wind flow.
They identify a toroidal volume of dusty gas that they associate to the torus, providing a scaling law between
the mass accretion rate and dust production. Remarkably, the predicted torus sizes match those obtained from
Very Large Array (VLA) imaging \cite{carilli2019}. 

As a general conclusion, these studies point towards dust production in QSOs as a non-dominant process. 
However, the amount of dust in the torus is currently not well constrained, even in local systems \cite{esparzaarredondo2021}. 
Further future investigations of the circumnuclear dust content of AGN hosts will be necessary to shed further light
on the issue of dust production in QSOs and AGN.

\subsection{Intergalactic Dust}
\label{sec_intergalactic}
The possible presence of dust in the intergalactic space has gained attention since its obscuring effects have been proposed
as one possible reason for the dimming of distant Type Ia SNe \cite{aguirre1999,croft2000}. 
Dust grains have also been proposed to play an important role in the chemical enrichment of the IGM \cite{ferrara1991} 
and an unknown fraction of intergalactic dust might be required to account for the cosmic dust budget \cite{gioannini2017}. 

\textls[-15]{A few studies have addressed the fate of the dust grains ejected from galaxies into the intergalactic space \cite{montier2004}. 
  Bianchi \& Ferrara \cite{bianchi2005} followed the dynamics of radiation pressure-ejected dust from high-redshift galaxies to the IGM.
Calculating the motion of grains from galaxies to the IGM requires taking into account a multitude of 
processes, including grain size and composition, the dust opacity of the multiphase gas, Coulomb and collisional drag forces, sputtering rates and magnetic forces. 
A fraction of the dust residing in the ISM  can escape the halo of the galaxy in a few hundred Myr, typically with escape \linebreak velocities $>100$ km/s
\cite{ferrara1991,aguirre2001}, thus overcoming the gravitational attraction of massive halos.}

Grains collide with gas particles, with a kinetic energy comparable to the thermal energy of a gas with temperature of T $\sim$~10$^6$ K.
Non-thermal erosion of grains can strip off refractory material and deposit metals even in the lowest temperature IGM. 
With respect to the coldest regions the ISM, in the IGM grains are more frequently charged because of frequent collisions with ions and photoionisation from UV background, 
and their charge has strong influence on the Coulomb drag forces that slow them down, on 
the Lorentz force that deviates them and on the overall sputtering efficiency. 
The distribution of sputtered metals appears inhomogeneous, and grains are never completely destroyed, supporting
the persistence of a low amount of grains in the IGM as indicated by other evidence, such as the intergalactic dust extinction of distant QSOs \cite{menard2009}
and dimming of Type Ia \linebreak supernovae \cite{aguirre1999}.

Protective mechanisms may be in place that keep dust from being completely destroyed even in the hot intracluster medium \cite{gjergo2020,shchekinov2022}.
Direct evidence of the presence of intracluster dust in Virgo has been achieved from the reddening of background \linebreak sources \cite{longobardi2020}
and from observations of infrared dust emission performed with the Infrared Space Observatory in Abell \cite{stickel2002} and, among others,
in various clusters with Planck \cite{planck2016}. 
In the intracluster medium the temperature of the gas is higher than in the IGM and more uniform; in this case, the dominant dust erosion mechanisms
are thermal and collisional sputtering \cite{shchekinov2022}, in which energetic, heavy ions remove material from the grain surface. 
An open question concerns how dust can survive the hostile intracluster environment. 
It has been proposed that dust particles can survive thermal destruction, because during their transport they are shielded in dense, cold clouds or streams
\cite{daines1994,shchekinov2022}. 
It is possible that some fraction might be embedded in dusty clouds in hot galactic winds \cite{farber2022} and that
efficient radiative cooling might enable the survival of dust grains in such dense medium. 
A similar shielding might work also in stripped clouds,  as indicated by the signatures of molecular gas and dust detected in the tails
in a sample of 'Jellyfish' galaxies \cite{giunchi2023}, typical structures present primarily in galaxy clusters.
In such clouds, it has been proposed that a density larger than $\sim$0.3 cm$^{-3}$ can be sufficient to protect dust from destruction  \cite{shchekinov2022}, 
and that magnetic fields on scales of $\sim$a few  kpc (comparable to cloud sizes) can mitigate thermal evaporation \cite{rosseland1958}. 
The conditions for dust survival in the ICM is particularly favourable in the peripheral regions of clusters, where the ambient density is lower and 
the stripped clouds are more likely to survive in isolation, without being disrupted by merging with high-velocity clouds \cite{mihos2017}.


{

\subsection{Dust Evolution in Multiphase Models: Hydrodynamical Simulations}\label{sec:hydro_dust}

\label{sec_simulations}
Over the past few years, an increasing number of studies were focused on the 
explicit treatment of interstellar dust in hydrodynamical simulations. 
The principal advantage of this approach is that hydrodynamics naturally follows phenomena like advection, mixing and
compression of astrophysical gases.

The assumption of spatial coupling between dust and gas is frequently adopted in simulations; this allows one 
to adopt a one-fluid approach to model dust, or to treat dust as a passive scalar perfectly coupled to gas motion \cite{mckinnon2018,granato2021}.  
In other approaches, the dust component is modelled by means of live simulation particles, allowing a more realistic treatment of dynamical
and thermal interactions between dust, gas, and radiation \cite{mckinnon2021}. 

When dust is treated as an additional fluid or set of particles, the same framework can track the transport
of grains through multiphase environments, 
from the coldest ISM component into galactic-scale outflows, and back to the ISM via accretion flows,
while simultaneously applying local source- and sink-terms for condensation, growth \linebreak and destruction. 

The codes used for this purpose integrate the compressible Euler equations—sometimes augmented with magneto-hydrodynamics—to
conserve mass, momentum and energy, while carrying additional continuity equations for metals and, when implemented, one or more
tracers, that can also be dust grains.

Hydrodynamic codes generally fall into two broad numerical families:  
Eulerian schemes, which solve the equations on a stationary or moving mesh and therefore conserve quantities 
within fixed volumes \cite{teyssier2015}, and Lagrangian schemes,
which follow mass elements such as smoothed-particle hydrodynamics (SPH) particles  \cite{springel2010}.
Additionally, two innovative, numerical schemes have  been recently introduced: 
AREPO \cite{Springel2010b} adopts a moving-mesh strategy in which the simulated volume is distributed
among a discrete set of particles, or cells, using a Voronoi tessellation and 
GIZMO \cite{Hopkins2015} instead follows a mesh-less finite-mass approach,
in which every particle carries a smoothing kernel
that assigns it an overlapping control volume with smoothly varying boundaries. 
These hybrid schemes were devised to bypass key drawbacks of the previous methods, 
such as grid-alignment errors for Eulerian schemes, and
inadequate treatment of shocks and fluid mixing for the Lagrangian ones. 

In terms of physical set-up, the models can target isolated galaxies,  ideal for dissecting
the interplay between star formation, feedback and radial dust transport in, for example, Milky-Way analogues or other local
systems (e.g., \cite{bekki2015,relano2022,vandergiessen2024}). 
While isolated galaxy simulations including dust offer the advantage of high spatial resolution and allow detailed testing and calibration
of individual physical processes, such as dust growth and destruction, 
they lack the cosmological context required to model the hierarchical assembly of galaxies.
As a result, they cannot capture environmental effects, merger histories, or the statistical diversity of galaxy populations, making them unsuitable
for predicting dust-related quantities on cosmological scales in a fully self-consistent way. 

Other models target full cosmological volumes and zoom-ins, that embed the same physics in a hierarchical growth framework \cite{mckinnon2018,aoyama2018,parente2022}
as in FIRE \cite{Choban2024}, SIMBA \cite{lorenzon2025}, IllustrisTNG-Dust and COLIBRE \cite{Trayford2026}. 
A complementary strategy is to inject dust in post-processing, coupling pre-computed gas densities and metallicities to grain-evolution networks
to produce synthetic observables at a fraction of the computational cost, in general using radiative-transfer codes \cite{trcka2020}. 
One drawback of the post-processing approach is that it does not enable a self-consistent modelling of the 
properties of interstellar dust, such as the evolution of the mass and composition, which cannot be tracked directly from
the star formation and chemical evolution history of the simulated systems.  
The most robust solution is to run fully self-consistent simulations in which dust is evolved on the fly;
only then can the changing mass, composition and spatial distribution
of grains be traced directly from the underlying star-formation and chemical-enrichment histories. 


Over the past decade, large-scale numerical simulations have moved from treating interstellar dust as a {\em post--processing} diagnostic to modelling it as a live,
advected component that is created, destroyed, and dynamically coupled to the gas. 
These efforts complement simple chemical evolution models by (i) resolving the spatial inhomogeneity of star formation and feedback, (ii) capturing radial gas flows,
fountains, and outflows, and (iii) tracking how dust grains are processed in different phases of the multiphase interstellar medium (ISM).  
Fully self-consistent simulations confront additional challenges—notably the limited resolution that keeps grain growth and destruction on sub-grid scales,
and a stringent timestep requirement, because the interval must be short enough to resolve the fastest physical processes and the most rapidly accelerated fluid elements.
In this Section we summarise the current state of the field, emphasising the physical ingredients that distinguish the major simulation suites and the insights
that can (and cannot yet) be gained from them.


Regardless of hydrodynamic method, the majority of simulations implement a common set of dust source and sink terms on a per--timestep basis: 
\begin{equation}
  \frac{\mathrm{d}M_{\rm d}}{\mathrm{d}t}=\dot{M}{\rm d,*}+\dot{M}{\rm d,grow}-\dot{M}{\rm d,destr}-\dot{M}{\rm d,astr} - \dot{M}{\rm d,out},
  \label{eq_dust_hydro}
\end{equation}
where $\dot{M}{\rm d,*}$ and $\dot{M}{\rm d,grow}$ represent stellar grain production and gas-phase grain growth, respectively; 
$\dot{M}{\rm d,destr}$ is the generic destruction rate as due to shocks from SN remnants,  thermal and non-thermal sputtering and collisions; 
$\dot{M}{\rm d,astr}$ the astration term as dust is locked into new stars, and $\dot{M}{\rm d,out}$ advection in galactic winds \cite{mckinnon2016,aoyama2018,parente2025}. 

Implementation details---for example whether growth is tied to the local molecular fraction,
or whether destruction scales with resolved shock energy---vary between groups 
and strongly influence the emergent scalings. 
In contrast to classical single-phase chemical-evolution models, cosmological hydrodynamic simulations 
can follow the multiphase ISM and evolve the grain-size distribution self-consistently.  
Models accounting for grain size distribution or two-size approximation include in Equation (\ref{eq_dust_hydro}) additional terms for coagulation and
shattering processes
\cite{mckinnon2018,gjergo2018}.  


Table~\ref{tab:hydro_sims} provides a non--exhaustive list of recent cosmological or zoom--in simulations with explicit dust evolution. 
Box sizes range from $\approx$10~kpc zooms of dwarf galaxies to $100,\mathrm{cMpc}$  
boxes capable of capturing the growth of cosmic structures in various environments. 
In the following, we discuss some of the main features of each set and most remarkable differences of the approaches. 



\begin{table}[H]

\caption{Representative hydrodynamical simulations that track dust as a live quantity. Columns list hydrodynamic code, simulated volume or zoom type, typical gas--particle mass resolution, included dust evolution processes, and key references.}
\label{tab:hydro_sims}

\begin{adjustwidth}{-\extralength}{0cm}

\begin{tabularx}{\fulllength}{m{3cm}<{\raggedright}m{2.5cm}<{\centering}m{3cm}<{\centering}m{3cm}<{\centering}m{3cm}<{\centering}m{4cm}<{\raggedright}m{4cm}<{\raggedright}}
\toprule
\textbf{Simulation} & \textbf{Code} & \textbf{Volume} & $\textbf{m}_{\rm \textbf{\emph{gas}}}$ & \textbf{Dust Physics} $^{\textbf{\emph{a}}}$ &\textbf{ Ref.}\\
\midrule
GADGET3-OSAKA & \textsc{GADGET}      & (50 h$^{-1}~$cMpc)$^3$       & $1.3-10\times10^{7},M_{\odot}$ & Sd,
Gr,Ds & \cite{aoyama2018} \\
FIRE-2-Dust & \textsc{gizmo} (MFM) & Zoom--in & $\sim$7 $\times$ 10$^{3},M_{\odot}$ & Sd,Gr,Ds & \cite{Choban2024} \\             
SIMBA & \textsc{gizmo} (MFM) & (100 h$^{-1}$~cMpc)$^3$ & $1.8\times10^{7},M_{\odot}$ & Sd,
Gr,Ds,AGN & \cite{li2019,Dave2019} \\
MUPPI     &  GADGET3    & 5 $\times$ (26 cMpc)$^3$     &   $7 \times10^{6}   M_{\odot}$ & Sd,Gr,Ds   &  \cite{ragonefigueroa2024}        \\
IllustrisTNG--Dust & \textsc{arepo} & 75 
 h$^{-1}$~cMpc & $1.4\times10^{6},M_{\odot}$ & Sd,Gr,Ds (post-p.) & \cite{Shen2022} \\
\bottomrule
\end{tabularx}
\end{adjustwidth}
\footnotesize{$^{a}$~Sd: stellar dust condensation; Gr: grain growth; Ds: destruction in SN shocks; AGN: dust sputtering in AGN--driven outflows; (post): dust added in post--processing.}
\end{table}

The GADGET3-OSAKA code implements a self-consistent treatment of dust evolution within a cosmological framework, specifically designed
to follow different dust-related properties across  various galactic environments.
The model is built upon the GADGET-3 Smoothed Particle Hydrodynamics (SPH) code and includes the following key features: 
it follows the dust mass evolution using an implementation of the fundamental mass-exchange equation similar to \ref{eq_dust_hydro}, 
accounting for stellar production, interstellar grain growth, destruction and other processes, including shattering and coagulation.
In fact, it features a two-size approximation, which simplifies the grain size distribution into \linebreak two distinct populations: small grains (a $\le$ 0.03 $\upmu$m)
and large grains (a > 0.03 $\upmu$m). This allows the code to follow size-dependent processes like shattering and coagulation without the computational cost
of a full grain-size solver. 

Choban \cite{Choban2024} presented a subset of cosmological zoom-in
simulations of MW to dwarf-halo mass galaxies from the Feedback in Realistic Environments (FIRE)-2 project that include
an integrated dust evolution model, previously calibrated by means of runs of isolated systems. 
The FIRE-2 runs including dust use the cosmological zoom-in technique,  
in which the target halos  are selected from a large, dark matter only simulated periodic volume
(typically of the order of $\sim$100 Mpc on a side \cite{wetzel2016}). 
For each halo of interest, the `Lagrangian region' is traced back to the initial conditions,
and that region is regenerated at finer mass- and spatial resolution, while keeping the rest of the box low-resolution.
This allows one to concentrate computing power on the target halo, yet preserving the physical large-scale tidal field.

The simulations include the `Species' dust model \cite{choban2022}, implemented in the FIRE-2/3 simulations.
This model treats dust as a collection of distinct chemical species (including silicates, carbonaceous dust and metallic iron).
It calculates dust destruction and grain growth is confined to cold environments ($T\le300$ K) and destruction is coupled
to individual, time-resolved SN feedback events. 
As for the grain-size distribution, a fixed power law is adopted, along with
a single effective grain radius when computing accretion, sputtering and SN-shock destruction,  
updating the mass of each dust species. 

The SIMBA simulation set is characterised by a cosmological run of a (100 h$^{-1}$ cMpc)$^3$ volume with 1024$^3$
gas particle elements \cite{Dave2019}, evolved with the mesh-less-finite-mass (MFM) variant of GIZMO.
Clearly, the gas-particle mass resolution  of $1.8\times10^{7},M_{\odot}$ is much lower than the one of the FIRE-2 runs.
While sacrificing small-scale resolution, SIMBA's large periodic volume provides the environmental diversity and statistical power
necessary to investigate the co-evolution of dust and AGN. This large-scale approach is essential for capturing a representative sample of
high-mass systems presenting AGN activity, allowing for a direct assessment of how black hole feedback regulates the galactic dust budget. 
The feedback sources are different, as SIMBA features both stellar feedback and AGN jets that deposit in a sub-grid fashion
momentum that heat and directly alter the fixed-size dust grain budget. 

Thanks to their high resolution, zoom-in simulations like FIRE 2 allow one to probe
a large dynamic range in ISM density, 
tracing the interplay between star formation, feedback, and dust processing,
with insights on local galactic properties, such as how dust depletion varies with local ISM density \cite{choban2022}.  
By contrast, the coarser but box-wide SIMBA suite evolves thousands of galaxies simultaneously, making
it the better tool for the assessment of the cosmic dust-mass function and the redshift evolution
of the comoving dust density. 

Another recent set of cosmological simulations \cite{ragonefigueroa2024} includes an advanced description of a multi-phase ISM,
 the MUPPI sub-resolution model \cite{murante2015}, that is used to account for the
mutual evolution of the dust content and molecular gas in the star-forming gas. 
The simulations of Ragone-Figueroa et al. \cite{ragonefigueroa2024} consist of five independent, full-box cosmological runs—each covering a comoving volume of 
($26$ Mpc)$^3$, performed at uniform resolution without employing zoom-in techniques.
The authors adopt a dust evolution model based on the two-size approximation developed and calibrated in previous \linebreak works \cite{gjergo2018,granato2021},
where the grain size distribution is simplified into two representative populations, 
small grains (with radius a < 0.03~$\upmu$m) and large grains (with a > 0.03~$\upmu$m), each followed as a separate dust mass variable for every gas particle.
This framework enables the model to capture the dominant processes that govern grain evolution—such as growth, shattering, coagulation, and sputtering—while
remaining computationally feasible in large-scale cosmological simulations. 
Dust is produced by stellar sources, including core-collapse SN, AGB stars, and Type Ia SNe,
with yields depending on progenitor type and metallicity. Once injected, large grains can grow by accreting
metals in the cold and dense ISM, while coagulation allows small grains to join into larger ones.
Shattering due to grain-grain collisions in more diffuse and turbulent environments can break large grains into smaller fragments.
Thermal sputtering in hot gas further erodes both populations. These processes are applied locally to each gas particle,
allowing the dust-to-gas ratio and the grain size balance to respond dynamically to environmental conditions across cosmic time. 
This dust model is self-consistently coupled to the non-equilibrium chemistry module used in the simulations,
particularly affecting the formation of molecular hydrogen on grain surfaces,
a process that strongly depends on both the dust abundance and the grain size distribution.

As an example of simulations where dust is treated in post-processing, the IllustrisTNG–Dust model \cite{Shen2022} applies dust physics
to the standard TNG runs without evolving dust live during the hydrodynamical simulation.
\textls[-15]{The flagship IllustrisTNG simulations were run with the AREPO moving-mesh code on different volumes,
from (35 h$^{-1}$ Mpc)$^3$ to  \linebreak (205 h$^{-1}$ Mpc)$^3$, with gas particle resolution ranging from $5.7\times10^{4}~M_{\odot}$
to $7.4\times10^{6}~M_{\odot}$ \cite{Vogelsberger2020}. 
Rather than evolving dust on-the-fly, the model assigns dust to cold, star-forming gas based on a redshift-dependent dust-to-metal ratio (DTM)
$\propto z^{-1.92}$ and Monte-Carlo radiative-transfer calculations
with a modified version of the SKIRT code \cite{Camps2015} produce attenuated SEDs and IR emission \cite{Shen2020}. 
The radiative-transfer step adopts a multigrain mixture with 10 logarithmic size bins for graphite, 
silicate and polycyclic aromatic hydrocarbon (PAH) grains, but these sizes evolve only inside SKIRT, while the hydrodynamic run itself carries no live grain physics.  
This approach allows predictions for dust mass, extinction curves, and IR luminosities across cosmological volumes from $z \sim 0$ to $z \sim 8$, }
without modifying the original hydrodynamic run. 
While this approach provides good statistical power and radiative realism, it cannot capture the time-resolved cycling or destruction of dust
in multiphase environments. 

Figure~\ref{fig:dtg_dts} provides 
a comparative overview of how some key physical quantities —such as the dust-to-gas ratio, the dust-to-stellar mass ratio,
and their scaling with metallicity, stellar mass, and redshift—are captured across recent simulation models and against available observations. 
As visible in the left panel of Figure~\ref{fig:dtg_dts}, 
observations reveal a strong correlation between the dust-to-gas ratio (DTG) and the gas-phase metallicity,
here traced by the O abundance and expressed as 12+log(O/H), although with a varying degree of dispersion according to different datasets (see the references
in Figure~\ref{fig:dtg_dts} caption). 
In general, some care must be taken when interpreting the observational data due to significant systematics
in metallicity calibrations and gas mass estimates. 
The DTG quantity has often been studied and accounted for with chemical evolution models \cite{gioannini2017a} that, in very few cases,
can account also for the multiphase ISM structure \cite{millanirigoyen2020}. 

A correlation between DTG and metallicity is captured also by the simulations, although in some cases with an offset with the data, as due both to
the said systematics and uncertainties in the model prescriptions. 
The overall increasing trend of this relation is recovered by a significant range of models, including
cosmological simulation \linebreak Sets \cite{aoyama2018,vijayan2019,parente2025}.


Another key diagnostic is the dust-to-stellar mass (DTS) ratio (central panel of Figure~\ref{fig:dtg_dts}),
which serves as a valuable tracer of the dust content integrated over galactic lifetimes and provides insight into its star formation and
chemical enrichment history \cite{calura2017,Donevski2020,relano2022}. 
Its dependence on stellar mass and redshift makes it particularly useful for tracking
the buildup of dust across cosmic time and comparing galaxies at different evolutionary stages.

Observations of local and distant galaxies show a decreasing behaviour of the DTS as function of stellar mass, with a slope
sometimes depending on redshift or galaxy type, or with a flatter slope
in local spiral galaxies -as found in the SINGS \cite{Kennicutt2003} and KINGFISH~\cite{Kennicutt2011} samples- as compared, e.g.,
to the submillimetre galaxies (SMGs) at redshift $z > 1$ \cite{Santini2010}.
As visible in the central panel of Figure~\ref{fig:dtg_dts}, chemical evolution models are able to capture this overall trend,
showing a weakly decreasing or flat behaviour
-after an initial, rapid increase-in discs and a steeper decrease in high-redshift proto-spheroids. 
This trend remains challenging to reproduce in some cosmological simulations, which sometimes show a monotonically
increasing DTS  with stellar mass across a broad redshift range, leading one to interpret the
median DTS ratio as a proxy for the stellar mass \cite{Zimmerman2024}, albeit in a few cases
with a hint of flattening at the highest masses at low redshift \cite{relano2022}.

Despite substantial progress, several open challenges remain in modelling dust in cosmological simulations.
Some current frameworks still lack a fully self-consistent treatment of grain-size distributions,
dust dynamics, and composition-dependent processes. The interplay between dust and radiation—especially radiative
feedback and reprocessing in dusty environments—is often simplified or omitted entirely. Furthermore, limited resolution
prevents simulations from fully capturing dust growth and destruction in cold, dense molecular regions, where much of the dust evolution takes place.

Magnetic fields also represent an open frontier: if grains are charged, they can experience Lorentz forces, potentially altering their transport
and coupling to the gas—yet this is rarely included in current large-scale models. 
However, it is worth stressing that fully coupled magneto-hydrodynamic runs with Lorentz forces on dust grains still count few attempts,
even on molecular-cloud scales \cite{Hopkins2016,Commercon2023}. 

Altogether, these limitations affect predictions of extinction, dust emission, and molecular gas formation,
and highlight the need for continued development of sub-grid models and high-resolution benchmarks to anchor dust physics in a cosmological context.


\begin{figure}[H]

\begin{adjustwidth}{-\extralength}{0cm}
\centering 

\includegraphics[width=17 cm]{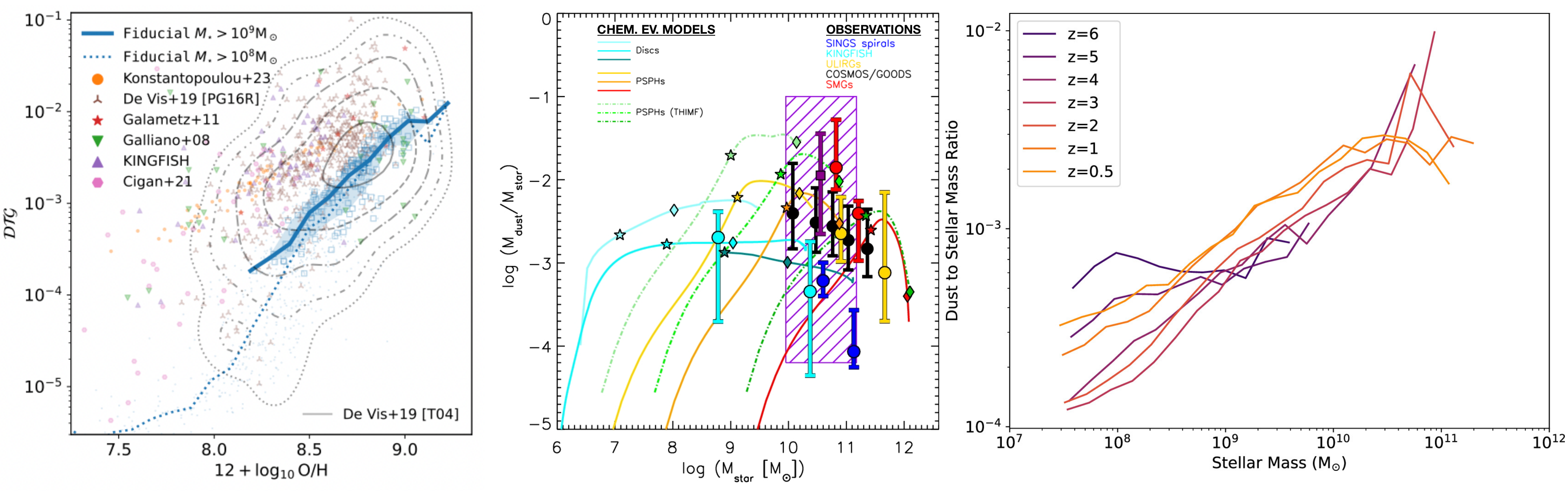}

\end{adjustwidth}
\caption{\textbf{Left:}
  dust-to-gas ratio as a function of metallicity in observed and simulated star-forming galaxies
  (figure adapted from \cite{Trayford2026}). 
  The symbols represent a compilation of observations in local galaxies, while the blue lines show the median values in simulated
  galaxies from the COLIBRE suite. \textbf{Middle}: dust-to-stellar mass as a function of stellar mass from various observational samples (see \cite{calura2017})
  and in chemical evolution models representing proto-spheroids and disc galaxies of various mass (see the text for further detail; figure adapted from\cite{calura2017}). \textbf{Right panel}: 
  median dust-to-stellar mass ratio as a function of mass at various redshifts from the SIMBA cosmological simulations (figure adapted from \cite{Zimmerman2024}). } 
\label{fig:dtg_dts}
\end{figure}

}


\section{How Dust Depletion Shapes Chemical Abundances at High Redshift}
\label{sec_howdust}
The evidence that dust was present in large quantities in high-redshift galaxies dates back to several decades,
along with the notion that dust grains have strong impact on the 
interstellar abundances of distant systems detected both in absorption and emission.
These samples include the DLA  systems (DLAs), discovered in the '80s in the lines of sight of distant QSOs \cite{Wolfe1986}. 
In this section, I review how chemical evolution models have been used for the interpretation of the abundances of refractory and non-refractory elements
measured in high-redshift galaxies, namely in DLAs (detected both in the lines of sight of QSOs and Gamma Ray Bursts) and in other star-forming systems, 
such as Lyman Break Galaxies and Lyman-$\alpha$ emitters.

\subsection{Dust-Depleted Interstellar Abundances in QSO DLAs}
The analysis of spectra from quasi-stellar objects (QSOs) has resulted in the identification of particular categories  of systems
that exist along numerous lines of sight, whose investigation has significantly advanced our understanding of galactic evolution.
Among these systems stand out DLAs, allowing us to gain valuable insights into the early stages of the Universe,
shortly after the formation of galactic structures. 
DLA systems are notable for their substantial presence of neutral gas, characterized by a column density of neutral hydrogen (N(HI) $\ge$ 2 $\cdot$ 10$^{20}$ cm$^{-2}$) 
Additionally, their metal abundances can vary widely, ranging from approximately 1/100th of the solar value to (or even greater than) the solar value \cite{pettini1994,wolfe2005,bashir2019}. 

The occupation of the damped region of the growth curve in DLA systems enables us to achieve a high level of accuracy in determining the hydrogen column density.
This, in turn, allows for precise assessments of chemical abundances for various low-ionisation species like SiII, FeII, and ZnII \cite{bechtold2001}. 
The absorption lines produced by DLA systems are typically detectable with moderate resolution, and the damping profile of these lines facilitates 
precise measurements of the column density of neutral gas. This measurement is crucial in conducting studies on chemical abundances.

\textls[-15]{Due to their large number, optimal for statistical studies and to probe the cold gas phase,   
DLAs are particularly suitable for studying the chemical evolution of the high-redshift Universe, 
allowing us to have a deep insight into galaxies during the initial phases of their history.  
Chemical abundances in DLA systems have been measured in considerably large samples of DLAs---for one of the most recent datasets of DLA metallicity, see \cite{bashir2019}.  
Several works have addressed the determination of chemical abundances in DLAs \cite{pettini1994,pettini1995,pettini1997,lu1996,lu1998,prochaska1999,centurion2000,prochaska2002}.}
In such studies, one main challenge lies in accounting for ionisation correction effects. 
In fact, in principle, for an accurate conversion of the observables, represented by column densities, into elemental abundances, 
one should also account for the ionisation state of the gas. 
It is commonly assumed that, due to the high  
column density of the absorbers \linebreak (N(HI) > 10 $^{20}$ cm$^{-2}$ \cite{wolfe2005}, ionisation corrections are minimal. 
As a result, the total amount of an element can be estimated from its main ionisation state---specifically, the lowest-energy ion whose ionisation potential is above 13.6 eV.
For most elements detected in absorption (like silicon, iron, zinc, sulfur, nickel, titanium, and chromium), this corresponds to the singly ionised form.
Therefore, the total column density of an element X is often approximated as N(X) $\simeq$ N(X II) \cite{milutinovic2010}. 
An exception to the usual case involves elements like oxygen and nitrogen, because their first ionisation energies are very close
to that of hydrogen. In these cases, their ionisation states are not determined just by radiation but also by charge exchange
reactions with hydrogen, which tend to maintain them in their neutral form.
The robustness of this approximation has been assessed various times, however, barring a few exceptions, 
in general the literature agrees  
on the fact that the majority of DLA column densities do not require significant ionisation corrections \cite{milutinovic2010}. 
Results from detailed photoionisation models indicate that ionisation 
corrections are usually negligible for the most common elements that are usually investigated \cite{howk1999,izotov2001,vladilo2001}.

Initially, based on kinematic considerations, DLAs were linked to observed rotating proto-disks during epochs when significant gas consumption had not yet occurred \cite{prochaska1997}.
At later times, intermediate redshift imaging has revealed a diverse range of morphological types within the DLA population.
These include low surface brightness galaxies, dwarf galaxies, spirals \cite{lebrun1997}, and in some very rare instances,
even the progenitors of early-type galaxies \cite{prochaska2003}. 

Chemical evolution models can be very useful in shedding light on the nature of DLAs and have often been used to interpret their observed abundance patterns. 
These models can provide insights into various quantities, with abundance ratios being a primary focus of such studies.  
Abundance ratios such as [$\alpha$/Fe] and [N/$\alpha$], are particularly valuable as they can serve as cosmic clocks when they involve two elements
that were formed on timescales that are appreciably different \cite{matteucci2002}. 
When such quantities are examined in conjunction with metallicity tracers, like [Fe/H] or [Zn/H], they can shed light on the specific
history of star formation characterising the DLA host \cite{calura2003,dessauges2004}. 
In fact, in a high star formation rate regime, typical of spheroids (i.e., elliptical galaxies and bulges), it is expected that an overabundance of
$\alpha$-elements will be observed over a wide range of [Fe/H]. 
Conversely, in a low star formation regime, such as the one of dwarf galaxies or late spiral discs, the opposite outcome is expected, i.e.,
an [$\alpha$/Fe] declining steeply with metallicity \cite{matteucci2002}. 
This difference in elemental abundances is attributed to the distinct roles played by Type II and Type Ia supernovae.
SNe Ia are primarily responsible for the production of Fe and Fe-peak elements and operate on  $>$30 Myr timescales, longer than 
the ones that are typical of SNe II explosions (from a few Myr to $\sim$30 Myr), which are responsible for the production of $\alpha$-elements 
(such as O, Ne, Mg, Si, Ca). 
Therefore, analyzing the relative abundances of these elements can provide key insights into the nature and age of the (proto-)galaxies that give rise to DLA systems.
Moreover, the study of the abundance patterns is also very useful to probe the ages of the galaxies under study. 

Early studies exploring the nature of DLA systems using chemical evolution models~ \cite{matteucci1997,jimenez1999,mo1994,prantzos2000} 
agree on the result that some of these objects may represent spiral disks caught during their formation,
while others could be low-surface brightness objects or starbursting dwarfs resembling local,
highly metal-poor, gas-rich starbursting galaxies, such as IZw18 \cite{kunth1994}.

Figure~\ref{fig:cal03} shows the metallicity observed in a sample of DLAs (in this case traced by the non-refractory element Zn),
as a function of redshift, compared to the results of a set of
chemical evolution models describing galaxies of different morphological types, i.e., spirals, ellipticals and irregulars \cite{calura2003}.
In the models, the evolution of the metallicity (traced in this case by [Zn/H]) and other elements has been computed from
a set of equations as Equations (\ref{eq_chem}) and (\ref{system1}) and 
the morphological types are characterised by different star formation efficiencies and different infall timescales, therefore different gas accretion and star formation histories.
All the galaxies started forming stars at the same redshift $z_f=5$, assuming a $\Lambda CDM$ cosmology with $\Omega_M=0.3$,  $\Omega_{\Lambda}=0.7$ and $h=0.65$.
The scatter in the observed data is substantial and can be caused by several factors, including different object ages, LOS
and morphologies. Although the model results are partially sensitive to the adopted value for $z_f=5$,
Figure~\ref{fig:cal03} shows that the majority of the data points are consistent with the theoretical abundances computed for the spiral and irregular
models, whereas the elliptical model is characterised by a too high metallicity.

Further information on the properties of the observed systems can be achieved from the analysis of the abundances for more elements. 
However, it is important to note that the observable abundance patterns in DLA systems can only be effectively interpreted using chemical evolution models
if they are driven by pure nucleosynthesis processes.  
The presence of dust in DLAs can pose a significant challenge when attempting to interpret their abundance patterns,
as it leads to the depletion of certain chemical elements (e.g., Fe, Si) to a greater extent than others (O, Zn), thereby altering the derived abundances.  
There are various indications pointing towards the existence of dust within DLAs, such as the observed reddening of background quasars
in the presence of DLA absorbers \cite{fall1993,ellison2005,geier2019}. 
Additionally, evidence of elemental depletion similar to the one observed in
the local \linebreak ISM \cite{savage1996} was achieved from the abundance ratios of elements characterised by differential depletion, such as 
the Zn/Cr \cite{pettini1994} and Zn/Fe \cite{vladilo2002a} ratios. 
More abundance studies provided further evidence supporting the presence of dust depletion
\cite{hou2001,prochaska2002}. 

In the presence of dust, the observed abundances may not accurately reflect the real chemical composition of the system. 
This discrepancy arises with particular elemental transitions from the gaseous to the solid phase, as very frequently happens in the ISM. 
The elements with higher resistance to heat and erosion, 
such as Si, Fe, Chromium (Cr), and Nickel (Ni), tend to be predominantly incorporated into dust grains.
However, it is important to exercise caution when interpreting the relative ratios of these elements,
as the different degrees of depletion can create a misleading resemblance to the intrinsic nucleosynthetic abundance patterns. 
The above studies suggested that simply considering the observed abundances at face value was insufficient 
for understanding the chemical properties of DLA galaxies, and that  
the impact of dust depletion had to be taken into account when interpreting the observed abundance patterns. 
Some approaches were proposed to circumvent the problem of dust depletion in DLAs abundances,
which were however limited to some undepleted elements, such as N \cite{centurion1998}, O \cite{Molaro2000,levshakov2002}, S \cite{centurion2000} and \linebreak Zn \cite{pettini1999,vladilo2000},
or to particular DLA sub-samples with low dust content \cite{pettini2000,Molaro2000}. 
To conduct a comprehensive study of DLA abundances, it is necessary to assess directly the effects of dust depletion. 

\vspace{-6pt}
\begin{figure}[H]

\includegraphics[width=10 cm]{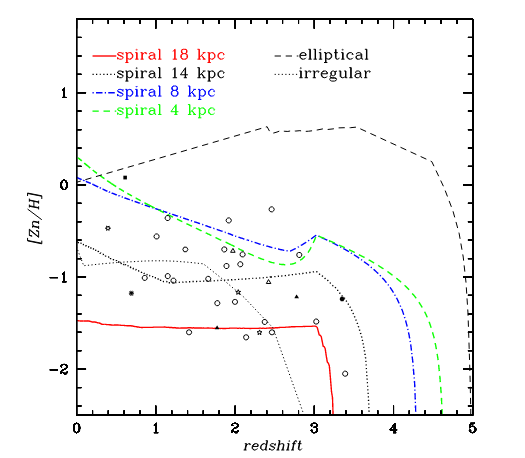}
\caption{Redshift  
 evolution of [Zn/H] observed in DLAs by various authors (black symbols) and as predicted by chemical evolution models for
ellipticals, spirals and irregulars \cite{calura2003}. 
The red solid, thick black dotted, blue dot-dashed and green dashed lines show the results for the multizone model of a
Milky Way-like spiral at different galactocentric distances (see the legend in the top-left corner).
The thin black dotted and black dashed line shows the theoretical evolution of [Zn/H] for an irregular and
elliptical galaxy, respectively. Figure adapted from \cite{calura2003}.} 
\label{fig:cal03}
\end{figure}   

\subsubsection{Dust-Corrected Abundance Pattern Observed in DLAs}

Some authors have developed formalisms with the aim of quantifying the impact of dust depletion on DLA abundances.
These approaches typically assume that the dust in DLAs is similar to the interstellar dust of our own Galaxy, scaling the depletion
by means of the dust-to-metal ratios \cite{kulkarni1997,vladilo1998}. The main limitation of these methods is that they do not account for
potential variations in the chemical composition of dust among \linebreak different absorbers.  

Vladilo \cite{vladilo2002a} derived an analytical expression to compute interstellar depletion
by accounting for variations in the physical conditions and chemical abundances of the medium. 
This method takes into account a set of interstellar parameters and provides a means to obtain dust-corrected abundances.
Such corrections have been applied to abundance ratios between various elements, with a specific focus on the [Si/Fe] 
a noticeable example of [$\alpha$/Fe] ratio \cite{vladilo2002b}. 
The abundances of these two elements can be easily measured, but their interpretation is complicated by the fact that they are
incorporated into solid grains in different proportions, a phenomenon called `differential depletion'. 
Some studies have noted that the [$\alpha$/Fe] ratios observed in DLAs are somehow consistent with the abundances measured
in metal-poor stars within the Milky Way \cite{lu1996,prochaska2001}, 
therefore showing some level of $\alpha$-enhancement,  with typical values of [Si/Fe] $>$ 0.25 \cite{vladilo2002b}. These values are 
similar to the ones observed in Galactic halo or thick disc stars with metallicities (at face value) comparable to
those found in DLAs, supporting a chemical evolution pattern dominated by Type II supernovae
\cite{mcwilliam1997,matteucci2001}. 
In principle, if the observed enhancement of [$\alpha$/Fe] is real and not affected by dust depletion, it could be indicative of the halo phase of a forming disc galaxy, or the formation of a spheroid. 
This also implies that the chemical evolution processes in DLAs are consistent with the early phases of the formation of a galaxy similar to the Milky Way, as well as their associated star formation history. 
In contrast to the cases mentioned earlier, there are instances where measurements of [Si/Fe] in DLAs show little evidence of
$\alpha$-enhancement \cite{ellison2001,pettini1999}.  
Pettini et al. \cite{pettini1999} specifically studied a group of absorbers at redshift $z < 1.5$ and found no indications of 
an enhancement of $\alpha$-elements relative to iron. 
They argue that in low surface brightness galaxies or in the outer regions of a disc where star formation occurs at a slower rate,
one could observe nearly solar values of [$\alpha$/Fe] even at low metallicities. 
This suggests that the presence or absence of $\alpha$-enhancement in DLAs may depend on the specific galactic environment and the pace of star formation within it.
However, even in cases where a clear $\alpha$-enhancement is not observed in DLAs, it does not necessarily rule out a connection between DLA sites and galactic disks,
as long as the observed DLAs represent the outer regions of galaxies~\cite{calura2003}.  
In fact, it is possible that the observations are predominantly sampling the external regions of galaxies.
In these outer regions, where star formation is slower and the chemical enrichment process may differ,
the $\alpha$/Fe ratios may resemble those found in the solar neighborhood or even show a lack of $\alpha$-enhancement.
In fact, the differential cross-section for DLA absorption can be approximated as $dA \approx 2\pi r , dr$ when considering a disk seen face-on \cite{vladilo1999}.
This implies that galactic regions with larger size, or radial distance from the center  (represented by $r$) have a higher probability of being detected as DLAs. 
In other words, the chances of observing a DLA increase for galactic regions farther away from the center,
as the cross-sectional area available for absorption by the DLA increases with increasing radial distance. 

The [S/Zn] ratio has been proposed as a reliable diagnostic tool and a valid alternative to study abundances in DLAs because both
sulfur (S) and zinc (Zn) are less affected by dust depletion
compared to other elements. 
However, estimating accurate S abundances is challenging due to the contamination caused by the Lyman-$\alpha$ forest,
which can have significant effects on the measurements.  
The available data on the [S/Zn] ratio do not exhibit the typical enhancement observed in Galactic metal-poor stars at similar metallicities~\cite{centurion2000}.
Globally, this suggests 
that there are different chemical evolution patterns observed in DLAs, characterized by either halo-like (i.e., in some cases $\alpha$-enhanced)
or solar abundance ratios. 
This could indicate the presence of a heterogeneous population of progenitor objects, more than a single type of object observed at different phases of their evolution.

Figure \ref{fig:cal03_2} illustrates the distribution of [Si/Fe] ratios as a function of redshift (above) and [Fe/H] (below)
based on models of galaxies with different morphological types, 
both neglecting (left panels) and taking into account (right panels) dust depletion corrections.
The advantage of using the [Si/Fe] ratio is that, in this case,  we can rely on a larger data sample than with the [S/Zn] ratio.    
The model predictions in Figure \ref{fig:cal03_2} (lines) are compared with DLA data (points with error bars) from various sources \cite{vladilo2002b}. 
The predictions in the figure are specifically related to different types of galaxies., i.e., a spiral, an elliptical and irregulars, including
a Large Magellanic Cloud-type and a dwarf galaxy that has undergone a single burst of intense star formation \cite{calura2003}. 
The results for a typical spiral are shown at various galactocentric distances, representing different regions within the galaxy. 
By comparing the model predictions with the observed DLA data, the figure provides insights into the potential morphological types
and evolutionary stages of the galaxies associated with the DLAs. 

\begin{figure}[H]

  \includegraphics[width=10 cm]{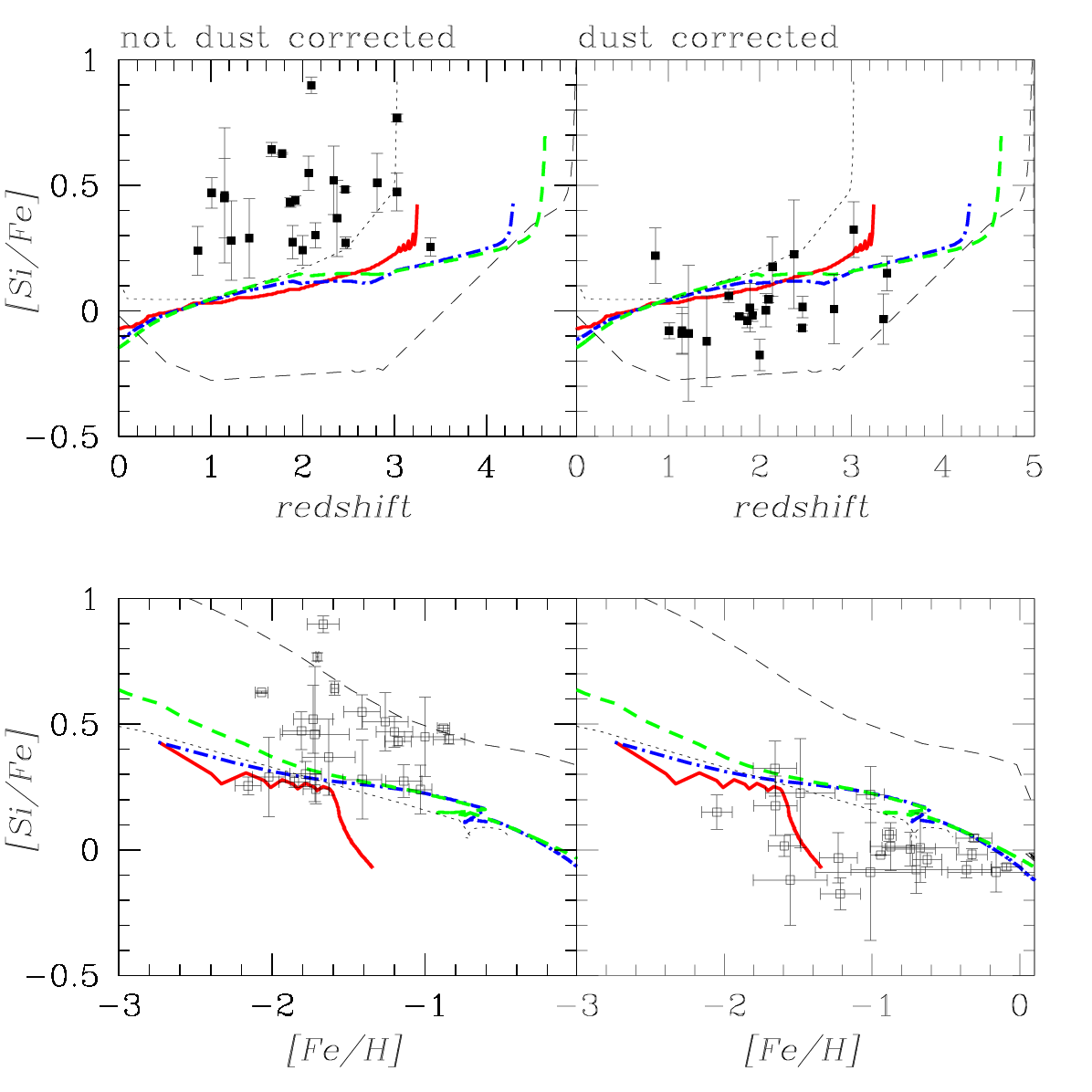}
\caption{[Si/Fe] vs redshift (\textbf{upper panels}) and [Fe/H] (\textbf{lower panels}) observed in
  DLAs (black symbols with error bars, see Calura et
  al. 2003 \cite{calura2003} and references therein) in the case of data not corrected (\textbf{left panels}) and corrected (\textbf{right
 panels}) for dust depletion and compared to chemical evolution models. 
 The red solid, blue dot-dashed and green dashed lines show the results for the multizone model of a
 Milky Way-like spiral at different galactocentric distances, i.e., 18 kpc, 8 kpc and 4 kpc, respectively.
 The thin black dotted and black dashed lines show the theoretical evolution of [Si/Fe] for an irregular and
 elliptical galaxy, respectively.}
\label{fig:cal03_2}
\end{figure}

The evolution of the [$\alpha$/Fe] ratio as a function of redshift, as shown in the models, reveals a common trend.
Initially, there is a phase characterized by an overabundant [$\alpha$/Fe] ratio. 
This phase is driven by the predominance of type-II SNe, which produce the $\alpha$-elements through their explosive nucleosynthesis.
This overabundance phase eventually comes to an end when type-Ia SNe start to occur.
The exact timing of this transition depends on the lifetimes of the type-Ia SN progenitors 
and the star formation history specific to each galactic morphological type. 
Once type-Ia SNe start exploding, they contribute significantly to the iron-peak elements, such as Fe, thus causing a decrease in the [$\alpha$/Fe] ratios. 

The disc models at different radii exhibit variations in the timing and magnitude of the initial $\alpha$/Fe enhancement. 
The innermost regions, characterized by higher gas densities, experience stronger star formation, leading to an earlier and more pronounced peak in the $\alpha$/Fe ratio. 
This pattern reflects the 'inside-out' formation of the disc and, in the innermost regions, resembles the star formation history of elliptical-like galaxies. 
On the other hand, at larger radii, where the gas density is lower, the star formation is less intense. Consequently, the peak in the $\alpha$/Fe ratio occurs later
and is weaker, similar to the behavior observed in irregular galaxies. 

The $\alpha$/Fe versus [Fe/H] plots exhibit a different behavior compared to the $\alpha$/Fe versus redshift plots.  
While the latter show the evolution of the $\alpha$/Fe ratio over cosmic time, the former represent the relationship between
the $\alpha$/Fe ratio and the metallicity of the system 
and is clearly independent of any choice of the cosmological parameters.

The behavior of the $\alpha$/Fe ratio in the [$\alpha$/Fe] versus [Fe/H] plots and the $\alpha$/Fe versus redshift plots can indeed appear contradictory. 
In the [$\alpha$/Fe] versus [Fe/H] plots, elliptical galaxies tend to exhibit higher $\alpha$/Fe values over the entire range of [Fe/H], indicating an overabundance
of $\alpha$-elements relative to iron. This is because in galaxies with intense star formation, such as ellipticals, the rapid production
of iron by type II supernovae results 
in high [Fe/H] values while the production of $\alpha$-elements by the same supernovae continues.

On the other hand, in the $\alpha$/Fe versus redshift plots, the situation is reversed.
This is because this plot illustrates the evolution of the $\alpha$/Fe ratio with cosmic time, and 
in a short time (or redshift) interval there can be a wide range of [Fe/H] values.  
On the other hand, since type Ia SNe produce iron but not $\alpha$-elements, at late times 
the $\alpha$/Fe ratio decreases over time, leading to a lower $\alpha$/Fe ratio at decreasing redshifts.
This results in the track of the elliptical galaxy exhibiting the highest $\alpha$/Fe over a very small redshift interval (corresponding to the duration
of the star formation period), followed by the lowest $\alpha$/Fe over most of the redshift range.  
In this picture, The [$\alpha$/Fe]--[Fe/H] diagram is the real diagnostic plot to infer the star formation history of the observed systems, whereas
the $\alpha$/Fe - redshift diagram depends mostly on  their age, and can be useful when used in conjunction with other plots (see Section \ref{sec_individual}). 

Based on the comparison between the model tracks and the data, in the [Si/Fe]--[Fe/H] plot uncorrected for dust 
the majority of the DLAs shows strong Si enhancement.

In the [Si/Fe] versus [Fe/H] diagram, the majority of the systems with very low metallicity ([Fe/H] < $-$2) 
are consistent with the expected pattern for spiral and irregular galaxies. 
As we move to higher metallicities ([Fe/H] > $-$2), we observe several data points that align with the predictions of the elliptical galaxy model.
There are also some points that fall in-between the elliptical and spiral/irregular curves. 
However, it is important to note that the observed trend indicating an overabundance of Si relative to iron-peak elements 
is robust only in the complete absence of dust contamination.

The right panels of Figure \ref{fig:cal03_3} show the comparison between the model predictions and observational data for the [Si/Fe] versus redshift and [Fe/H] plots,
after correcting all the observed abundances for the effects of dust depletion.  
The corrections were determined using a dust model described in Vladilo \cite{vladilo2002b},
which incorporates a scaling law for interstellar depletion.  
This scaling law allows for variations in the chemical composition of dust based on changes 
in the dust-to-metals ratio and/or the abundances in the medium.
In Figure \ref{fig:cal03_3}  one specific assumption regarding two dust-related parameters is presented, i.e., 
(1) the [Zn/Fe] ratio and (2) the percent variation of the abundance in the dust for a certain element, related 
to a variation of the abundance of the same element in the medium. 

For both parameters various values have been tested within reasonable intervals  \cite{calura2003} and, in all cases,
the $\alpha$-enhancement is considerably less pronounced than without the depletion correction. 
In general, the application of the dust corrections results in a reduction of the [$\alpha$/Fe] ratios by approximately 0.3--0.5 dex.
This reduction is significant enough to bring the ratios closer to the solar value, which improves
the agreement between data and models. 
Specifically, the corrected data indicate now a good agreement with models of spiral and irregular galaxies 
and not consistent anymore with the elliptical galaxy model. 
This discrepancy arises because elliptical galaxies exhibit excessively high [Si/Fe] ratios in the metallicity interval considered in Figure \ref{fig:cal03_3}. 
A similar conclusion regarding the improved agreement between the corrected [Si/Fe] ratios observed in DLAs and the values shown by
spirals and irregulars can be drawn from the [$\alpha$/Fe]-z plot. 
Distinguishing between typical low, smooth star formation patterns in irregular galaxies and spiral disks based solely on the redshift evolution
of abundance ratios such as [Si/Fe] can be challenging. As shown in the two figures, an irregular galaxy with a large gas content and
a very low star formation rate can exhibit a similar evolution of [Si/Fe] as the outer regions of a spiral galaxy. 
This similarity arises from the fact that both scenarios involve low-intensity star formation. 
Additionally, when considering the chemical evolution histories of a spiral galaxy at different radii, they can exhibit striking similarities.
The main distinction lies in the magnitude and timing of the initial peak of $\alpha$-enhancement,
which tends to be relatively brief in time. 
In contrast, the differences in evolution patterns become more pronounced when examining the plots as functions of [Fe/H].
This is true in particular for the red curve, representing the outermost regions of the spiral model that, due to their peculiar
star formation history, show the steepest decline in the right panels. 
The [$\alpha$/Fe]--[Fe/H] diagrams demonstrate their capability to discern between distinct chemical evolution histories 
and, more specifically, to identify the morphological types that most accurately represent the DLA population.\\

\vspace{-16pt}

\begin{figure}[H]
\includegraphics[width=10 cm]{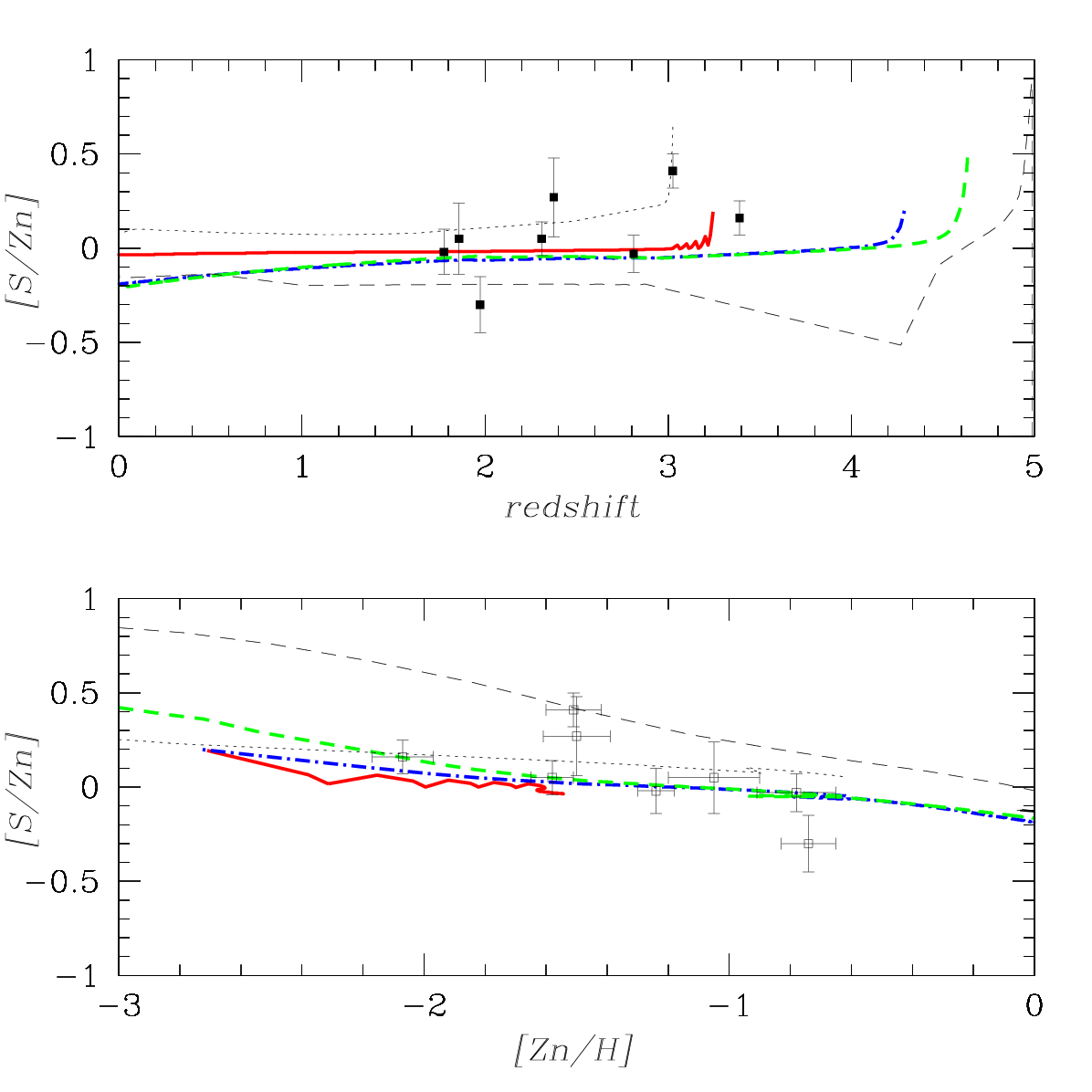}
\caption{[S/Zn] vs redshift (\textbf{upper panel}) and [Zn/H] (\textbf{lower panel}) observed in
  DLAs (black symbols with error bars, see Calura et
  al. 2003 \cite{calura2003} and references therein) and compared to chemical evolution models.
  The red solid, blue dot-dashed and green dashed lines show the results for the multizone model of a
 Milky Way-like spiral at different galactocentric distances, i.e., 18 kpc, 8 kpc and 4 kpc, respectively.
 The thin black dotted and black dashed lines show the theoretical evolution of [S/Zn] for an irregular and
 elliptical galaxy, respectively. } 
\label{fig:cal03_3}
\end{figure}  

The results discussed above, supporting mostly solar or sub-solar [$\alpha$/Fe] ratios in DLAs obtained
after the dust depletion corrections are applied to the observed abundances, are supported also
by the analysis of the abundance ratios between non-refractory (or more volatile) elements produced on different timescales.
A typical abundance ratio useful for this purposes in the [S/Zn],  
often regarded as a good measure as $\alpha$-enhancement, although with a few important caveats on Zn production \cite{nissen2007} and
on the refractory nature of S (see later on). 
Figure \ref{fig:cal03_3} shows the distribution of [S/Zn] as a function of redshift (above) and [Zn/H] (below), 
comparing models for galaxies of different morphological types with measurements in a sample of DLAs \cite{calura2003}. 
Despite the wide range of redshift values, in this limited sample a noticeable trend is evident from the top panel,
with the majority of data points aligning with the curves representing spiral and irregular galaxies.
In contrast, the evolution of the [S/Zn] ratio for elliptical galaxies exhibits a rapid decline, 
as it starts at high values ([S/Zn] > 0.5) in the initial phase and decreases to very low levels
even at redshifts below 4. 

Once again, the plot of [S/Zn] versus [Zn/H] diagram (lower panel of Figure \ref{fig:cal03_3})
provides with a more reliable indicator of chemical evolution
compared to the [S/Zn] versus redshift plot. It is noteworthy that both irregular and spiral galaxies 
 can effectively reproduce the majority of observed values.
 There is only one data point which exhibits the highest enhancement, that aligns
exclusively with the evolutionary trend of elliptical galaxies. Additionally, there is another DLA
system that exhibits a [S/Zn] value lower than what is predicted by any of the considered models. 
With the exception of these two systems, the [S/Zn] vs redshift and [Zn/H] plots confirm the same conclusions of the
dust-corrected [Si/Fe] vs redshift and [Fe/H] plots, i.e., that intrinsically, DLAs systems show little sign of
$\alpha$-enhancement and that their pattern is compatible with the one of mildly star-forming systems, such as spiral
discs and irregular galaxies. 
Although the results shown in \mbox{Figure \ref{fig:cal03_3}} pertain a limited sample of DLAs,
the same conclusions, i.e., a low ($\le$0.2) 
enhancement of S with respect to Zn, have been confirmed from other studies where different, more extended samples
were considered \cite{nissen2004,nissen2007,kulkarni2010,quiret2016}. 

It is important to stress that such conclusion is valid only if the following two underlying assumptions hold, i.e., that
(1) Zn nucleosynthesis is similar to the one of Fe,
i.e., these two elements trace each other; (2) both S and Zn are not severely affected by dust depletion. 
As for the first hypothesis, 
Zn is an Fe-peak element but, because of the observed trend of Milky Way stars in the [Zn/Fe]--[Fe/H] diagram
at very low metallicity in particular, i.e., below [Fe/H] $\sim$ $-$3, where [Zn/Fe] rises to values $\sim$ 0.3--0.5, 
refs. \cite{nissen2007,hirai2018} 
it is sometimes assumed that its production sites do not include type Ia SNe and this excess 
is explained by other contributors, such as electron-capture SNe \cite{hirai2018}.

As for DLAs, in very rare cases the measured metallicity (as traced by S and Zn) is lower than 0.001 solar \cite{bashir2019}. 
This is true also for [Fe/H] \cite{vangioni2015}, 
therefore  it is perhaps not unrealistic 
to assume that, in the metallicity range of DLAs Zn and Fe trace each other.
This implies that the conclusion regarding the little $\alpha$-enhancement indicated by the analysis of the [Si/Fe] and [S/Zn] ratios is valid.

\subsubsection{Study of Individual System}
\label{sec_individual}
In early studies, the DLA galaxy population was analyzed as a whole, and chemical evolution models
were optimised to interpret the observed abundance patterns in DLAs as an ensemble,
treating them as an evolutionary sequence \cite{matteucci1997,jimenez1999,hou2001,calura2003}. 
However, in order to gain a more detailed understanding of the individual DLA galaxies and their unique characteristics,
it is important to examine them on an individual basis and consider their specific chemical evolution histories.
However, various evidence supported the notion that DLAs trace galaxies with different evolutionary histories.
Besides low and intermediate-redshift deep imaging studies \cite{lebrun1997,nestor2002}
the large scatter observed in the $\alpha$/Fe peak element abundance ratios at a given metallicity
further support different progenitor populations and diverse evolutionary paths.
This emphasized the need for individualized analysis and consideration of diverse SFHs 
when studying the single DLAs, including quiescent discs, irregulars and 
proto-spheroids.

A few works attempted to constrain the SFH of DLA hosts focusing on the abundance pattern
of individual systems. 
The limited number of ions and elements typically detected in single DLAs makes it particularly difficult 
to assess photoionisation effects and the interplay between dust depletion and nucleosynthesis. 
To determine the intrinsic chemical abundance patterns in DLAs, unaffected by ionisation and dust depletion effects,
it is necessary to analyze various column density ratios of adjacent ions of the same element and of
different ionisation levels and consider the relative abundances of multiple \mbox{elements \cite{dessauges2004,dessauges2007}. }
In order to provide a comprehensive picture of these high-redshift galaxies,
Dessauges-Zavadsky et al. (DZ07) \cite{dessauges2004,dessauges2007} used measured elemental abundances and chemical
evolution models to constrain the nature of a sample of nine DLAs at \linebreak $z_{\rm DLA}$ = 1.7-2.5. 

These systems were chosen on the basis of: (i) a bright background quasar,
(ii) the detection of HIRES/Keck spectra, and (iii) minimal Lyman-$\alpha$ forest contamination to maximize access to different chemical species.
In order to satisfy the latter requirement, \linebreak $(z_{\rm QSO}$--$z_{\rm DLA}) < 0.6$ was required, but a velocity separation 
\mbox{$\Delta v(\rm QSO-DLA) > 3000 $ km s$^{-1}$} was also necessary to ensure that the two are not physically connected. 
These systems created a sample with a broad range of dust depletion levels, metallicities, and HI column densities, therefore
quite representative of a fair sample of the population of DLA galaxies. 
DZ07 produced a unique sample of DLA galaxies with thorough sets of constituent abundances
by merging high quality UVES/VLT spectra with pre-existing \mbox{HIRES/Keck spectra.}  

They quantified the column densities of 30 ions from 22 different elements, 
including C, N, O, Mg, Al, Si, P, S, Cl, Ar, Ti, Cr, Mn, Fe, Co, Ni, Cu, Zn, 
by detecting 54 metal-line transitions. This is quite remarkable, as in most DLAs just a few elements
(Si, Fe, Cr, Zn, and Ni) are often detected \cite{lu1996,prochaska1999,prochaska2001}.  

Again, to circumvent the problem of depletion and 
differentiate between dust-driven  and intrinsic nucleosynthetic patterns, different approaches
may be used. 
In the study of Z07, standard approaches were used, in which, in some cases, abundances between
weakly refractory elements were considered, or the contamination of dust was estimated to be low and
no correction was applied and, in most cases, 
depletion effects were quantified by applying dust corrections from existing studies. 
Also in this case, the observed abundances were corrected for dust depletion following the method developed by
Vladilo \cite{vladilo2002a,vladilo2002b}. 
Significant variations can be observed when comparing the absolute abundances of different iron-peak elements
(such as Cr, Fe, Ni and Zn) in some DLAs of the DZ07 sample. 
In contrast, Galactic stars exhibit a coherent tracking of Cr, Fe, Ni, and Zn, with relative values to Fe being solar-like
(except for the already mentioned enhancement of Zn at low metallicity). 
These variations in absolute abundances can be interpreted with the local interstellar depletion pattern \cite{savage1996}.
One DLA, the one toward Q1331+17 presenting [Zn/Fe] = +0.75, exhibits one of the largest depletion levels ever recorded.
Another system (DLA towards Q0100+13) shows [Zn/Fe] = +0.24, presenting therefore modest depletion.
Another approach involves the comparison of the absolute abundances of different  $\alpha$-elements.
In Galactic stars, the $\alpha$-elements generally exhibit similar trends within a range of approximately $\sim$0.20 dex. 
Also in this case, when comparing $\alpha$-elements with varying levels of dust depletion,
any deviation from their solar values can serve as an indication of the presence of dust.
For example, the DLA system towards Q0100+13 \mbox{presents [S/Mg] = $-$0.10 $\pm$ 0.12},
where Mg is known to undergo significant depletion in the Galactic ISM, while in this case S is considered a non-refractory element.
This ratio is consistent with the [Zn/Fe] and supports a small presence of dust in this system. 
On the other hand, the DLA toward Q1331+17 shows [S/Si] = +0.16, that would indicate significant presence of dust,
if it was not for the measured [S/Mg] = $-$0.07$\pm$ 0.18, that supports no substantial depletion. 
The latter result can be attributed to the limited accuracy of the Mg II column density measurement, in
particular due to blending effect that is frequent in the Lyman-$\alpha$ forest. 
This highlights the caution required in the interpretation of such abundance ratios. 
The method used by DZ07 to calculate the dust corrections groups together a series of 
models based on different assumptions regarding the relative abundances of key elements, such as the [Zn/Fe] ratio. 
The assumptions on this ratios are differentiated by 0.1 dex and can lead to variations in the calculated
intrinsic abundances of the same magnitude. 
The observed absolute abundances between different elements are calculated by summing the contributions of various transitions (e.g.,
the Fe abundance includes various transitions, including the FeII 2234, Fe II 2260 and other).
Particular care is taken to avoid the overestimation of the abundances obtained from strong metal-line profiles
compared to those derived from weaker metal-line profiles. 
For instance, as for Fe, the strongest Fe II transitions observed in DLA systems are typically associated with UV multiplet systems
and include Fe II 2344, Fe II 2382 and Fe II 2600. 
On the other hand, the weakest Fe II transitions are often associated with higher energy levels and lower oscillator strengths, such as
Fe II 2249, Fe II 2260 and Fe II 2374. 
While these transitions are weaker, they can still be measured in DLAs with high-resolution spectroscopic observations.
It is important to note that the strengths of these transitions can vary from system to system, depending on factors such
as the metallicity and other physical conditions, such as kinematics.
When dealing with very weak lines, such as, e.g.,  Ti II lines, it is possible to underestimate the [X/Fe] ratios by as much as 0.3--0.4 dex.
This occurs when the total Fe abundance is estimated based on strong Fe II lines alone.   
The weaker lines, like Ti II, may not exhibit the same strength and therefore may not accurately reflect the true abundance ratios.
In some of the DLAs studied by DZ07, this effect is
particularly important, in particular in the few cases showing complex metal-line profiles, with a large number of components.
One example of the measured abundance pattern in one of these systems is shown in Figure~\ref{Q0100+13}, for the DLA at $z$ = 2.309
toward Q0100+13, useful to visualize all the available abundances as a function of the atomic number and the
deviations from the solar values. 
The dust-corrected abundances are represented on a logarithmic scale and 
the reference point for this scale is hydrogen, which is defined to have an abundance of $\epsilon(\text{H}) = 12$.
Similarly, for any element X, its abundance is given by $\epsilon(\text{X}) = \log(\text{X/H}) + 12$.
These abundances are then compared to the solar abundance pattern, in this case the one of by Grevesse \& Sauval~\cite{Grevesse1998}, 
represented by a red solid line in Figure~\ref{Q0100+13}. 
The solar abundance pattern is scaled to match the observed sulfur metallicity of the DLA system.
The analyzed system exhibits a slight enhancement in the abundances of $\alpha$-elements,
specifically S and Mg, relative to the abundances of Zn and Fe.
The measured values indicate a ratio of \linebreak \textls[-15]{[S/Zn] = +0.19 $\pm$ 0.09 and a corrected ratio of \mbox{[Mg/Fe] = +0.29 $\pm$ 0.11}.
This enhancement of  $\alpha$-elements relative to Fe-peak elements is indicative of enrichment by massive stars,
suggesting their contribution to the elemental composition in this system}.
After dust correction, the ratio between Mg and S  is \mbox{+0.10 $\pm$ 0.12}, indicating that these elements closely track each other. 
On the other hand, other $\alpha$-elements such as argon (Ar) exhibit a slightly lower abundance compared to S.
In contrast, the Fe-peak elements, including Fe, Ni, and Cr, demonstrate solar values relative to each other, 
similar to Galactic stars with similar metallicities.

In this particular system, an interesting phenomenon known as the odd-even effect is observed. This effect manifests
as an underabundance of elements with odd atomic numbers compared to elements with even ones 
that originate from the same nucleosynthetic processes. 
Specifically, this system has [P/S] = $-$0.37 $\pm$ 0.12 when \mbox{[P/H] = $-$1.85 $\pm$ 0.13},
indicating an enhanced odd-even effect. This value is similar to measurements performed in other
DLAs \cite{levshakov2002,molaro2001}, 
tentatively suggesting the possible influence of pair-instability supernovae/hypernovae
in the nucleosynthetic processes responsible for these elements~ \cite{cayrel2004,fenner2004}. 
It is nevertheless worth stressing that the measurements of phosphorus (P) abundances in DLAs
hold significant importance in elucidating the nucleosynthesis processes responsible for this element. 
This is noteworthy since its detection is particularly problematic in cool Galactic stars \cite{caffau2011}. 

The DLA under examination exhibits an [N/S] ratio that is below the solar value, 
specifically measuring at $-$0.83$\pm$ 0.14 when [S/H $] = -1.48\pm 0.11$. 
Nitrogen (N) is a particularly intriguing element due to its complex nucleosynthetic origin.
It is generated through the CNO cycle in stars, and its production 
can be categorized as either secondary or primary. In the secondary scenario, N is produced in proportion to the initial carbon
(C) and oxygen (O) content within the star. 
In the primary scenario, N can be produced ``in situ'' using C and O synthesized within the star itself. 

In the N/$\alpha$ versus $\alpha$/H diagram, the ratio of secondary N to $\alpha$-elements is expected to increase steeply with increasing metallicity,
while the ratio of primary N to $\alpha$-elements remains relatively constant as metallicity increases
\cite{talbot1974,matteucci1986}.
The exact progenitors responsible for primary N production have not been fully constrained yet. However,
stellar models  incorporating stellar rotation indicate that intermediate-mass stars play a significant role in N production 
\cite{meynet2002}. 
The measured [N/S] ratio DLA towards Q0100+13 corresponds to a ``high'' value, meaning it is close to the primary N ``plateau''
at [N/$\alpha$] $\approx$ $-$0.75, as measured in blue compact dwarf galaxies from N and O abundances \cite{izotov2004}. 
This value falls towards the upper end of the range of N/S ratios observed in DLAs and is in good agreement with typical values
observed in H II regions at similar metallicities \cite{pilyugin2002,izotov1999}.  
\begin{figure}[H]

  \includegraphics[width=12 cm]{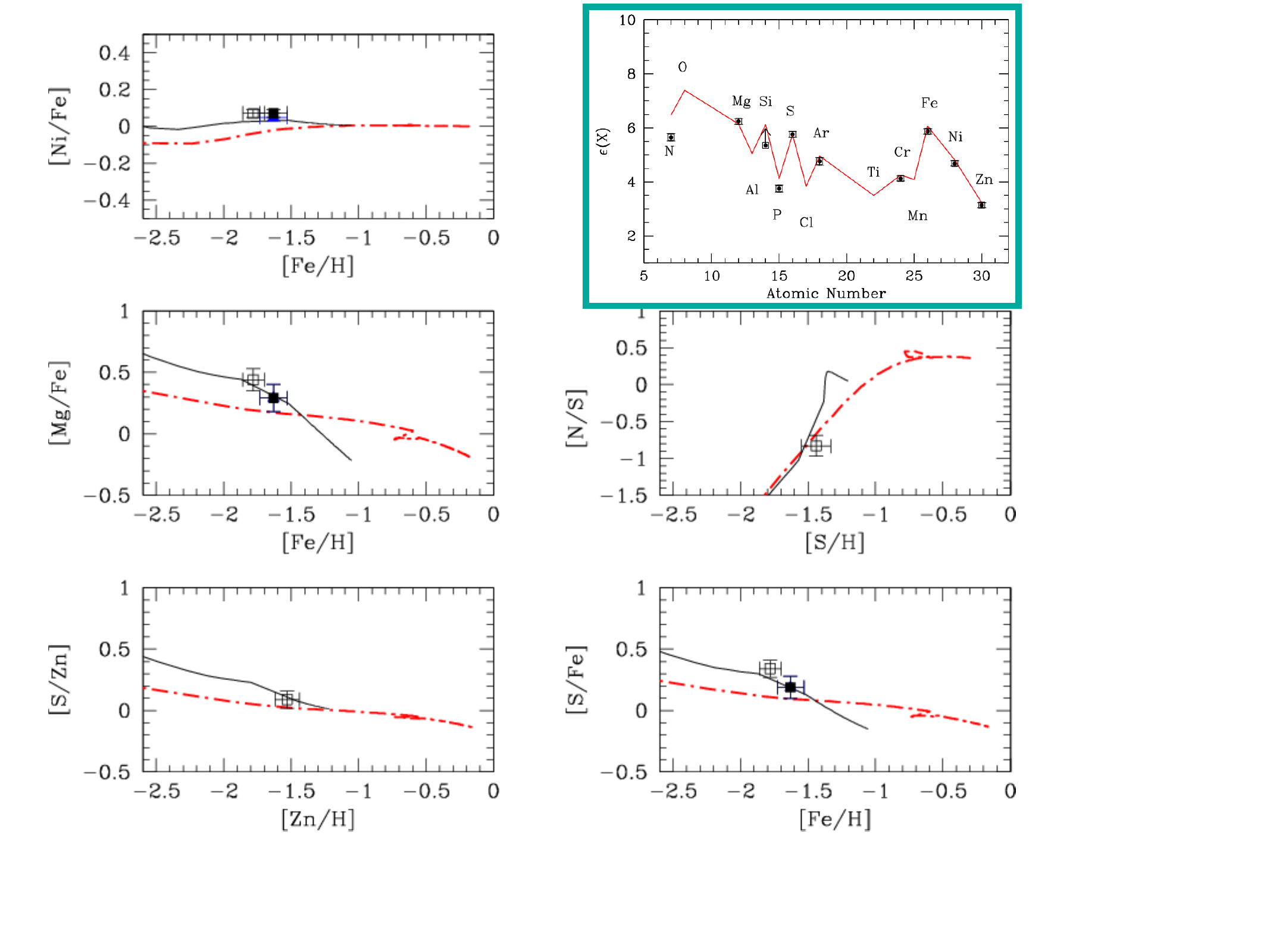}
\caption{Abundance ratios of the DLA at $z_{abs}$ = 2.309 toward Q0100+13.
  The plot on the top-right, enclosed by a solid dark-cyan box,
  is  the nucleosynthetic abundance pattern (i.e., the dust-corrected abundances) of the DLA, defined as
  $\epsilon(X) = log(X/H)+12$, in which hydrogen is defined to have $\epsilon(H)$ = 12. The red solid line
  represents the solar abundance pattern \cite{Grevesse1998}, normalized to the same sulfur abundance value.
  The filled circles and are the dust-corrected abundances obtained with the E00 model of \cite{vladilo2002b}.
  In the other panels, the thick red dash-dotted lines and thin black solid lines are
  for a Solar Neighbourhood and dwarf irregular model, respectively, whereas the open squares, filled triangles and 
  filled squares are the measured abundances, dust-corrected with the E00 and with the E11 models of \cite{vladilo2002b}, respectively.
  Figure adapted from Dessauges-Zavadsky et al. (2004)\cite{dessauges2004}.}
\label{Q0100+13}
\end{figure}

The analysis of various abundance ratios shown in Figure~\ref{Q0100+13} is useful to constrain the SF history of the system
under consideration, in particular when compared with results from chemical evolution models. 
In general, from the theoretical point of view, while the absolute abundances depend on the most fundamental
model parameters regulating the SFH (SF efficiency, gas accretion or outflows),
the relative abundances are sensitive to the nucleosynthesis, the IMF and the stellar lifetimes.
When the relative abundances are examined together with the absolute abundances such as the metallicity, in this particular
case traced by [Fe/H] and [Zn/H], they turn out as  particularly sensitive to the SFH. 
In this DLA system, the dust corrections are minimal, as [Zn/Fe] $\sim$ +0.25 $\pm$  0.04. 

Dessauges-Zavadsky et al. \cite{dessauges2004} compared the measured abundances with the ones calculated with chemical evolution models,
assuming a variety of star formation histories and describing spiral, elliptical and irregular galaxies.
{To assess the accuracy of the model in reproducing the data, they used a statistical method that considers, for each model,
the minimal distance between the data point and the model predictions. 
Once the minimal distances were computed for all the abundance diagrams, a weighted mean was determined using
the 1-$\sigma$ errors of the measures as weights. The best model was finally derived from the comparison of the weighted
means obtained for each model.  }

For Q0100+13, Dessauges-Zavadsky et al. (2004)\cite{dessauges2004} examined one multi-zone model for a MW-like spiral galaxy, computed at various galactocentric radii,
finding the best agreement for a Solar Neighbourhood-like SFH, i.e., for the model run at a galactocentric distance R = 8 kpc.
However, the data could be reproduced also by a ``dwarf irregular'' model used previously to analyse the abundances of
local blue compact galaxies \cite{lanfranchi2003}. 
Such a model has a similar SF efficiency as for the spiral model, but a much shorter SF duration, i.e.,  $\Delta t = 0.07$ Gyr. 

The fact that two models with rather different SFHs reproduce the data with a similar degree of accuracy
indicates a substantial degeneracy between the model parameters.   
Despite this inconvenience, more useful information come from the study of abundance ratios as
a function of redshift. By assuming various values for the formation redshift of the system,
that is associated to the epoch where star formation began, and using again a statistical
method similar to the one mentioned above to break, or limit the parameter degeneracy,
Dessauges-Zavadsky et al.\cite{dessauges2004} found for this DLA a young age of $\lesssim$250 Myr.
Therefore, the enrichment of this system is primarily driven by the products of massive stars, with a minimal contribution from type Ia SNe. 

These works confirmed that, in most cases, DLAs are young galaxies and 
may either be associated to the outer regions of spiral galaxies, 
or to dwarf irregular galaxies characterised either by low-intensity, or continuous or bursty star formation activity. 
This is in line with what found by most authors; overall, the cases where DLAs could be associated with strong starbursts were rare \cite{levshakov2002}. 
At variance with other high-redshift systems such as Lyman-break galaxies, DLAs were never found to represent significant contributors
to the total comic SFH \cite{hopkins2005}.

\subsubsection{A Chemical Evolution Model with Dust: Results for Damped-Lyman Alpha Systems}
After the development of new chemical evolution models with dust and the extension 
to a variety of environments, new insights came from the application of such models to the study of real
systems.  The use of the models for the interpretation of rich datasets of systems with a complex dust composition
and carrying a lot of information, such as DLAs, turned out as 
particularly fruitful.
 
Gioannini et al. \cite{gioannini2017a} used a chemical evolution model for dwarf
irregular galaxies which accounts for the presence of dust to
interpret the abundances observed in a large set of 34 DLA systems with associated column density measurements
of various species. 
They focused on two refractory
elements, iron and silicon, that are frequently measured and used in DLAs 
to probe their chemical evolution and dust content. 
To fine-tune the parameters of their DLA model, they also used volatile elements such as zinc and sulfur. 

Local interstellar observations \cite{jenkins2009} and condensation temperatures estimated for large sets of elements \cite{lodders2003}
indicate that zinc and sulfur are primarily volatile, although the case of sulfur remains somewhat ambiguous \cite{calura2009,jenkins2009}. 
In the highest-density gas, i.e., in molecular gas, some amount of sulfur may be incorporated into dust \cite{cazaux2022}; 
however, this is not a cause for concern since the molecular fraction is generally very low in DLA systems
\cite{ledoux2003}. 
They first tuned the parameters of their galactic chemical evolution model
from the analysis of the S/Zn ratio, which was assumed to be undepleted in both models and data. 
They used S and Zn as references to measure the relative abundances of refractory elements. 
As an initial step in their procedure, they adjusted the model to account for the observed S/Zn ratio. 
They conducted tests on the input parameters by varying factors such as the wind and  
star formation efficiency, the infall mass and the IMF.

Figure~\ref{fig_gio17} presents the relative abundance ratios of refractory elements (Si and Fe) to volatile elements (S and Z)
in relation to the absolute abundances of the corresponding volatiles. They analyzed various possible combinations of these elements,
i.e., Si/S and Fe/S as a function S/H  and Si/Zn and Fe/Zn vs Zn/H, shown in the upper and bottom panels, respectively. 
For all the dust production and destruction processes, they adopted improved prescriptions with respect to the ones of previous models.

\begin{figure}[H]

\begin{adjustwidth}{-\extralength}{0cm}
\centering 

\includegraphics[width=15 cm]{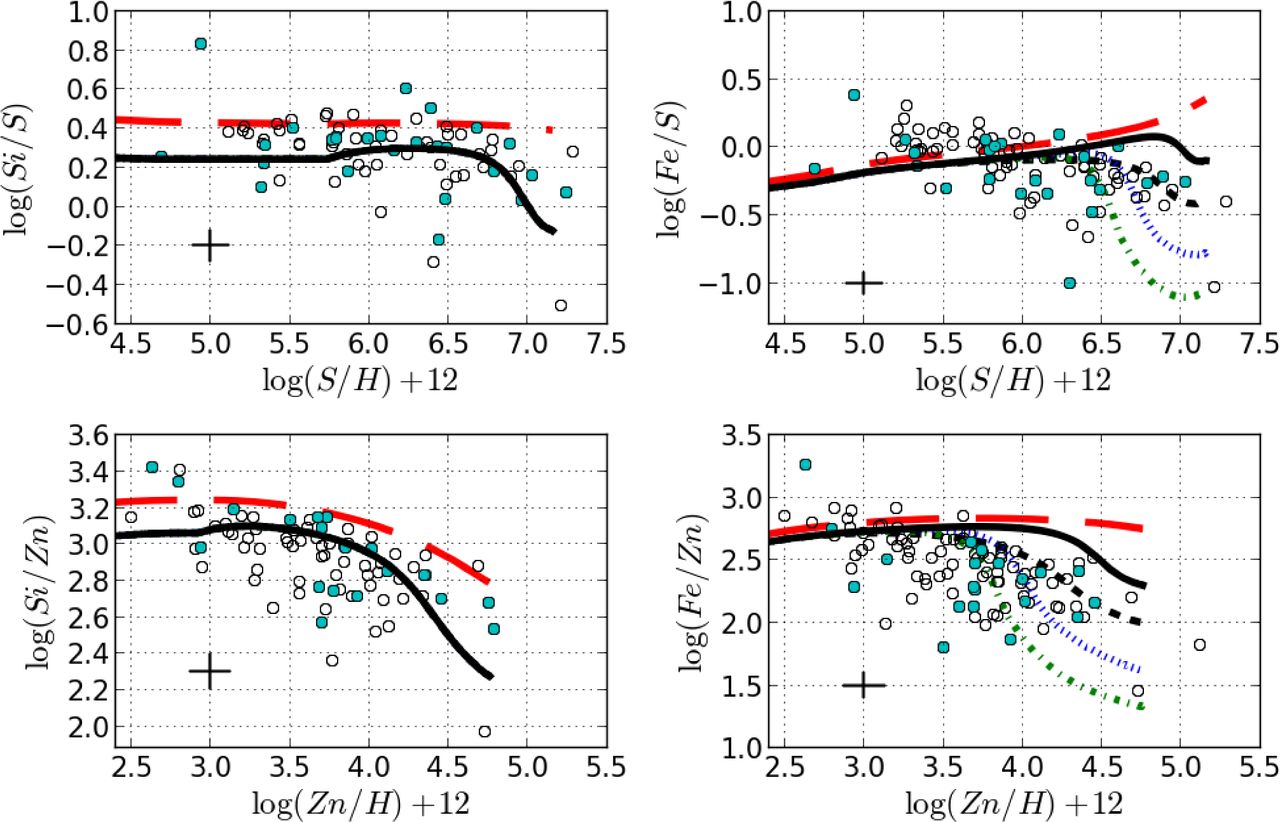}
\end{adjustwidth}

\caption{Abundance ratios of the refractory elements Si (\textbf{left}) and Fe (\textbf{right}) over volatile elements S (\textbf{top panels}) and Zn (\textbf{bottom panels}) as a
  function of the volatile absolute abundances observed in DLAs (black and dark-cyan symbols, see the compilation in Gioannini et al. \cite{gioannini2017a})
  and calculated in chemical evolution models that incorporate dust evolution.  
  The red long-dashed lines represent the 'cosmic' ISM abundance ratios of a model for a typical irregular galaxy,
  whereas the black solid lines represent the abundances in the gas, i.e., computed after the removal of a fraction of the element in dust.
  The black short-dashed lines represent a model including iron dust production from Type Ia SNe.
  The blue dotted and green dashed–dotted lines show the predictions computed assuming a 5 and\mbox{ 10 times} more efficient interstellar iron accretion,
  respectively. The figure is from \mbox{Gioannini et al. \cite{gioannini2017a}.}}
\label{fig_gio17}
\end{figure}

For the dust condensation efficiencies in stellar sources, Gioannini et al. \cite{gioannini2017a} used updated mass- and metallicity-dependent
models \cite{piovan2011}. Moreover, they adopted metallicity-dependent 
prescriptions for the swept-up mass, a key parameter regulating the destruction rate \cite{asano2013}, and
for the accretion rate. 
The model results are compared to a compilation of (S/Zn) vs (Zn/H) measured in DLAs \cite{vladilo2011}. 
The difference between the undepleted (red dashed lines) and depleted (black solid lines)
abundances, representing the ones measured in the gas phase, is clearly visible. 
In particular, the intrinsic abundance ratios are significantly higher than the depleted ones. 
This confirms that it is legitimate to expect the refractory-to-volatile abundances
to lie above the measured ISM abundances, as shown by the fact that the predictions
for the ISM models for the abundances in gas+dust lie above most of the data. 
The instances where the measurement ratios exceed the model may result from either the natural variation
within the DLA sample or the uncertainties associated with the stellar yields.
that are particularly significant at low metallicity \cite{romano2010}. 
It is interesting to analyse the different trends show by the elements considered in this figure, and in particular
the differences of Si (left panels) and Fe (right panels).
The predicted gas-phase (Si/S) shows a nearly flat trend with metallicity. 
On the other hand, the abundances measured in DLAs and the models taking into account depletion
show a flat trend at low metallicity, followed by a decrease at higher metallicity.
A similar behaviour is shown also by the (Si/Zn) ratio, even though this ratio
is slightly less constant, i.e., weakly decreasing, at increasing metallicity.  
Both diagrams suggest that the amount of silicon in dust tends to increase
over the course of galactic chemical evolution. 
These authors achieve a good agreement between the model and the data for both Si/S and Si/Zn. 
Therefore, combined together, the prescriptions for accretion, dust production by SNe 
and AGB are sufficient to account for the observed depletion pattern of Si in DLA systems. 
It is noteworthy that Si shows traces of depletion even at the lowest metallicities, 
which supports a scenario where Type II SNe contribute rapidly to Si dust production.

The right-hand panels of Figure~\ref{fig_gio17} show the pattern of iron abundance
in relation to volatile elements. 
The effect of iron depletion tends to disappear at the lowest metallicities;
this is suggestive that the processes responsible for the production of iron-rich dust occur over
longer time scales than those typical of Type II SN explosions. 
The interstellar (Fe/S) shows a light increase, 
that disappears at the highest metallicity values once the effects of dust are
incorporated in the models. 
Additionally, the difference between the total ISM model and the gas-phase one increases with metallicity. 
However, unlike the case of Si, the gas-phase model (black solid lines) does not seem to match the observed data.  
This is visible at high metallicity (log(S/H)+12 $\gtrsim$ 6.5), where the model predicts significantly
higher values than the observed data. 
In the interpretation of the current models, this 
indicates that it underestimates the amount of iron present in dust. 
In this model, only the accretion process is considered to make a significant contribution,
while the dust production by Type II SNe and AGB stars
contributes a negligible fraction to the overall Fe abundance in the dust. 
The authors notice that, 
even if they considered enhanced contributions from either Type II SNe or AGB stars,
the total amount of iron in dust would be negligible compared to the vast quantity of Fe ejected by Type Ia SNe in the gas phase. 
Essentially, their model cannot predict any significant iron depletion until the metallicity 
reaches a level high enough for the accretion process to become important. 
Therefore, they postulated that an additional source of iron dust production 
needs to be included to resolve this discrepancy.
To enhance the production of iron dust, the authors considered two possibilities.
First, they tested other dust prescriptions independent on metallicity and used in previous works \cite{dwek1998,calura2008}. 
The results obtained in this case are shown  with black short-dashed lines in Figure~\ref{fig_gio17}. 
Despite the  arbitrariness and substantial lack of support of the previous prescriptions, they offer 
an improvement in the fit to the data. 
As a second possibility, they assumed that most of the iron is incorporated into a solid component,
which is different from silicates and for which they assume an enhanced accretion efficiency.
In particular, they reduced the Fe accretion timescale to increase the efficiency of Fe dust production in the ISM.
The blue dotted lines and green dot-dashed lines show the results for an Fe dust timescale decreased by a factor 5 and 10, respectively.
This assumption leads to a significant improvement in the agreement between data and models and is motivated
by a series of independent studies. 
Some works provided evidence for an iron-enriched population of compounds present in dust grains in various forms, such
as triolite (FeS), kamacite (FeNi) or oxides (FeO), whose presence is detected in local interstellar clouds \cite{kimura2003}.
Moreover, Fe and Si are commonly thought to be locked together in silicate compounds, such as olivine and pyroxene \cite{zeegers2017} 
and are often regarded as characterised by a similar interstellar depletion level,  
but they could undergo a different degree of incorporation into dust grains.
In addition, Small Magellanic Cloud observations indicate that the depletion of Fe and Si may diverge \cite{sofia2006}, suggesting that 
Si may not always tied to the same grains as Fe.
A study on dust depletion across 196 different sight lines in the Milky Way \cite{Voshchinnikov2010} 
supports the idea that silicate grains cannot consist solely of olivines and pyroxenes and that, instead, some portion of iron
must be found in another population of dust. 
Mid-IR observations indicate that a population of 'iron needles' contribute substantially to the extinction of the Galactic Centre \cite{dwek2004a}
and can be significantly present in local SN remnants, such as SN 1987A \cite{wickramasinghe1993} and 
Cas A \cite{dwek2004b}.
It could be that these  Fe-rich grains constitute a particularly resistant population of dust that, at variance with other species,
might be capable of surviving even in the harshest ISM conditions \cite{gioannini2017a,vladilo2018}. 

As final note, the two different approaches to study the abundances measured in DLAs of
(i) applying depletion corrections to the data, and use models that do not take dust into account, and (ii)
consider uncorrected data and interpret it with models that incorporate dust depletion, lead to the same
conclusions regarding a primary role of differential depletion in these systems and, in particular,
supporting a stronger effect for Fe with respect to other elements, such as Si. 

\subsection{Dust-Depleted Interstellar Abundances in GRB DLAs}
Gamma-ray bursts (GRBs) are other important tools for studying the early universe at high redshifts.
They originate at cosmological distances, and their typical duration ranges from a few milliseconds to $10^3$ seconds.
GRBs that last more than 2 seconds are generally referred to as long GRBs (LGRBs \cite{kouveliotou1993}). 
The extreme brightness of their afterglows-namely, the prolonged, lower-energy radiation
(in the X-ray, optical, and radio bands) that is visible for several days following the GRB event---has enabled detailed studies of the physical properties of the ISM in their host galaxies.
Thanks to various space facilities,  hundreds  of  GRBs  have
been detected even up to very high redshift \cite{tanvir2009,salvaterra2009} (at the time of the writing, the highest estimate is
is GRB 090429B with a photometric redshift of z $\sim$ 9.4 \cite{cucchiara2011}). 
In fact, as most GRBs come with  associated UV/optical afterglows,  
a subset of these exhibit apparent magnitudes that are sufficient for high-resolution spectroscopy using telescopes of 10 meters or larger. 
LGRBs are often associated with the death of massive stars, i.e., to CCSNe of Types Ib and Ic. 
Various studies of GRB afterglows have clearly indicated a connection between tens of LGRBs and these types of \linebreak supernovae \cite{hjorth2012,cano2017}.  
In this framework, the afterglow spectra turned out as powerful probes of distant star-forming galaxies.  
The constraints on the underlying progenitors collected so far favor both single progenitor, i.e., the  Collapsar \cite{macfadyen1999,woosley2006}
or the Millisecond Magnetar \cite{wheeler2000,bucciantini2009} and Binary progenitor models \cite{fryer2005,detmers2008}. 
Present-day GRB progenitor models suggest that these events originate from massive, fast-rotating, and low-metallicity objects
\cite{hjorth2003,woosley2011}. However, this paradigm is questioned by the  discovery of a few GRBs
that may have occurred
in environments with solar or even supersolar metallicity \cite{prochaska2009,kruehler2012}. 
The association of LGRBs with  the  explosion  of  massive  stars may be of multifold, cosmological
value. In principle, they can be useful to 
reveal even the death of the first stars occurring at very high redshift \cite{petitjean2011} 
and to probe the cosmic epoch when the IGM was reionised \cite{kistler2009}.
Afterglow absorption spectroscopy provides a unique tool to probe some fundamental features of the GRB environment. 
In fact, the power-law afterglow spectrum contains features that reflect the gas present in the ISM of the host galaxy, 
in contrast to studies of quasars, which emit integrated photon output that extends out to several tens of kiloparsecs. 
Additionally, quasar sight lines often pass through foreground star-forming galaxies such as DLAs 
and showcase  different characteristics compared to the afterglow spectra.
As DLAs detection is strongly linked to the neutral gas cross section, the QSOs lines of sights are expected to intersect
only rarely the compact and dense active star-forming regions. 
In this regard, GRB afterglow spectra provide complementary information to QSO-DLAs and insights into various ISM phases in star-forming galaxies.
These phases include 
the neutral ISM of the host galaxy and diffuse gas present within the galactic halo, 
the H II region created the progenitor itself and nearby OB stars and the circumstellar material
surrounding the progenitor \cite{prochaska2008}. 

GRBs are initially identified based solely on their high-energy emissions. Consequently,
the host galaxies can be studied in great detail after the afterglow emission fades, enabling
access to samples of star-forming galaxies that are 
unbiased with respect to their intrinsic luminosity. Therefore, GRBs and their host galaxies can be used to trace
cosmic star formation independently of magnitude-limited samples, such as Lyman-break galaxy surveys \cite{jakobsson2011,hjorth2012},
although also GRB hosts are likely affected by dust extinction \cite{watson2012} and metallicity biases \cite{cucchiara2015}. 

High-resolution GRB spectroscopy performed with Echelle spectrographs 
has allowed the determination of the 
chemical abundance pattern \cite{savaglio2003,prochaska2004,penprase2006,wiersema2007} of the local ISM, 
as well as 
the determination of a few quantities related to their host 
galaxies, such as their dust content, their stellar mass and their 
 star formation rate \cite{bloom1998,christensen2004,savaglio2009}.\\
 This required rapid localization by the Swift satellite of the hard X-ray emission within a few arcminutes 
  and of the soft X-ray component to a few arcseconds \cite{osborne2005},
 which have enabled observers to obtain high signal-to-noise (S/N), high-resolution spectra 
 of GRBs with 10 m class telescopes \cite{prochaska2007}. 
 Chen et al. \cite{chen2005} were among the first to report on the chemical abundances measured in one 
DLA system associated with the host of one GRB discovered by Swift, namely GRB 050730 at
z = 3.96855. 
The analysis of the Echelle spectra of GRB 050730 revealed a forest of narrow absorption lines
due to heavy ions including O I, C II,  Si II, Fe II, C IV, Si IV, allowing them to derive the column
density and the abundance of several chemical species.
Following a standard practice for the analysis of interstellar lines observed in DLAs, 
the column densities have been converted into gas-phase chemical abundances,
presented in Prochaska et al. \cite{prochaska2007a}, assuming negligible ionisation corrections. 
The very large H I column density values and the analysis of their abundances indicated the occurrence of
the absorptions in dense  systems, 
generally enriched to metallicity values higher than the cosmic average at the sampled 
redshifts ($z\gtrsim 2$) and typically hinting for an $\alpha$-enhanced abundance pattern \cite{prochaska2007a}.
From these first studies it became immediately clear that GRB DLAs 
preferentially probed dense gas, associated with the innermost, most metal-rich regions of galaxies. 
For these particular properties, one fundamental aspect which concerns the use of GRBs as tools to investigate
the properties of high-redshift galaxies is that they represent probes which are completely independent from 
and complementary to other systems, such as QSO DLAs.

Cucchiara et al. \cite{cucchiara2015} compared large samples of detected GRB- and QSO-DLAs and studied the evolution of their
metallicity across a broad redshift range (0 < z $\sim6$).
Their GRB-DLA sample was drawn from all the GRB afterglows observed during the 2000–2014 time span with S/N $\ge$ 3
and for which HI and metallicity measurements could be obtained.
They also used a complete, up-to date list of high-resolution, high S/N QSO-DLAs.
They analysed the H I column density distributions of their samples and presented very clear evidence that
GRB-DLAs typically trace a very dense phase of the ISM, likely associated  to regions of on-going star formation
such as molecular clouds, whereas QSO-DLAs probe a lower column density medium, possibly farther from the densest galactic regions. 
The redshift revolution of the metallicity of the two samples is visible in Figure~\ref{fig_cucchiara}, which shows that at $z\gtrsim 2$ 
GRB-DLAs probe more metal rich regions than QSO-DLAs (with a maximum difference in the average trends of $\sim$0.5 dex)
and that the GRB-DLA average metallicity seems to decline at a lower rate than the one of QSO-DLAs.

\begin{figure}[H]

\includegraphics[width=13 cm]{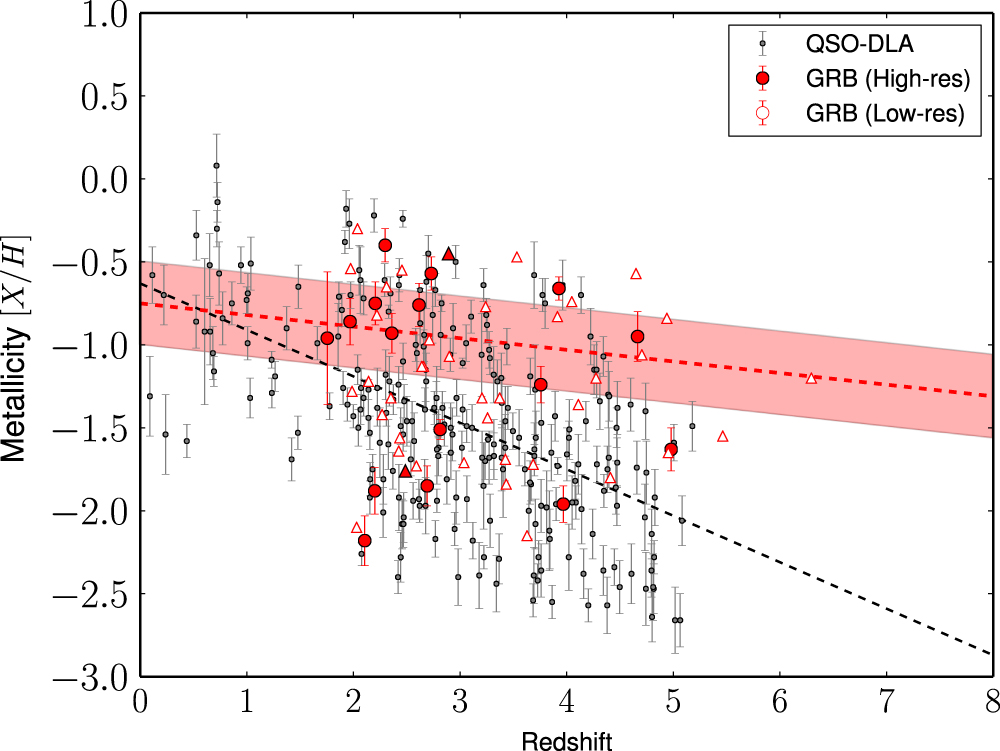}
\caption{Metallicity (as traced by the elements S, Fe, Zn, Ni) evolution of QSO- (gray data) and GRB-DLAs (red data) as a function of redshift
  (see Cucchiara et al. \cite{cucchiara2015}), where no depletion correction is applied to the measured metallicities.  
  The red upward triangles, filled and open symbols indicate lower limits, high- and low-resolution measures, respectively.
  The dashed black line is a linear fit of the QSO-DLA data. 
 The red dashed line is a linear regression fit of the GRB-DLA data, whereas the
 shaded area represents the 1-$\sigma$ error in the fitting parameters, obtained using 500 bootstrap iterations.} 
\label{fig_cucchiara}
\end{figure}

\subsubsection{Abundance Pattern  of Individual GRB-DLAs}

In principle, the affinity of GRB-DLAs with actively star-forming galaxies 
with sites containing young, massive stars should correspond to an abundance pattern different from the 
one of the QSO DLAs. 
The first attempt to derive the SFH of a set of individual GRB DLAs by means of their
abundance pattern was presented in Calura et al. \cite{calura2009}. 
From the simultaneous analysis of the abundance ratios between various elements as a function of metallicity
(traced by multiple elements such as [Fe/H], [S/H], or [Zn/H]), and by means of chemical evolution models for galaxies of different morphological types and
including dust depletion, Calura et al. \cite{calura2009} were able to derive constrains on the nature and age of 
some of the host galaxies of the sample, which included 4 GRB-DLAs.  
They tested models both with and without dust evolution.
Overall, the study of the abundance pattern did not indicate strong differences between GRB- and QSO-DLAs. 
Their analysis supported the hypothesis that long-duration GRBs occur preferentially in low metallicity, star-forming galaxies, typically characterized by low SF efficiencies (of the order of \mbox{$\nu\le 0.1$ Gyr$^{-1}$}). 
The study also evidenced a dust content in agreement with the one derived using different diagnostics, i.e., estimates
of the dust-to-gas ratios based on metallicity-dependent relations calibrated on local galaxies, such as the
Small Magellanic Cloud \cite{prochaska2007a}.  
A few systems show peculiar abundance ratios, such as very high [Zn/Fe] values, which,
despite the uncertain origin of Zn, are difficult to explain by means of stellar nucleosynthesis and, therefore, reveal a high depletion level for Fe.
This is at variance with some expectations that, despite they are likely to originate in nearby molecular clouds,
the environment of GRBs is unlikely to show large molecular or dust content, mostly because the X-ray and UV components
of the afterglow can destroy dust and molecular hydrogen out to 10$-$100 pc \cite{fruchter2001,prochaska2007a}.
One possible explanation is that in some cases, absorption  may occur at larger distances from the GRB.
D'Elia et al. \cite{delia2014} reported that it is not uncommon to have components located farther than 200 pc from the GRB.
In such cases, the properties of the ISM are likely to be different than the ones of the region in which the GRB originated
and even from the one affected by the afterglow \cite{petitjean2011}. 
Indeed, the possible bias effects of dust in GRB samples are still under debate.
{The high metallicity presented by some systems suggests that, in some cases, highly 
  dust-depleted regions are likely to be probed by GRB-DLAs \cite{cucchiara2015}.
  Moreover, due to their star-forming character, it is not unlikely that GRB host galaxies will present a significant amount of dust. 
In fact, by means of mid-IR and radio observations it was showed that 
some GRB host galaxies are characterized by a large amount of star formation obscured by dust \cite{berger2005,lefloch2006}. 
Other, separate studies point towards significant fractions of GRBs occurring in dusty environments \cite{perley2013,hatsukade2012}. 
A high level of dust extinction characterising the host galaxy may also be one likely explanation for the so-called dark GRBs,
namely the events in which no optical/near-infrared afterglow is detected, or where
the optical afterglow emission is lower than that expected from the X-ray afterglow \cite{fynbo2009}. 
If a fraction of dark GRBs explode in regions dusty enough to make their afterglow undetectable in the optical band, 
then the current samples of hosts are likely biased against dusty and high-metallicity galaxies \cite{petitjean2011}.}

By means of chemical evolution models for ellipticals, spirals and irregular galaxies and including the effects of dust depletion,
Grieco et al. \cite{grieco2014} extended the previous studies to three more GRB-DLAs with detected abundances for several elements,
namely GRB 081008 \cite{delia2011}, 120327A \cite{delia2014} and 120815 \cite{kruehler2013} and comparing model results with the measured abundances. 
The three models were meant to represent an elliptical, a spiral and an irregular galaxy with average properties and were
calibrated in order to account for some basic observables of each morphological 
type, such as their present-day SFR, as well as their dust abundance pattern, where available, and the observed dust-to-gas ratios. 
Once again, the aim was to constrain the history of SF and have some insight on the nature of the host from the analysis of the
abundance pattern for a set of chemical elements (C, N, O, Mg, S, Si, Ni and Zn). 
In the models used in this study, among these species, only N and Zn are not affected by depletion effects.

In Figure \ref{fig_grieco14}, Grieco et al. \cite{grieco2014} present the results for the host of GRB 081008 at a redshift of $z = 1.968$ \cite{delia2011}. 
To evaluate the role of dust in this analysis, Figure \ref{fig_grieco14} illustrates predictions for  $\alpha$-elements, Zn and Ni in relation to Fe,
both in models that consider the incorporation of elements into dust (left-hand column) and  without it (right-hand column). 
It is important to note that the [Si/Fe] ratios are weakly affected by dust because in these models these two elements 
are depleted to a similar degree. 
On the other hand, for other ratios, such as [Zn/Fe]-[Fe/H], in the two cases the outcome may differ significantly. 
In fact, while in the diagram without dust depletion, the curve for the irregular and elliptical galaxy show the lowest and highest [Zn/Fe] values,
respectively, this condition is reversed when the dust effects are taken into account. 
In the irregular model, the amount of Fe in dust is high due to a less efficient dust destruction
compared to other galaxy types due to its lower star formation rate. 
As a consequence, at any level of metallicity the [Zn/Fe] ratio with dust is much higher than without dust depletion.
In the spiral and elliptical models, the [Zn/Fe] ratio is less affected by dust depletion than in irregulars. 
This is because these models are characterized by a dust destruction rate that is at least comparable to the accretion rate.
In scenarios without dust, various models present abundance ratios much higher than  the observed data.
This confirms that the removal of Fe by dust is necessary to account for the observations. 
When examining all the abundance ratios together in Figure \ref{fig_grieco14}, 
it is hard to reach a definitive conclusion on the nature of the host galaxy for this GRB. 
The [Si/Fe] and [Zn/Fe] ratios seem to point towards a spiral galaxy, although there are some data, such as the [Ni/Fe],
that cannot be reproduced by any of the theoretical curves. 
Although many uncertainties still exist regarding the adopted stellar yields and observational data, 
two of the three GRBs considered by Grieco et al. \cite{grieco2014} appear to favor either a spiral or a spheroidal host,
while the third GRB (GRB120815) seems to be more confidently associated with a proto-spheroid. 
This is at variance with what is generally found for GRB hosts, i.e., that they are metal-poor \cite{christensen2004,fruchter2006}. 
This apparent discrepancy may stem from the galaxy sample used to infer the properties of the host galaxies, which mainly consists of galaxies at z $\gtrsim$ 2. 
Studies based on high-redshift surveys suggest a complex picture for GRB hosts, 
indicating that the majority appear to experience strong star formation \cite{hunt2011,perley2013}.
This result is not surprising, considering that GRBs are associated with the death of massive, short-lived stars that can be regarded as
tracers of SF inside their host galaxies, 
and can be understood considering the evolution of star-forming tracers as a function of redshift. In particular, in the local Universe, 
star formation is primarily traced by low-mass, blue galaxies. 
This emerges from the luminosity function in various 
bands, including those that trace star formation, that indicate a progressive decrease of the characteristic 
luminosity with redshift \cite{heyl1997,gabasch2006}. A decrease in the characteristic SFR is observed also in the SFR function \cite{katsianis2017} which, 
matched to the evolution of the Main Sequence relation that indicates the correlation between stellar mass and SFR, 
supports an increasing dominion of dwarf galaxies of the cosmic SFR density at lower and lower redshift.

At higher redshifts, the contribution to star formation of more massive systems increases. 
This implies that, at higher redshift, the mass range of 
galaxies that could host a GRB would increase significantly, including more massive, star-forming galaxies. 

Further evidence supporting this hypothesis comes from the study of so-called dark GRBs \cite{perley2013}, 
whose hosts present high degrees of extinction ($A_v>1$). 
This peculiar category of GRBs are typically observed at high redshifts and often exhibit high metallicity. 
Surveys aimed at the search of dark GRBs and unbiased searches of GRB host galaxies like the TOUGH survey \cite{hjorth2012a} 
indicate that GRBs at intermediate redshifts are typically associated with star-forming galaxies that exhibit a great diversity
in mass, morphology, and dust content. 
This also suggests that high-redshift GRBs may be more accurate indicators of cosmic star formation
than their low-redshift counterparts \cite{hunt2014,perley2014}. 

This intriguing fact support the idea that the GRBs studied with the chemical evolution models by Grieco et al. \cite{grieco2014} 
might originate from active star formation in massive spheroids at high redshift.

\begin{figure}[H]

\begin{adjustwidth}{-\extralength}{0cm}
\centering 

\includegraphics[width=15 cm]{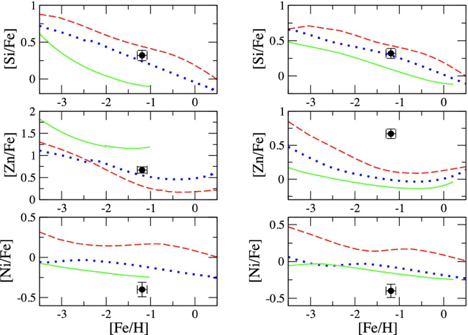}

\end{adjustwidth}
\caption{Observed and predicted abundance pattern as a function of metallicity for GRB081008~\cite{delia2011} (solid circles with error bars)
  compared to the theoretical abundances for an elliptical (red dashed line), a spiral (blue dotted line)
  and a dwarf irregular (green solid line) galaxy.
  Figure from  \linebreak Grieco et al. \cite{grieco2014}}
\label{fig_grieco14}
\end{figure}

The peculiar behaviour of the [$\alpha$/Fe] ratios in spheroids reveals a prolonged plateau,
indicating that $\alpha$-elements are overabundant compared to Fe. 
This is explained by the time-delay model \cite{matteucci2001,matteucci2021}, 
which describes the behavior of abundance ratios based on the production of different elements in various stars with distinct time-scales.
In particular, $\alpha$-elements are primarily produced on short time-scales by CCSNe. Conversely, Fe is mainly produced in SNe Ia
and is restored over a wide range of time-scales, from 30 Myr to a Hubble time. 
This suggests that if star formation is highly efficient (like in proto-spheroids), CCSNe can produce a significant amount of  $\alpha$-enriched
ejecta that keeps being incorporated in new stars. The interstellar $\alpha$/Fe ratio 
starts to decrease by the time SNe Ia begin to contribute significantly to the pollution of ISM producing the majority of Fe.  
This can produce high [$\alpha$/Fe] ratios even at high metallicity. 
The opposite scenario occurs when the SFR is low, such as in dwarf galaxies.
In this case, the enrichment from core-collapse supernovae (CCSNe) is modest. 
When Type Ia supernovae begin to contribute significantly to the overall iron content, the metallicity is still low, as the overall
interstellar $\alpha$/Fe ratio. As a result, this leads to low [$\alpha$/Fe] ratios at low metallicity. 
This is another confirmation of the predictive power of the [X/Fe]--[Fe/H] 
diagrams posed as unique tools to constrain the nature of high-redshift objects.

Through the use of the abundance ratios 
between elements produced on different timescales (such as the
[$\alpha$/Fe] ratio), from the comparison between the observed and
theoretical values vs redshift diagrams, Grieco et 
al. \cite{grieco2014} could also constrain the formation redshift of the studied GRB hosts and therefore their age, finding values between 15 Myr and 0.3 Gyr.
This suggested that GRBs could be hosted by galaxies that contained very young stellar populations or
that were characterized by very active star formation, and in some cases undergoing their first major star formation episode. 
For other GRB-DLAs samples analysed with a similar method, 
older ages were found, between 0.6 and \mbox{1.5 Gyr \cite{calura2009}}. The large dynamical
range of the ages of similar systems and with comparable models
confirms the broad variety of GRB DLAs and the fact that their hosts
are likely to include different morphological types and objects with
a variety of star formation histories. This is also confirmed by the broad
range of metallicities showed by the high-redshift sample analysed by
Cucchiara et al., which include systems with metallicity between $\sim$1/100 $Z_{\odot}$ and nearly solar metallicity, at variance with QSO
DLAs at the highest redshifts, which show more frequently low metallicity values.

Following on the methods developed by Calura et al. \cite{calura2009} and Grieco et al. \cite{grieco2014} to
constrain the nature of GRB hosts, Palla et al. \cite{palla2020} (P20) continued the 
exploitation of chemical evolution models for galaxies of different
morphological types to constrain the nature and the SF history of GRB
DLAs. 
As for the sample of GRB hosts, P20 
considered systems with a quite large number of observed abundances
and which include the five afterglow spectra studied previously with chemical evolution models
of galaxies of various morphological types \cite{calura2009,grieco2014},
plus a couple of systems never considered before, namely GRB 120815A \cite{kruehler2013} and GRB 161023A \cite{deugarte2018}.
Their decision to re-include data already 
used in previous studies is to compare the results obtained with older
prescriptions with their new, more up-to-date ones.  

Motivated by the growing evidence of LGRBs detected in 
dusty hosts \cite{hunt2014,perley2017} the models of P20 included
also an improved treatment of dust depletion, including more accurate 
prescriptions for dust yields and more physically-motivated recipes for accretion and destruction \cite{asano2013}.
 As for the set of 
chemical elements considered in the models, the one of P20 includes H,
He, C, N, $\alpha-$elements, Fe, Ni, Zn.  P20 excluded from their study the
observed abundances of C and O as they were lower or upper limits or
estimates affected by biases, such as line saturation or 
blending. The models had already been tested in previous papers \cite{grieco2014,gioannini2017a},
where it was shown that they could reasonably reproduce some of the main features of each morphological type. 
Since real galaxies show a typical spread in their basic properties, they considered extended ranges of values for
various parameters, such as galactic mass, infall timescale and star formation efficiency, 
chosen to effectively reproduce the main chemical properties of local systems. 
As for the uncertainties related to stellar nucleosynthesis and their effects in
the host identification, P20 focused on N in particular, for which they tested
various sets of yields. 

{ As for the GRB host identification, P20 adopted a statistical test similar to 
the one used for the identification of QSO DLAs \cite{dessauges2004,dessauges2007}, consisting, in each abundance diagram and for each GRB,
in the evaluation of the minimal distance between the data points and the model curves and of the weighted mean.
A similar procedure was used to compute the ages of each systems.}

The comparison between the results from models of different morphological types and the observed abundances is shown in 
Figure \ref{fig_pal20}. 
From the figure, it appears that the host galaxy exhibits a SFH typical of an irregular galaxy.
In particular, the [S/Fe] and [Ni/Fe] ratios support this hypothesis. 
The consistency of the results with the lower and upper limits for silicon and magnesium support further these findings. 
The authors analysed in detail N production, for which they studies various cases:
(i) constant primary N produced by massive stars and nucleosynthesis models and (ii) tabulated stellar yields that do not include primary N.   
In the first case, they considered the simple assumption of a fixed amount of primary nitrogen produced by massive stars,
regardless of stellar metallicity\cite{matteucci1986}. 
This assumption allows one to reproduce successfully the [N/Fe] ratios in a series of systems that include the Solar Vicinity,
particularly the observed plateau at low metallicities in Milky Way halo stars \cite{israelian2004}, 
in low-metallicity star-forming galaxies \cite{james2015} and even in low-metallicity QSO-DLAs \cite{pettini2008}.
This behaviour contrasts with what standard nucleosynthesis models predict, i.e., 
that only secondary N is produced by massive \linebreak stars \cite{nomoto2013}. 
While primary production of N has been predicted in rotating, very low-metallicity massive stars
\cite{meynet2002,frischknecht2016}, no models exist that predict primary nitrogen production for higher-metallicity rotating stars,
although it is required by a large amount of observational data. 
For these reasons, they have chosen to rely solely on the couple of assumption (i) and (ii) 
to perform a comprehensive analysis and gain a better understanding of the nature of nitrogen production.

\begin{figure}[H]
\centering

\begin{adjustwidth}{-\extralength}{0cm}
\centering 

\includegraphics[width=15 cm]{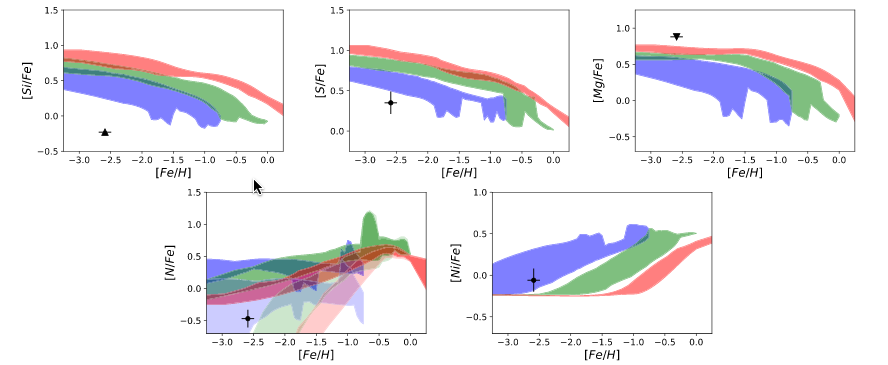}

\end{adjustwidth}
\caption{Observed and predicted abundance pattern for the GRB 05073 host. The observed abundances \cite{prochaska2007a} are the
black solid circles with error bars, whereas the upward- and downward-pointing triangles indicate lower and upper limits, respectively.    
The blue, green, and red shaded regions show the theoretical abundances from families of models for irregular, spiral, and 
spheroidal galaxies, respectively. The figure is from Palla et al. (2020) \cite{palla2020}.}
\label{fig_pal20}
\end{figure}

Regarding the [N/Fe] ratio, the observed value is in agreement with irregular galaxy models that consider only secondary production
by massive stars 
\cite{nomoto2013} (lighter shaded areas). In contrast, yields based on primary production produce values that
are too high compared to the observations for all the adopted models. 

In this particular case, the model which described best the observed abundance pattern is the one of an irregular
galaxy with moderate SFE ($\sim$0.1~Gyr$^{-1}$).

Another case of the ones considered in P20, namely GRB 050820, is shown in \mbox{Figure \ref{fig_pal20_2}}.
This figure outlines the considerable difficulties of the models in accounting for the observed
abundance pattern, in particular concerning the abundance ratios between two refractory elements (such as [Si/Fe] and [Mg/Fe]). 
The three upper panels show the observed and theoretical [$\alpha-$/Fe] ratios, supporting an abundance pattern typical of
a star-forming spheroid for the GRB host.
At the same  time, the panel showing [Ni/Fe] versus [Fe/H] support 
late-type (irregular or spiral) galaxy models, although the authors stress out
that the element Ni cannot be considered a discriminant because of its considerably \linebreak uncertain nucleosynthesis.  

In the case of [Zn/Fe], the measured value is significantly higher than all the model predictions.
However, the overabundance of this element is common, particularly for hosts identified as spheroids, compared to what the models predict.
In this and other systems Zn seems to behave like an $\alpha-$element \cite{palla2020}, in that
it is enhanced with respect to Fe when other $\alpha-$elements (i.e., Si, S, Zn) show enhanced patterns.  
This similarity between $\alpha$-elements and Zn is an important observational fact, as it casts doubt on the assumption
that Zn traces Fe, as suggested by some authors \cite{mcwilliam1997,decia2016}. 
The large difference between the data and the theoretical abundances ($\sim$0.5 dex) can hardly be explained in terms of dust depletion. 
However, also the [S/Fe] ratio is underestimated by $\sim$0.2 dex, that supports the requirement of a stronger
Fe depletion than what assumed in the models. 
The authors conclude that the host galaxy of  GRB 050820  is a young ($\sim$0.15 Gyr) and massive ($\sim$10$^9$M$_{\odot}$)
proto-spheroid caught in an intense star-forming phase.

\begin{figure}[H]
\centering

\begin{adjustwidth}{-\extralength}{0cm}
\centering 

\includegraphics[width=15 cm]{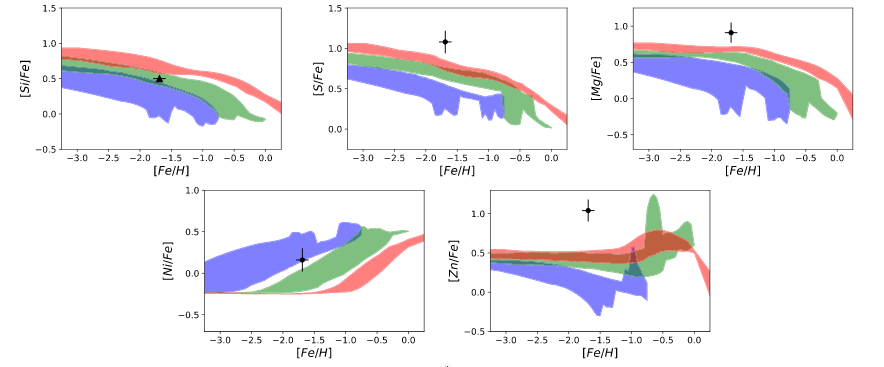}
\end{adjustwidth}
\caption{Observed and predicted abundance pattern for the GRB 050820 host.
The observed abundances \cite{prochaska2007a} are the 
black solid circles with error bars, whereas upward-pointing triangles indicate lower limits.  
The blue, green, and red shaded regions show the theoretical abundances from families of models for irregular, spiral, and
spheroidal galaxies, respectively. The figure is from \linebreak Palla et al. (2020) \cite{palla2020}.}
\label{fig_pal20_2}
\end{figure}

Palla et al. \cite{palla2020} also showed the effects of the assumed IMF on the computed abundances.
In particular, they test two different assumptions for spheroidal galaxies, 
i.e., a standard Salpeter \cite{salpeter1955} IMF ($\Phi(m)\propto m^{-2.35} $) and a top-heavy IMF expressed as:
\begin{equation}
  \Phi_{TH} \propto m^{-1.95}. 
\end{equation}
\indent This choice is motivated by the necessity of explaining a few important scaling relations found in local elliptical galaxies, 
such as the colour-luminosity relation plus intracluster abundances \cite{Gibson1997} and, most of all, the
observed [$\alpha$/Fe]-velocity dispersion relation \cite{demasi2018}. 
The slope of the IMF determines the relative number of massive stars, which are the main producers of most of the heavy elements 
and important sources of dust, therefore it impacts strongly the interstellar abundances.
As for the [Si/Fe] and [S/Fe] abundance ratios computed for two different IMFs, compared with the ones observed in the GRB 050820 host   
and shown in the left panel of Figure~\ref{fig_pal20_3}, 
a top-heavy IMF (THIMF) produces a higher overabundance of $\alpha$-elements with respect to Fe. 
This occurs because Si and S are produced mainly by massive stars, whereas the majority of Fe is synthesised by Type Ia SNe explosions, 
whose progenitors are low- and intermediate-mass stars. 
As visible in the right panel of Figure ~\ref{fig_pal20_3}, a  THIMF causes
a small (of the order of $\sim$0.1 dex) enhancement also of the [S/Fe] ratio. 
The effects of dust is to increase further the overabundance of 
the $\alpha-$ and non-refractory elements with respect to Fe, which is more easily incorporated into dust grains.
However, in the models of Palla et al. \cite{palla2020}, the enhancement caused by dust is very small for both S and Si.
Even with the adoption of a THIMF, in one case (GRB 120815A) the measured [S/Fe] value is still by $\sim$0.2--0.3 higher than the model predictions, 
which indicates that the Fe depletion is likely underestimated. 

\begin{figure}[H]
\centering
\includegraphics[width=15 cm]{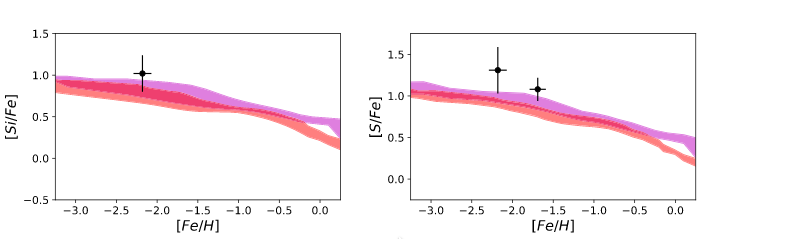}
\caption{Observed pattern for GRB 120815A and GRB 050820 host galaxies (solid circles with error bars)
  compared with the results from chemical evolution model for spheroidal galaxies obtained assuming a Salpeter IMF (red shaded regions)
  and a top-heavy IMF (purple shaded regions). The figure is from Palla et al. (2020) \cite{palla2020}.}
\label{fig_pal20_3}
\end{figure}

With the exception of one case,  the results of Palla et al. \cite{palla2020} indicate for six GRBs 
ages of the host galaxies much younger than 1 Gyr.  
Moreover, all three galactic morphological types-irregulars, spirals, spheroidals-seem to be present
in the considered sample of host galaxies, confirming previous results \cite{grieco2014} and that 
GRB-DLAs are found to occur in young proto-spheroids more frequently than  QSO DLAs, in particular in hosts of high-redshift $(z>2)$ GRBs. 
This confirms that at variance with QSO DLAs, at these redshifts GRB afterglows 
are more frequent probes of the ISM of early-type galaxies during their period of active SF, 
when the signature of CC SNe is still present in their interstellar \linebreak abundance pattern. 

\subsubsection{Systematic Effects and Uncertainties in GRB Abundance Studies}

{
Beside the already mentioned possible selection bias against highly obscured sightlines, 
more systematic effects need to be mentioned when analysing GRB abundances. 
One major source of systematic uncertainty arises from ionisation corrections. 

Fine-structure transitions and early-time (within the first hour) line profile variability in GRB afterglow spectra
are interpreted as signatures of an evolving radiation field arising directly from the GRB -and possibly nearby young stars-
affecting the gas around the progenitor (e.g., \cite{vreeswijk2007,cucchiara2015}).

Separating gas that is ionised or excited by the GRB itself (in the immediate vicinity of the progenitor)
from the more distant, generally ionised ISM of the host galaxy is challenging. If the absorbing gas lies very close to the GRB
(within a few hundred parsec scales), its ionisation state evolves with time, requiring detailed,
time-dependent photoionisation modeling to derive accurate corrections \cite{Krongold2013}, as in the case of the sub-DLA detection of GRB 080310 \cite{vreeswijk2013} - 
sub-DLAs absorption occur at column density values $19 \le$ log(N(HI)/\text{cm}$^{-2}) \le 20.3$, i.e., lower than DLAs \cite{wolfe2005} and partially overlapped with the range
of Lyman limit systems (17.2 $\le$ log(N(HI)/\text{cm}$^{-2}$) $\le$ 20.3 \cite{bechtold2001}.).
In most GRB-DLAs the data are acquired at much later time, at which  
the objects do not present line profile variation, suggesting that they are likely to probe the ISM at large  distances from the GRB, where  the gas clouds are
substantially unaffected by its radiation \cite{cucchiara2015}. Even in the case of the sub-DLA of GRB 080310 and in other
similar cases, the ionisation corrections are generally negligible \cite{schady2015} or minimal, of the order of $\lesssim$10\% , therefore
often not \linebreak applied \cite{cucchiara2015}.

As for dust depletion corrections, as already mentioned, long GRB occur preferentially in low-metallicity galaxies. Therefore, in most cases,
their dust content is, in principle, not expected to be high. 
Assessing the impact of dust depletion in DLAs requires measurements of both refractory and volatile elements,
allowing the reconstruction of the depletion pattern. 
Previous attempts to estimate depletion patterns in GRB environments through their dust-to-metal ratios
indicated depletion corrections in most cases between $\sim$0.1 and 0.3 dex and independent of metallicity \cite{decia2013,cucchiara2015}.
Depletion corrections are generally applied using the same scaling relations adopted for other classes of absorbers, as there is no compelling evidence
that the dust properties of GRB-DLAs differ systematically from those of other DLA populations \cite{vladilo2011,decia2013}. 
This consistency enables GRB-DLAs to be combined with other absorbers populations
to construct unified depletion-correction methodologies spanning cosmic time and different 
galactic environments \cite{Konstantopoulou2024}.

The analysis of dust properties summarised above relies on systems for which chemical abundances
of multiple elements have been measured, particularly both refractory and volatile species, in order to reconstruct the depletion pattern. However, simultaneous and robust abundance measurements for an extended set of chemical elements are currently available only in a very limited number of cases.
The buildup of more extended datasets of DLA samples with comprehensive elemental abundances
will be essential to provide deeper insights into dust depletion patterns in galaxies at early cosmic times.}


\subsection{Effects of Dust on the Abundance Pattern of Distant Starburst Galaxies}
\label{sec_starburst}
Another interesting class of high-redshift systems which allow one to probe different physical
conditions than DLAs are Lyman-break galaxies (LBGs). 
The idea that LBGs are probes of different systems than DLAs stems from the observational evidence that
they are characterised by different kinematic properties \cite{shapley2003}, metallicity \cite{pettini2004}, 
clustering \cite{adelberger2003} 
and typical SFR values \cite{wolfe2003,hopkins2005} from the ones of DLAs. 
In the most common interpretation of the properties of these systems, while DLAs are thought to originate in quiescent, low-metallicity outskirts of
disc galaxies or in dwarf galaxies, the properties of LBGs, in particular their strong clustering, intense SF activity and
outflow kinematics suggest that 
they might represent the progenitors of spheroids and preferentially located in high-density environments \cite{wolfe2003}. 
The idea of LBGs as the progenitor of present-day ellipticals found further support with the first attempts to constrain the chemical abundance pattern
in high-quality LBGs spectra.  
The first measure of a set of chemical abundances performed for a sizeable number of elements in a Lyman-break galaxy (LBG) was the one of Pettini et al. (2002) \cite{pettini2002},
who observed MS 1512-cB58 (cB58), at the time brightest known LBG owing to its gravitationally lensed nature.
cB58 was found to be a $\sim L_{*}$ galaxy at z = 2.7276, magnified by a factor of $\sim$30
by the foreground cluster MS 1512+36 at z = 0.3. This implies a significant improvement of the S/N of its spectrum which, if matched
with superior quality and wide wavelength coverage obtained with Echelle spectrographs, enables a detailed identification of
several heavy element absorption lines.
For this galaxy, \mbox{Pettini et al. (2002)} measured the abundances for several elements (O, Mg, Si, S, N, Fe), indicating a metal-rich
interstellar medium and an 
abundance pattern bearing significant marks of fresh type II SN explosions.
Moreover, absorption lines with velocities up to $\sim$1000 km s$^{-1}$ were detected, 
supporting the likely presence of a strong outflow.
Moreover, the significantly reddened UV continuum and the detection at sub-mm and mm wavelengths were clear indications that dust
had to be present in cB58, supported also by the underabundance of refractory Fe-peak elements, interpreted as evidence of dust depletion.
For this system, Pettini et al. derived abundances of O, Mg and Si of $\sim$0.4 of their solar values, whereas Fe was found to be >3 times 
underabundant with respect to the other elements.
\mbox{Pettini et al. (2002) \cite{pettini2002}} suggested that the effects of dust depletion on the abundances of the most refractory elements were of a factor of $\sim$2 .
Further constraints on the dust content of cB58 came from subsequent 3D Lyman-$\alpha$ transfer models, suggesting a gas-to-dust ratio lower than solar and than
typical values found in DLAs \cite{verhamme2008} but consistent with what found in other LBGs. 
The abundances measured for cB58 were interpreted by means of a multi-zone chemical evolution models describing the starbursting,
early phases of an elliptical galaxy by Matteucci \& Pipino (2002) \cite{matteucci2002a}. 
They concluded that this system is a small (i.e., characterised by baryonic mass of a few $10^{10}~M_{\odot}$) young (with an age of a few 10 Myr) elliptical  
undergoing a burst of star formation and a galactic wind.
The model of Matteucci \& Pipino did not include any treatment of dust depletion and failed to reproduce the 
high $\alpha$-enhancement ([$\alpha$/Fe] > 0.5) shown by this system.
Several years later,\mbox{ Pipino et al. (2011) \cite{pipino2011} }developed a model to describe young elliptical galaxies
which included dust evolution, with prescriptions similar to the ones of Calura et al. \cite{calura2008} but with a factor 10 decrease in the
SN II dust yields. 
Pipino et al. (2011)\cite{pipino2011} first consider the ratios between volatile and refractory elements, i.e., [N/Fe] and [N/O],
and for the abundances observed in cB58 they considered two different data sets \cite{Teplitz2000,pettini2002}.  
These two sets of abundances showed significantly different abundance ratios,
with $\sim$0.6 dex differences for both [N/O] and [N/Fe].
Such a large difference prevented Pipino et al. (2011)\cite{pipino2011}  from reaching firm conclusions and reproducing the observed abundances,
outlining serious discrepancies in the considered N yields and the degeneracy between yields and dust prescriptions.
As for the [Mg/Fe] and [Si/Fe] ratios, the observed values were still larger than the theoretical ones by $\sim$0.2--0.3 dex,
likely indicating a still low fraction of Fe incorporated into \linebreak dust grains.

These works represented strong incentives to improve the treatment of dust in chemical evolution models and to 
identify further cases of gravitationally lensed LBGs which could enable a characterisation as complete as possible
of the intrinsic abundance pattern measured in these systems.
One important ingredient which turned out as necessary to account for the abundances observed in high-redshift
systems is differential dust depletion, i.e., the fact that chemical species can
be incorporated in different amounts into dust grains.

The local ISM shows strong signs of differential depletion \cite{jenkins2009}, which deserves to be considered
in theoretical studies of high-redshift galaxies as well. 
A significant improvement in the analysis of the abundance pattern in LBGs and in other lensed systems was
the one of Palla et al. \cite{palla2020b}, which took into account the effects of differential depletion in chemical evolution models of starburst galaxies.
Palla et al. \cite{palla2020b} performed a theoretical study of the abundance pattern observed in a sample of high-redshift starbursts, where
the measured abundances were compared with models including dust evolution and where also the effects of different IMFs were tested.
The latter quantity is known to have a strong impact on the galactic abundance pattern, in particular concerning several elements
known to build the bulk of the dust budget in galaxies.

A clear example of how both differential dust depletion and the IMF affect the abundances in distant starbursts is shown in Figure ~\ref{fig_sife_pal20}. 
The models shown in Figure ~\ref{fig_sife_pal20} were aimed to describe starburst galaxies of various baryonic masses,
ranging from $10^{10}~M_{\odot}$ to $10^{12}~M_{\odot}$. 
The models shown in the leftmost panels are computed without taking into
account dust depletion and are compared to abundances observed in three lensed starbursts at redshift between $z \sim 2.4$ and 
$z \sim 2.7$ (see the Caption of Figure ~\ref{fig_sife_pal20} for further details).

The upper panel shows the results for one model at fixed baryonic mass ($10^{10}~M_{\odot}$) and the effects of the IMF on the computed abundance ratios.
Clearly, the assumption of a THIMF is to increase the [Si/Fe] ratio at fixed metallicity. This is caused by the increased fraction of massive stars
achieved with a THIMF, which leads to an increase in the production of $\alpha-$elements (such as Si) with respect to Fe, produced mostly by Type Ia on much longer timescales. 
In general, the top-heavier the IMF, the higher the $\alpha-$enhancement of the computed abundances; however, the upper panel of  Figure ~\ref{fig_sife_pal20}
shows that even with 
the most extreme assumption (the $\beta$ = 1 case of the \cite{weidner2011} THIMF), it is impossible to reproduce the observed abundance ratios with this particular model.
Clearly, also the star formation history of the modelled galaxy plays an important role in the computed abundances. 
The star formation efficiency is known to increase with galactic mass, as indicated by arguments related the `Downsizing' character shown by galaxies
of various masses,
in which more massive galaxies need to form the bulk of their stellar mass on shorter timescales than low-mass galaxies \cite{cowie1996,spitoni2020}. 
One possible, major implication of Galactic Dowsizing is that massive galaxies show a higher star formation efficiency \linebreak (e.g., \cite{matteucci1994}). 
In the models shown in the lower panel of Figure ~\ref{fig_sife_pal20}, the SF efficiency increases with mass, which explains why the [Si/Fe] ratio increases
with mass at fixed metallicity and at fixed IMF - in this case, this is visible with a Salpeter IMF.  
In summary, \mbox{Figure ~\ref{fig_sife_pal20}} show two important facts, i.e., (1) that the highly $\alpha-$enhanced abundance pattern of starburst galaxies cannot  be
accommodated by increasing the `Top-Heaviness' of the IMF and that (2) even an increase of the SF efficiency to extreme values does not suffice to this purpose,
if a 'normal' IMF is assumed in the models. Only massive models with extreme assumptions regarding the IMF (i.e., models with an IGIMF and $\beta \le 1.6$)
produce results barely in agreement with the measured abundances. 

\begin{figure}[H]

\begin{adjustwidth}{-\extralength}{0cm}
\centering 

\includegraphics[width=15 cm]{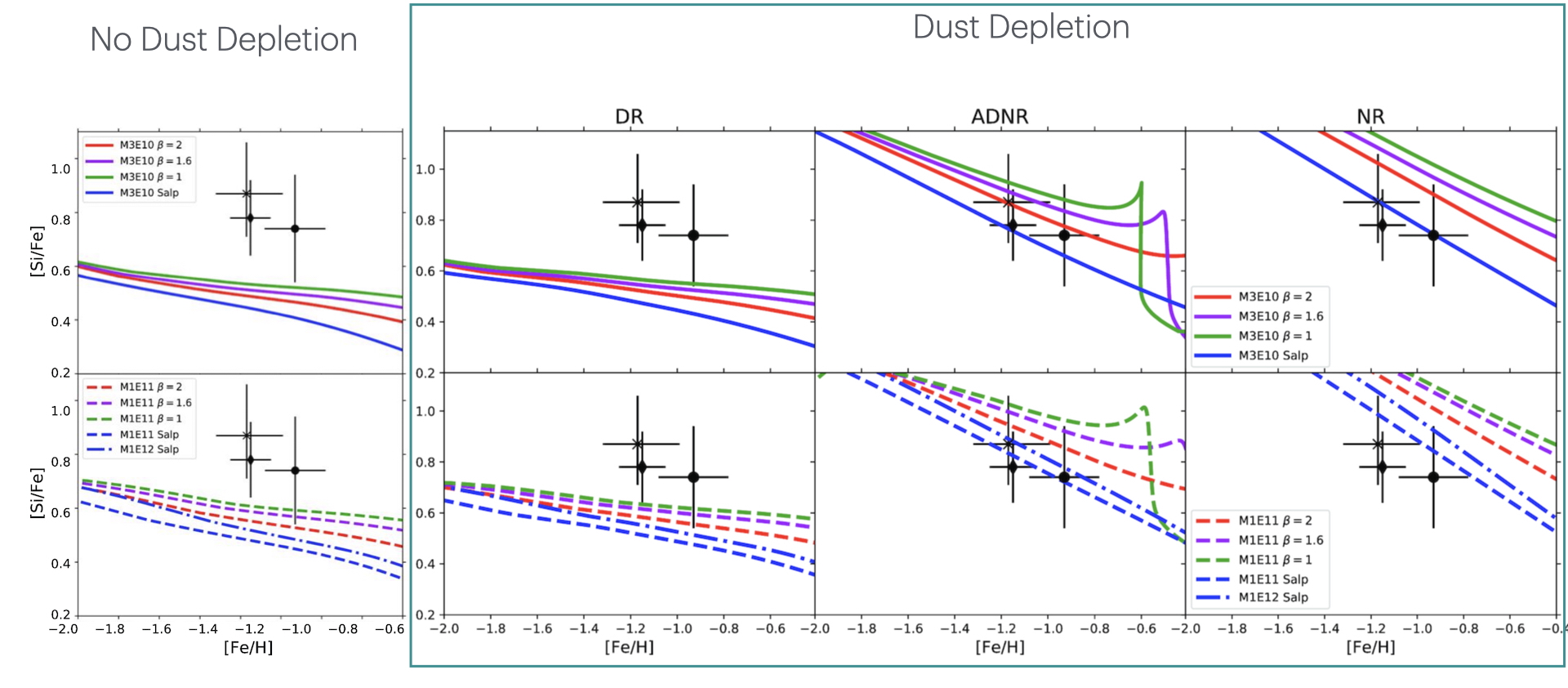}
\end{adjustwidth}

\caption{Effects of the stellar IMF, of the SF efficiency and of differential dust depletion on the abundance ratios of distant starburst galaxies. 
  The two leftmost panels show the results of models in which the effects of dust depletion are not taken into account, therefore 
  they are useful to highlight the effects of the IMF (\textbf{upper panel}) and of the assumed SF efficiency (\textbf{lower panel}). 
  The upper panel shows theoretical abundances for a proto-spheroid with baryonic mass of $3~\times~10^{10}~M_{\odot}$ and assuming an IGIMF characterised by
  $\beta=2$ (red line), $\beta=1.6$ (magenta line), $\beta=1$ (green line) and with a Salpeter (1955) IMF (blue line).
  The observed abundances (black symbols with error bars) are from Pettini et al. (2002) \cite{pettini2002} (diamond);
  Quider et al. (2009) \cite{quider2009} (cross) 
  and Dessauges-Zavadsky et al. (2010) \cite{Dessauges2010} (filled circle). 
  The bottom panel on the left shows the model results for proto-spheroids of mass $10^{11}~M_{\odot}$ (dashed lines),
  characterised by a higher SF efficiency than the  $3~\times~10^{10}~M_{\odot}$ model, and for a model with mass $10^{12}~M_{\odot}$ and with a Salpeter IMF (dash-dotted line).
  The panels in the box enclosed by the solid, dark-cyan line show the abundances corrected for dust depletion and 
  computed assuming different prescriptions for dust evolution; the symbols in each couple of upper and lower panels 
  are as in the previous panels, but on the left ones  
  dust destruction in the reverse shock in SNe dust yields is taken into account, 
  while all the other processes (interstellar destruction and growth) are switched off (DR model of \cite{palla2020b}).
  In the middle panels the models include dust condensation in stellar ejecta, destruction, but are without the reverse shock (ADNR model).
  Finally, in the panels on the right only dust production in stars is considered, with no reverse shock in SNe (NR model).
  Figure adapted from \cite{palla2020b}. }
\label{fig_sife_pal20}
\end{figure}

The effects of dust depletion on the [Si/Fe]--[Fe/H] diagram as computed by P20 
for different assumptions regarding the model for dust evolution are enclosed in the dark-cyan box in Figure~\ref{fig_sife_pal20}. 
Note that the effects of dust depletion increase as one moves from the left to the right of Figure~\ref{fig_sife_pal20}.
Within the box, the upper and lower left plots show the results of  ‘minimal dust’ models (DR), in which
dust destruction in the reverse shock in SNe dust yields is taken into account, while   
all the other processes (interstellar destruction and growth) are switched off. 

The two middle plots shows instead the results obtained with intermediate models, including 
dust condensation in stellar ejecta, destruction, but without the reverse shock (ADNR model).
Finally, the panels on the right of Figure~\ref{fig_sife_pal20} show the results of a family of maximal dust models, in which
only dust production in stars is considered, with no reverse shock in SNe (NR).

The minimal dust models show small differences from the ones without any dust depletion of Figure \ref{fig_sife_pal20}, in which all the corresponding
curves appear slightly shifted upwards, still underestimating the observed abundances. 
This implies that, as expected and already seen previously,
the net effect of dust is to cause an apparent increase of the $\alpha-$enhancement in the abundance pattern.
This effect increases further in the middle panels,
where all the curves are shifted to the point that even the results obtained with a normal IMF (the one of Salpeter)
appear consistent with the observational data. 
This becomes even more extreme in the rightmost panels, where the abundances obtained with the Salpeter IMF and the most massive
models overpredict by $\sim$0.1 dex the $\alpha-$enhancement with respect to the observations.
Similar effects were seen also in other abundance diagrams considered by P20, including [Mg/Fe]--[Fe/H] and [O/Fe]--[Fe/H].
The main conclusion of this study was that the intermediate ‘ADNR’ models account satisfactorily for the observed
abundances, in particular the ones obtained with a standard, Salpeter IMF and/or a minimally or mildly THIMF,
such as the IGIMF with $\beta \ge 1.6$.
A non-standard, THIMF in high-redshift starburst galaxies is supported by other studies
based on the use of chemical evolution models  to interpret the relative abundances of rare isotopes, 
such as 13C, 15N, 17O and 18O detected in the infrared band with the the Atacama 
Large Millimetre/submillimetre Array \linebreak (ALMA) \cite{romano2017}.

\section{On the Redshift Evolution of Dust Production and of Its Global Budget Up to High Redshift}
\label{sec_ontheredshift}
In this section we will briefly review the current knowledge regarding the
overall dust content of high-redshift systems and its origin. We will focus our attention on the observational
determination of the dust mass in galaxies and on a few, important 
relations which have been exploited in the last few years to infer some fundamental galactic properties. 
We will discuss the redshift evolution of some of these properties, important to put their
derivation in a broad context and also valuable in perspective of future instruments. 

It is important to stress the importance of these quantities, of the considerable
amount of knowledge that they carry along and their strong value as direct, unique probes of the 'hidden'
components of galaxies, built mostly by the coldest, dust-embedded regions where stars originate or
where obscured AGN are located.

\subsection{Observational Derivation of the Dust Mass}
Despite the ground-based studies based on absorption systems provided unique insights on the properties
of dust grains in high-redshift galaxies, they were not suited to quantify one fundamental aspect, i.e.,
the amount of mass in the form of dust in galaxies. Such information required access to the dust FIR thermal 
spectra, which became possible with the construction of the first large infrared telescopes, with 
the Kuiper Airborne Observatory and the InfraRed Astronomical Satellite (IRAS) as some first notable examples.

The measure of the dust mass in high-redshift galaxies enabled the use of new diagnostics
to interpret the properties of galaxies at low- and high-redshift.

Pairing dust–mass measurements with far-infrared fine-structure–line observations—which now yield the masses of individual ISM phases,
the star-formation rates and the internal kinematics of distant galaxies \cite{decarli2025}—
has opened a new set of diagnostics for interpreting galaxy properties from the nearby Universe to the highest redshifts. 

Sizeable samples of galaxies with measures of the dust mass started to be collected and analysed,  combining such estimates 
with other available quantities measurable at other wavelengths, such as stellar and gas masses and metallicity measures.
This led to novel, fruitful studies where new scaling relations were analysed \cite{cortese2012,calura2017,casasola2020,casasola2022,parente2022,shivaei2022,pastrav2024,Sawant2025}. 

The first, significant improvements in the assessment of the dust mass in galaxies were possible 
thanks to FIR observations of the thermal emission performed with the IRAS instruments.
The dust mass in the Galaxy and in external galaxies were estimated from the measure of the
60 $\upmu$m and 100 $\upmu$m flux densities \cite{lonsdalepersson1987,young1989}. 
These estimates allowed to probe the so-called {\it warm} dust
component, commonly referred to dust temperatures in the  30--50 K range.
As these values are larger than the temperatures predicted since the earliest studies
for the interstellar dust in `normal' galaxies, 
of the order of 10--20 K \cite{spitzer1978}, there was general consensus on the fact that only
little information could come from these observations. The use of the 60 $\upmu$m and 100 $\upmu$m fluxes
lead to severe underestimations of the total amount of dust, as the bulk was expected to be in the form
of a cold component, characterised by a thermal emission peaking at significantly larger wavelength, presumably
between $\gtrsim$100 $\upmu$m and 300 $\upmu$m and beyond the capabilities of-at the time-current telescopes, such as IRAS \cite{kwan1992}. 
In this regard, it  became clear that the effect of redshift on the spectra of distant galaxies
could lead to positive conditions for the observations at submillimetre and millimetre wavelengths. 

The use of dedicated submillimetre telescopes lead to the
detection of the thermal dust continuum emission from several objects at high redshift
\cite{clements1992,dunlop1994,ivison1995}. 
The submillimetre detections of high-redshift objects brought immediately attention to 
their very high dust masses, that were significantly larger  
than the values measured in local galaxies \cite{eales1996}. 
Since their discovery, there was general agreement that these high dust masses implied
that the objects were caught in an early  
stage, even though  such conclusion was highly qualitative and based on approximate arguments
that lead to speculations about its origin and its strong link to massive star formation   
\cite{dunlop1994,eales1996}.

The Submillimetre Common-User Bolometer Array (SCUBA) observatory played a fundamental role to 
advance the knowledge of dust in high-redshift galaxies. 
SCUBA enabled a very quick mapping of the sky, since 
individual bolometers turned out as significantly more sensitive than any 
previous instrument. 
Deep submillimetre surveys performed with the SCUBA bolometer array 
revealed a population of extremely  
dusty and obscured objects with high bolometric luminosities and residing at $z >2$
\cite{barger1998,dunlop2001,ivison2002,dunne2003}. 

For the extremely high SFR values inferred from their IR luminosity (up to \linebreak 100--1000 $M_{\odot}/yr$),
the SCUBA sources were almost immediately associated to proto-elliptical galaxies,
caught during the intense starburst events characterising the formation of the bulk of their stellar populations
and likely residing in the greatest overdensity regions of the Universe. 
Such findings corroborated the idea that the bulk of the stars of local ellipticals were formed 
at high redshift through rapid (occurring on timescales $<$1 Gyr) and violent starbursts
was motivated by several evidences, gathered through the years, based on the tightness of known and widely used
scaling relations, such as  the colour-magnitude diagram \cite{bower1992} and others, such as the correlation between the
integrated [$\alpha$/Fe] and the velocity dispersion \cite{matteucci1994}. 

In the following years, it was thanks to the higher sensitivity of FIR space instruments that
also fainter sources could be detected. This opened up our view to the diversity of systems
detected at high redshift and characterised by a broader variety of star formation regimes, including
also quiescent objects \cite{kaneda2007,panuzzo2007}. 

The galactic dust mass can be observationally determined through fits to multi-band observations in the far-infrared and 
sub-millimetre bands. The quantities that are measured in submillimetre continuum observations useful for characterising the spectra of high-z galaxies
are the flux densities in one or more  bands in which the atmosphere is partially transparent.
For instance, bands that can be used for dust mass measures at high-z are  centered  
approximately at 350, 450, 600, 750 and 850 $\upmu$  \cite{eales1996}. 
The quantities that are routinely estimated from these measurements are the bolometric luminosity 
and the dust temperature; combined together, these quantities lead to estimating the dust mass (see later on). 
The best conditions to perform an accurate determination of the dust mass is to be able to sample
the peak of the dust emission in the measured spectra; obviously, 
for each facility, this is possible only in a limited redshift interval and can be severely limited by source
brightness and instrument sensitivity.  

The most common and simple approach to measure dust masses is through the assumption
of a single temperature dust component. This consists in the assumption that the bulk of the dust
is described by a single, diffuse medium in thermal equilibrium which permeates the entire system. 
In conditions of optically-thin emission, the mass of
dust in a galaxy at redshift $z$ can be estimated from the observed
submillimetre flux $S_{\nu_{obs}}$ at  the observed frequency $\nu_{obs}$ 
as
\begin{equation}
  M_{\rm d} = \frac{D_{\rm L}^2 S_{\nu_{obs}} }{(1+z) \kappa_d(\nu)B_{\nu}(T_{\rm d})},
  \label{eq_mdust}
\end{equation} 
where $z$ is the redshift, $D_{\rm L}$ is the luminosity distance, 
whereas $\nu$ is the rest-frame frequency, i.e., the frequency at which the 
radiation is emitted. This equation represents the so-called
modified black body approximation, where the thermal emission from dust grains is
modelled through the Planck function multiplied by the dust absorption coefficient $\kappa_d(\nu)$. 
For this quantity, generally assumed independent of temperature \cite{eales1996}, a power-law form is adopted: 
\begin{equation}
	\kappa_d (\nu) =  \kappa_{d}(\nu_{\mathrm{0}}) \left(\frac{\nu}{\nu_{\mathrm{0}}}\right)^{\beta'},
\label{eq_kappa}
\end{equation}
(e.g., \cite{eales1996,gall2011}), where $\beta'$ is the emissivity index.

The critical frequency at which a source becomes optically thin is typically
$\nu_{\mathrm{0}}$ =\mbox{ 2.4 THz} (corresponding to $\lambda_0$ = 125 $\upmu$m) \cite{gall2011}; 
the dust absorption coefficient 
at this frequency is  $\kappa_{\mathrm{d}}(\nu_{\mathrm{0}}) = 18.75$ cm$^{2}$ g$^{-1}$ \cite{hildebrand1983}. 

In principle, the absorption coefficient is also a function of the grain size \cite{gall2011}. 
Assuming spherical grains, it can be expressed as 
\begin{equation}
  \kappa_{d}(\nu,a) \equiv (3/4) \frac{Q(\nu,a)}{a \rho}, 
\end{equation}
where $a$ is the grain radius, $\rho$ is the dust  density and
$Q(\nu,a)$ is the dust absorption efficiency~\cite{gall2011}. 
It is clear that in this formulation one main uncertainty is the grain \linebreak size distribution. 

A wide range of values for $\kappa_d (\nu)$ have been estimated using various techniques,
most of which require assumptions about the physical properties of dust grains \cite{clark2016}.
Useful information can also be extracted from the UV or optical dust extinction curve (e.g., \cite{gall2014}).
The Comparison of the FIR/submm emission and the UV/optical extinction in local Galactic nebulae 
are sometimes used to estimate $\kappa_d (\nu)$ that require assumptions regarding
the cloud geometry \cite{bianchi2003} and with little flexibility, considering that
the geometric properties dusty regions can vary significantly with the environment,
as the radiative transfer properties of the dust. 
Laboratory studies based on pre-solar dust grain composition offer an alternative method for
determining $\kappa_d (\nu)$ \cite{mutschke2013}.
The literature values of $\kappa_d (\nu)$ suggested by various methods can vary enormously.
Clark et al. \cite{clark2016} analysed a compilation of determinations of the dust mass absorption coefficient
performed at a wavelength of \mbox{500 $\upmu$m}, showing that the determinations obtained
with various methods span over 3.5 orders of magnitude in total (see also \cite{alton2004}). 
Other estimates at longer wavelengths (800 $\upmu$m) report a somewhat more limited range
of a factor of 25 \cite{hughes1993}. 
The obvious implication of such large uncertainties is
that estimates of absolute dust mass values tend to have significant errors.

However, in several cases, it may be convenient to be 
interested in the ratio of dust masses between different galaxies
rather than the absolute values themselves.
In this context, as long as dust mass is consistently estimated from a homogeneous set of
flux measurements, e.g., with the same value of $\nu$ (or $\upmu$m), 
the uncertainty in $\kappa_d (\nu)$ becomes unimportant.  
For instance, it is possible to fix arbitrarily $\kappa_d(\nu)$ to some value at a
given emission wavelength and consider the relative fluxes between different datasets
(e.g., a local galaxy and some others at high $z$) to estimate the dust mass ratios,
in a way that is unaffected by the lack of knowledge of the absolute  $\kappa_d(\nu)$ value \cite{eales1996}.

Another troublesome parameter in Equation~(\ref{eq_kappa}) is the emissivity index $\beta'$,
for which an increase with increasing wavelength (decreasing frequency) has been found.
In particular, for $\lambda$ $\lesssim$ 200 $\upmu$m a value $\beta' \sim 1$ was derived,
whereas for $\lambda$ $\gtrsim$ 45 $\upmu$m, \linebreak  $\beta'$ $\sim$ 2 \cite{erickson1981,schwartz1982,dunne2001}.

Another fundamental uncertainty in the derivation of the dust mass (Equation  (\ref{eq_mdust})) concerns the Planck function $B_{\nu}(T_{\rm d})$. 
An accurate estimate of $B_{\nu}(T_{\rm d})$ requires a good method to determine $T_{\rm d}$,
that is most of the times a crucial issue \cite{pozzi2021}. 
As long as the frequency at which the radiation is emitted is significantly
smaller than the one of the peak of the spectral energy distribution,
to first order, the Planck function can be estimated as
\begin{equation}
B_{\nu}(T_{\rm d})  \simeq 2 k_B T_{\rm d} \nu^2 c^{-2}, 
\end{equation}
where $k_B$ is the Boltzmann constant and $c$ is the speed of light. 
This approximation represents the so-called Rayleigh-Jeans limit of the Planck function and, besides
being a simple formula, it has another advantage, i.e., that it implies that 
the fractional uncertainty in the obtained dust mass is no larger than the 
fractional uncertainty in the dust temperature.

For a simultaneous determination of $T_d$ and other parameters, such as the emissivity index $\beta'$,
flux measurements at different wavelengths are required. 
This is a hard task in particular for high-$z$ systems, that often require one to assume
bona-fide values for \linebreak such quantities.

Still today, one primary issue with deriving the bolometric luminosity of dust is something  
that has complicated our understanding of galaxies with extremely high far-infrared luminosities
since the times of the earliest surveys. Essentially, it is unclear whether the dust is heated by stars or a hidden active nucleus. 
This uncertainty poses a general problem when using bolometric luminosity as 
a measure of galactic evolution, especially for high-redshift objects detected 
in the submillimetre waveband, as many of these are known to host active nuclei. 

The cold dust is typically characterised by a $T_d$ between 25 and 40 K, with an emission peaking
at around 100–120 $\upmu$m \cite{pozzi2020}. This implies that moving in redshift, 
different rest-frame parts of the spectrum are sampled by submillimetre observations, i.e., the Rayleigh–Jeans region at low $z$ and
the Wien region at high redshift (typically $z\gtrsim 1$ considering  observations at 160 $\upmu$m \cite{pozzi2020}). 

In most galaxies, it is safe to assume an optically thin submillimetre emission. 
This is indicated by two key observations: 
(i) the spectral energy distribution in the submillimetre wavelength range
is steeper than what one would expect from the Rayleigh-Jeans tail of a black body, 
and (ii) the size of the galaxy's disc is significantly larger than what would
be predicted if we assumed that the emission were optically thick \cite{eales1996}. 
Another common argument to support a small dust opacity at long wavelengths is related to the typical grain size. 
Since the dust grains are small 
($\lesssim$1 $\upmu$m ), the dust opacity is highest in the UV and optical bands, but
decreases strongly at far-infrared and submillimetre wavelengths \cite{scoville2017}. 
However, in the case of some particularly
luminous objects, the compact size of the regions that are producing the 
far-infrared emission implies that the  optically thin assumption 
may break down, as in the case of compact, dusty galaxies at very high z \cite{ferrara2022} or QSOs 
\cite{walter2022}.
In the case of optically thick emission, the dust mass is \cite{eales1996}
\begin{equation}
  M_{\rm d} = \frac{D_{\rm L}^2 S_{\nu_{obs}} }{(1+z) \kappa_d(\nu)B_{\nu}(T_{\rm d})} \frac{\tau(\nu)}{(1-exp[-\tau(\nu)]))}.
  \label{eq_thick}
\end{equation} 
\indent The dust optical depth $\tau(\nu)$ can be assumed is proportional the dust  surface density $\Sigma_d$ as
$\tau(\nu) = \kappa_{d}(\nu) \Sigma_d$ \cite{gilli2022}. 
In the case of spatially-resolved sources it is possible to obtain a minimum dust temperature condition $T > T_{min}$ expressed as 
\begin{equation}
T_{min} \equiv \frac{h \,\nu}{k_B}\frac{1}{{\rm ln}\left[ 1 + \frac{2h \nu^3}{c^2\,I_{\nu, \, min}} \right] }, 
\label{eq:tmin}
\end{equation}
\cite{gilli2022} where $h$ is the Planck constant, 
\begin{equation}
I_{\nu, \, min} \equiv \frac{(1+z)^3\,S_{\nu_{obs}}}{\Omega} + B_{\nu}(T_{CMB}), 
\end{equation}
$\Omega$ is the solid angle covered by the source and $T_{CMB}(z)=2.725(1+z)\,\rm{K}$ is the Cosmic Microwave Background temperature at redshift $z$ \cite{gilli2022}. 
From Equation (\ref{eq_thick}) it is also straightforward to derive the widely used optically thin formula of Equation~(\ref{eq_mdust}) in the case $\tau(\nu) << 1$.

In most cases, the SEDs of high-$z$ galaxies are best fitted with a single temperature 
model (e.g., \cite{priddey2001,hughes1997,pozzi2020} and many others). Bianchi \cite{bianchi2013} showed how such an approach is
safe to obtain a reliable fit to the SEDs for $\lambda \ge$ 100 $\upmu$m. 
However, in principle the emission can also be fitted using a two temperature-model able to a
account for a warm and a cold component. The equation used
for such two-component model is similar to Equation~(\ref{eq_mdust}), but it includes two separate terms for the Planck function, and it also
features the mass fraction of the two components. 
The dust masses derived with two-component models~\cite{ivison2010} are usually a factor $\sim$2 
higher than single temperature models \cite{dunne2001,vlahakis2005,gall2011}. 

The wealth of detections of unexpectedly large dust masses in the first billion years
revealed critical shortcomings in explaining the rapid emergence of dust in early galaxies and
introduced new theoretical challenges.

\subsection{Theoretical Studies of Dust Evolution at High Redshift}

From a theoretical point of view, it has always proven challenging to explain the origin of the large dust masses observed at very high redshift ($z>6$),
when the Universe was \linebreak $<$1 Gyr old and star formation in early galaxies had presumably just started \cite{dwek2007,valiante2009,pipino2011,palla2024}. 
To tackle the problem of the inferred large dust masses in early galaxies, 
chemical evolution models have often been one preferred method to track the evolution of several galactic physical properties.
This was because they allow one to easily and rapidly investigate a set of parameters, with the possibility of
studying the role of various processed and with a large predicting power for useful quantities,
such as dust masses, element abundances, SFRs and others.
Chemical evolution models were preferred over other approaches, such as cosmological models, in particular when studies
of individual galaxies were performed. In the following, I will list a few attempts to model various examples of dust-rich sources at high redshifts. One representative case which has raised interest from various authors is the
QSO J1148+5251 at z = 6.4 \cite{fan2003}. 
One of the initial effort to explain the large derived dust mass ($\sim$2 $\times$ 10$^8 M_{\odot}$)in this system was made by \cite{dwek2007}.
These authors used the observed dust mass to constrain the stardust production: 
assuming that SNe are the only sources, they find that an average SN has to produce at least 1 $M_{\odot}$ of dust.
The total mass of the QSO host galaxy is estimated to be around \(5 \times 10^{10} M_{\odot}\),
which is in agreement with the suggested dynamical mass for this system \cite{walter2004}. Valiante et al. \cite{valiante2009}
incorporate AGB stars in their models, finding that these sources are the primary producers of the 
few $10^8 M_{\odot}$ of dust in QSO J1148+5251. 
However, the estimated baryonic mass of the galaxy, $\simeq$10$^{12}~M_{\odot}$, is 
by more than an order of magnitude larger than the observed dynamical mass. 
The model incorporates a star formation history derived from a hierarchical galaxy merger tree \cite{li2007}.
They use a simple chemical evolution model with
dust that, for a given SFH, allows them to follow the time evolution of the gas and dust, and 
their equations do not account for gas inflows and outflows. In their framework, star formation
began at redshift $z = 15$, allowing approximately 550 million years for the stars to evolve,
with the SFR reaching values as high as \(10^{4} M_{\odot} \, \text{yr}^{-1}\).

Pipino et al. \cite{pipino2011} utilized models for elliptical galaxies \cite{calura2008} to 
account for the dust contributions from various stellar sources (including AGB and SNe), a QSO outflow,
and the growth of dust grains in the ISM. 
A model galaxy with a mass of \( \times 10^{12}~M_{\odot} \) was used to study J1148+5251.
The predicted SFR exceeds \( 3 \times 10^{3}~M_{\odot}~\text{yr}^{-1} \) and the observed large dust masses could only
be explained by including a significant contribution from interstellar dust growth,
alongside the dust produced by the considered stellar sources.
The results are significantly impacted by a high assumed rate of dust destruction caused by SN shocks in the ISM.
Additionally, the models incorporate dust contributions from Type Ia SNe, despite the lack of clear evidence supporting
substantial dust production from these sources.
Furthermore, the dust contribution from the quasar wind, estimated to be a few times \(10^7 M_{\odot}\), is negligible
with respect to the other formation channels. 

Calura et al. \cite{calura2014} presented an extended dataset of measured dust masses in a sample of
58 QSOs at $z>5.7$ and attempted to account for these observables and other scaling relations 
(such as the dust-to-gas vs mass) by
means of chemical evolution models for proto-spheroids of various masses. 
The models describe the starbursting phase of proto-elliptical galaxies and their study analyses the effects of
various quantities on the dust mass, including 
the assumption of the Larson \cite{larson1998}  THIMF, of an enhanced star formation efficiency and dust growth
efficiency, represented by the accretion timescale parameter $\tau_0$. 
The effects of these quantities were also studied in combination,
in order to test their overall impact and the most favourable conditions for achieving the sometimes extreme 
measured data, that include a few systems characterised by dust masses of $\sim$10$^9~ M_{\odot}$.
According to their results, these values could be reproduced either by means of models characterised by extremely high
baryonic masses ($\sim$10$^{12}~M_{\odot}$), or by models with lower mass but assuming simultaneously a  THIMF,
an enhanced star formation efficiency \linebreak ($\nu$ = 60 Gyr$^{-1}$) and an extremely low value for the growth timescale ($\tau_0 \sim$ 1 Myr).
As for the requirement of a high SF efficiency, Calura et al. \cite{calura2014} considered the observed
Kennicutt \cite{kennicutt1998}-Schmidt \cite{schmidt1959} relation for star-forming galaxies and compared the computed
SFR and gas mass surface densities of the QSO hosts at $z \gtrsim 6$ with the 
models which successfully accounted for the observed amount of dust to
a previous collection of data. They found that most of the observational data 
were lying  on the extreme upper edge of the universal SF law derived by Krumholz et al. \cite{krumholz2012}.
This finding confirmed a high SF efficiency in this systems which,
in the framework of models of large, star-forming spheroids,
could be explained by means of positive 
feedback due to the QSOs, attainable in the earliest stages of supermassive central 
BH growth \cite{silk2005,ciotti2007,pipino2009}. 
Before that, it was largely known since at least a couple decades that {\it negative} feedback from AGN may help
reconcile a few discrepancies between observations and theory within galaxy formation models, such as
the overcooling problem in massive galaxies, which manifests itself mostly with an overestimated bright end of the galaxy
luminosity function and, besides, AGN helps account for the BH mass-$\sigma$ correlation~\cite{silk1998,king2003}. 
Another presumed effect of AGN which was widely discussed  in the last few years is their positive feedback, i.e., their capability
of promoting SF formation in particular systems \cite{silk2013}. In this picture,
the AGN-generated central outflow compress dense clouds. The propagation of highly energetic jets into the ISM 
leads to the formation of peculiar structures such as expanding, over-pressurized cocoons, 
sweeping up cold clouds which can become gravitationally unstable, collapse and eventually be the site of newborn stars.
In principle, the interaction of the outflow 
with the surrounding gas stimulates SF on a time-scale shorter than the one required by negative feedback from SNe. However, as soon as new stars
form, `negative', pre-SN feedback from massive stars might be immediately active and locally mitigate the positive effect induced by the AGN.
An attempt to model the effects of positive AGN feedback in a starburst galaxy and
its complex interaction with stellar feedback may be found in Zubovas et al. \cite{zubovas2013}.

As for the requirement of a  THIMF to explain the large dust masses, some works have already shown that the
net effect of this assumption is to increase the specific dust productivity \cite{gall2011,dwek2011}.

The assumption of rapid dust accretion occurring in the coldest parts of the ISM is another viable, 
frequently invoked solution in order to have enhanced dust production.
The dust accretion timescale is a complex function of several additional parameters which are related to the microphysics of dust grains,
including the local thermodynamic state of the clouds, the 
grain size and the local metal content \cite{dwek1998,calura2008,kuo2012,palla2020b}.
It is not possible to have deep physical insights by means of chemical evolution models only,
where gas dynamics are not considered and the ISM modelling is generally single-phase. 
A deeper insight into the true dependence of $\tau_0$ on these quantities requires a dedicated modelling of the cloud physics, including also hydrodynamics,
heating phenomena such as thermal and radiative feedback and radiative cooling.
Chemical evolution models are nevertheless useful in order to gain constrains on this process based on phenomenological arguments.
If very high SF efficiencies need to be
assumed to reproduce the general properties of  QSO hosts, values for $\tau_0$ of the order of
a few Myr, need to be assumed to account for their dust content.
Remarkably, such value is one order of magnitude lower than the one typically
assumed to reproduce the dust content of local galaxies \cite{dwek1998,calura2008}.  
Such assumption implies a significantly enhanced grain growth and is supported by other arguments,
yet sometimes based on contrasting results, discussed later.

The early efforts to account for dust in various one-zone, non-cosmological
models favoured the development of a detailed formalism that allows one to account for dust evolution in more elaborate, physical models. 
One frequent way to model the evolution of galactic properties is by means of ab-initio cosmological
models that account for the evolution of galaxies based on the merger history of dark matter halos.
Broadly speaking, these approaches can be classified in two categories: one based on
semi-analytic models for galaxy formation, and another one based on hydrodynamical
simulations. 
The former approach follows self-consistently the evolution of the DM halos
and adopt semi-analytic recipes to model the properties of baryonic matter,
taking into account various processes such as radiative cooling, star formation and feedback
\cite{baugh2006}.  
In the second approach, the evolution of DM halos is modelled through cosmological
N-body simulations, whereas the gas evolution is tracked by resolving
the Euler equation set of gas dynamics, again taking into account all the most relevant baryonic processes
\cite{teyssier2002,Vogelsberger2020}.  
Many of these models now include prescriptions for dust production and destruction. 
There are several cases in the literature of the use of ab-initio cosmological models 
to study the evolution of both the dust-based scaling relation observed in galaxies and
the evolution of the dust budget in galaxies 
\cite{Mckinnon2017,aoyama2018,granato2021,parente2023}. 
Some results of such models will be discussed later,\linebreak  in Section \ref{sec_dmd}.

\subsection{On the Role of Interstellar Dust Growth in Galaxies}
\label{sec_growthrole} 
As interstellar dust can be both destroyed and produced,  
its survival requires that the continuous decrease in mass
to be compensated by a regeneration occurring at an adequate rate. 
As for the production mechanisms, it has been often advocated that stars can not be
the only ones contributing to its creation.  
Various studies indicate that interstellar silicates
are predominantly amorphous, whereas silicate samples from meteorites, likely of stellar origin,
are $\sim$20\% crystalline \cite{sarangi2018} (but see also \cite{li2001,kemper2004}).   
This is one argument often used to support the common notion that dust in the ISM might also 
have an origin different from stellar. 
Among the alternative creation processes proposed so far, the one of dust growth occurring in the coldest regions of the ISM, i.e., in molecular clouds, represents the best accredited. 
The role of dust growth and its contribution to the global dust budget has indeed been the matter of lively  
discussions, based both on  model results and low- and high-redshift galaxies observations. 


The standing question is whether accretion might be responsible for the large dust masses observed at high redshift in various sources.   
Grain growth by accretion is a challenging process hindered by several difficulties, 
which become more acute at high redshifts due to the rising temperature floor set by the Cosmic Microwave Background, meaning that grains are hotter \cite{ferrara2016}.
In the cold neutral medium (CNM), the density is too low and the dust is too hot for accretion to be efficient \cite{ceccarelli2018}.
In compact high-z systems, the thermal desorption is expected to be too strong and acting on too short timescales for small grains
to grow. 
Other processes that could hamper the growth are the Coulomb repulsive forces between positively charged ions (Si, C) that prevent them to
reach the grain surfaces and the strong UV radiation field \cite{ibanezmejia2019}. 
Molecular clouds offer more favourable conditions than the CNM for the growth to occur- for instance,
due to its high density (typically >10$^2$ cm$^{-3}$), molecular gas offers a natural shield against
UV radiation and the average grain temperature is lower.
{ In molecular clouds, also grain charge can play a role in the formation of large grains. 
  The charging due to collisions of dust grains with the plasma of thermal electrons and ions from the gas 
  leads to a strong Coulomb repulsion between grains.  
  However, the photo-emission induced by cosmic rays (see Section \ref{sec_charge}) 
  provides an approximately equal abundance of positively and negatively charged dust grains,
  giving place to optimum conditions for coagulation and growth of large aggregates~\cite{ivlev2015}}.  
In such conditions, it is presumably more likely for the refractory seed grains to develop icy mantles composed 
 primarily of water ice, plus  some more abundant molecules such as  NH$_3$, CO, CO$_2$ and others \cite{gall2018} 
However, one problem is that, after the dissolution of the natal molecular cloud, the grains are exposed again to
the erosion due to the above phenomena, therefore it is not clear how such mantles can survive in such conditions. 
The growth of graphite seems unlikely, as atomic carbon is very rare in molecular clouds as it is prone to be incorporated 
into CO molecules. Even the growth of silicates is challenged by the paucity of gaseous silicon presumably present in cold clouds \cite{ceccarelli2018}.
Although Coulomb barriers and other processes offer plausible brakes on grain growth,
in some cases the sensitivity of these conclusions to poorly constrained parameters—such as sticking efficiencies,
surface reaction rates and grain-charge distributions— may reveal how little we really 
know about the microphysics of interstellar grain surfaces \cite{hirashita2011,galliano2018}.

On the observational side, the role of growth has been examined through various studies of dust depletion in a variety of environments,
sometimes based on contradictory evidences. 
As for the local evidences for dust growth, besides 
the different depletion pattern of the multiphase medium directly observable
in our Galaxy \cite{savage1996}, indirect arguments are invoked. 
A commonly cited argument in favour of this process is that,
the total amount of dust produced by stars alone is insufficient
to account for the local interstellar dust budget, as it is significantly lower than the amount  
of dust destroyed in SN shocks \cite{dwek1998,calura2008}. 
However, one alternative possibility to consider is that the dust destruction rate measured in
SN remnants \cite{temim2015} or in models of dust formation in the ejecta of SNe 
\cite{bianchi2007,biscaro2016} may be overestimated. 
If this is the case and dust destruction plays a lesser role in the global dust budget, then dust accretion must also be less significant. 
This would help explain the observed pattern of dust depletion 
which, for some authors, can be adequately accounted for by stardust production \cite{gall2018}. 
A few more detailed, theoretical studies of dust accretion are based 
on models which take into account various physical processes,
such as the grain size distribution,  
grain shattering \cite{kuo2012} and coagulation \cite{asano2013}. 
In such models, the grain size plays an important role in the process of grain growth,
in particular with extremely high SFR values such as the ones commonly measured in high-z systems,
expected to generate a high degree of turbulence.
Numerical experiment indicate that turbulence in the ISM seems to enhance the compressibility of
the gas in cold clouds, enhancing the growth rate \cite{limattsson2020}
and possibly providing a solution to the standing questions about the timescale for dust formation at the highest redshifts \cite{mattsson2020}. 
In a highly turbulent medium, collisions between grains accelerated at high velocities may be particularly frequent,
rendering grain shattering very efficient and increasing the production of small grains.  
This effect can be understood considering the proportionality between  
the grain growth rate and the surface-to-volume ratio of the dust grains \cite{asano2013}, 
since a highly efficient shattering tends to 
increase the latter quantity, and in such an environment the growth is highly enhanced
(see, e.g., \cite{kuo2012,asano2013}. 
In this framework, a shorter accretion time-scale can be naturally explained by means of 
 a smaller average grain size and a large abundance of small grains. 

However, an overabundance of small grains is in contrast with the indications from another relevant observable,
 namely the extinction curve 
of high-z QSOs, generally obtained from their optical–near-IR spectra (e.g., \cite{gallerani2010})
and which appear significantly flat, with the implication of an overabundance of large grains (see also \cite{asano2014}).  
In principle, an enhanced dust growth could occur even in such conditions, possibly due to particularly 
large interstellar density values. 
Average molecular cloud and gas densities in star-forming galaxies at high redshift are typically found
to be larger than in local galaxies \cite{shirazi2014,dessauges2019,gilli2022}); this is another reason why
it is plausible that the high-$z$, cold ISM may be characterised by lower accretion time-scales than the local one. 

Also interstellar metallicity is expected to play an important role in dust growth. 
As both the formation and growth of dust grains are bound to the availability of heavy elements 
in the gas, 
\cite{asano2013} pointed out that a `critical metallicity' $Z_{cr}$ must be reached in galaxies 
for the contribution of dust growth to become dominant over the one of stellar production
(see also \cite{inoue2011}). 
The quantity $Z_{cr}$ was found to depend on some crucial galactic parameters, such as the star formation 
timescale, and to be typically solar or oversolar~\cite{asano2013}. 
In principle, this implies that accretion is less dominant in low-metallicity systems.
However, ref. \cite{hirashita2011} showed how the metallicity level at which 
the grain growth becomes dominant is strongly sensitive to 
the grain size distribution function. In particular, the adoption of a power-law grain distribution can decrease
$Z_{cr}$ to values as small as $\sim$0.1 $Z_{\odot}$, with the potential implication of
a significant role of grain growth even in local dwarfs or early galaxies. 

A more empirical indication on the role of dust growth in local and high-z systems 
comes from the observed evolution of the metal-to-dust (MTD) ratios, usually derived from the absorption
spectra of QSO or GRBs. 
\cite{zafar2013} 
combined extinctions and metal column densities from a large 
sample of GRB afterglows across a wide redshift range, from $z\sim~0$ to $z \lesssim 6$. 
The dataset includes HI and metal column densities measured in QSO absorbers and a few galaxy-lensed quasars.
The data collected by \cite{zafar2013} sample a large variety of lines of sight and galaxy types,
including different regimes of star formation (with dwarfs, `normal' galaxies and starbursts) and of stellar masses. 
Such variety is reflected in the broad range of measured column density values, spanning three orders of magnitude, and
metallicities, ranging from $\sim$1/100 Z$_{\odot}$ to supersolar.
The results of \cite{zafar2013} lend support to a universal MTD ratio, constant across a wide redshift and metallicity range 
and within a $\sim$ 0.3 dex scatter. 
The average MTD is also consistent with the values found in local systems, such as the Galaxy \cite{watson2011} 
and the Magellanic Clouds \cite{weingartner2001a,gordon2003,bernard2008} 
  using different techniques, as well as in the Andromeda galaxy \cite{smith2012}, 
and in other nearby spiral galaxies~\cite{issa1990} and dwarf galaxies \cite{lisenfeld1998}. 
The study of \cite{zafar2013} confirms also previous results obtained in a sample of QSO DLAs, 
 where the dust masses were constrained by means of the observed [Cr/Zn] depletion, which also found a
 roughly constant MTD in an extended redshift range ($0<z<3$, \cite{pei1999}). 
The most straightforward interpretation of these results is a high degree of synchronicity between dust and metal production. 
According to \cite{zafar2013}, the observed constancy of the DTM in systems characterised by a variety of properties (redshift, SFHs, 
metallicities) might indicate two possibilities. 
In the first scenario, 
in which dust is primarily of stellar origin, dust needs to be produced rapidly and by ubiquitous sources, 
in which case ideal candidates are core-collapse SNe, originating from the explosions of massive stars.  
The timescale of metal production is of the order of the lifetime of an `average' massive star exploding
as SN. This value depends on various parameters related to the choice of the IMF, that include both 
the mass range and the shape. In the reasonable case of a Kroupa (2001)\cite{kroupa2001} IMF, defined between $0.1$ M$_{\odot}$
and 40 M$_{\odot}$, and assuming that the SN progenitors are stars with mass $8 \le M/M{\odot} \le 40$,
the average progenitor has mass $<m_{MS}> \sim 20$ M$_{\odot}$, where
the average massive star value can be computed through the equation 
  \begin{equation}
    <m_{MS}> = \frac{\int_{8M_{\odot}}^{40M_{\odot}} m \phi_{K01}(m) dm}{\phi_{K01}(m) dm}. 
  \end{equation}
Moreover, a  20 M$_{\odot}$ star is characterised by a lifetime between $\sim$6 and 10 Myr, considering various standard stellar evolution libraries 
\cite{schaller1992,padovani1993,portinari1998}. Therefore, in the first scenario where massive stars exploding as SNe are
both factories of metals and dust, this value can be regarded as the typical production timescale for both components. 

If other dominant stellar producers are invoked (such as AGB), the significant delay between the formation
of the metals, having place mostly in SNe, and the production of dust, which must occur after these sources have moved off the main sequence, is, in principle, expected to lead to significant scatter in the 
dust-to-metal ratio. Therefore, an origin of dust in AGB stars, with lifetimes as long as 40 Myr or more, is not tenable
within a picture where the delay between the production of metals and dust has to be shorter than the typical timescale
of metal enrichment. 
One problem of this picture is that the stellar production of dust is dependent on metallicity, as
is the production of heavy elements in stars.  
The condensation of refractory elements in dust grains is unavoidably dependent on the element abundance, 
and this is to imply that dust production in stellar sources is inefficient below a critical metallicity \cite{mattsson2014a}. 
Such critical metallicity has been postulated to exist for grain growth, but the same could be true also 
in massive stars, in particular in the case in which, despite the large amounts of newly produced metals, dust production
depends on the presence of a few key elements necessary to have efficient nucleation, such as Si or Fe \cite{mattsson2014a}.
Considering that, according to the current knowledge, metal production is not universal and expected
to vary in environments with different metallicity, a nearly constant MTD of stellar origin
and a tight relation between metal column density and optical extinction 
might perhaps be achieved if the effective stellar dust yield is proportional to the metallicity
(\cite{mattsson2014a}; see also \cite{marassi2019}). 

Wiseman et al. (2017)\cite{wiseman2017} studied the DTM ratio in a sample of high-redshift GRB-DLA systems.
These authors discuss the different methods used to estimate the DTM in absorbing systems.
In the `traditional method', the extinction $A_V$ is compared to the equivalent metal column density, defined as
log N(H) + [M/H], where N(H) is the density of H and [M/H] is the metallicity of the gas computed
with respect to a reference value that, in most cases, is the solar metallicity \cite{chen2013,zafar2013}. 
In the alternative definition of the DTM, 
the dust fraction $F_d$ is determined from the dust depletion pattern 
of heavy elements in QSO and/or GRB absorbers \cite{vladilo2004,decia2013}.
This method is based on the evaluation of the depletion factor, an elaborate parameter that 
requires the knowledge of the column densities of both refractory (e.g., Fe) and volatile elements (e.g., Zn),
but that has a considerable practical and intuitive utility to express the degree of depletion.
Extinction-based methods could be affected by dust present along the line of sight, but not in the ISM of the
system under study \cite{decia2013}. 
This might lead to overestimate the dust in low-extinction systems that have dusty foreground
objects, and, in some cases to unreliable DTM estimates. 
At variance with the ones based on the optical extinction, studies based on the dust fraction $F_d$
find an increasing DTM as a function of metallicity. 
This is interpreted as 
a dominant role of growth in dust production with respect to stellar sources. 
This result is supported by a later dust depletion analysis of 18 metals in various 
environments, including the Milky Way, the Magellanic Clouds, and
DLAs toward GRBs and QSOs \cite{konstantopoulou2022}.
This analysis considered the relative abundances of heavy elements
with different refractory properties, including 70 new column density measures for some elements  (Ti and Ni)
in 70 QSO DLAs. 
Tight linear relations were found between the depletion of the metals (traced by the ratio between the abundance
and the non-refractory element Zn) and the global dust depletion strength, traced by the observed [Zn/Fe].
The slope of such dust depletion sequences were found to correlate
with the condensation temperature of the elements, in a way that  the more refractory elements showed steeper depletion sequences.
The strong correlations obtained in a variety of environments, SFHs and from low-metallicity systems
to other ones with solar abundances indicate a common origin for dust grains, outlining
the key role of growth in the buildup in cosmic dust \cite{konstantopoulou2022}.

Besides this set of empirical arguments based on the analysis of the depletion pattern in various environments,
there is further evidence of star-forming systems showing the presence of significant amounts of dust at very young ages
that precede the lifetimes of the most massive SN progenitors. 
One example is in the local dwarf galaxy NGC 5253, hosting a young star cluster with a stellar mass of $\sim$10$^6$ M$_{\odot}$ that has gather attention due to its 
very high efficient star formation \cite{turner2015}. 
The dust continuum emission observations performed with the Submillimetre Array 
suggest a dust mass of 1.5$\times$10$^4$ M$_{\odot}$ and the detection of the CO J = 3-2 rotational transition enabled
the estimate of the gas-to-dust, 
found to be lower than the Galactic value. 
Stellar evolution models support a 4.4 Myr old young stellar population, a too young age for a significant SN enrichment. 
This evidence indicates a young star cluster deeply embedded into dust, in which the dust was presumably produced locally.
This is not the only case of a local detection of dust-rich, young star clusters. 
High-resolution observations of the blue compact galaxy II Zw 40 performed with ALMA revealed the presence of massive star clusters
surrounded by various gas- and dust-rich clouds \cite{consiglio2016}. Despite the technical difficulties of mapping the isolated dust emission, that requires subtracting the 3 mm free-free continuum from the 870 µm continuum,
the dust continuum is enhanced towards the star-forming regions compared to the CO emission. 
 The interpretation is that the presence of 
 the dust is spatially associated to the young stars hosted by the star clusters, whereas 
considerations on the dust-emitting region size and the presence of radio emission support rapid dust production. 
 
More ALMA observations indicate the presence 
of 12--14 dusty star cluster in the central starburst of NGC 253 \cite{levy2021}. 
Also in this case, the clusters are very young (\mbox{$10^5$ yr}), therefore it is implausible
that they underwent significant SN enrichment. 
The list can continue, with obscured, young (with ages of a few Myr) star clusters detected in the Antennae galaxies \cite{gilbert2000} and in other starbursts
\cite{cohen2021,Rodriguez2023,Whitmore2023}.

A further proof of the ubiquity of young clusters in dusty environments even at high redshift is provided by the observation of the Sunburst
lensed star cluster at $z=2.37$ \cite{vanzella2022}, 
where the high Si depletion levels measured in the surrounding clouds indicate that they host dust grains \cite{pascale2023}.

Considering carbon-rich Wolf–Rayet binaries as possible sources, for these systems, models indicate dust production
rates $\dot{M}_d$ between $10^{-10}$M$_{\odot}$ yr$^{-1}$ and $10^{-6}$M$_{\odot}$ yr$^{-1}$ \cite{lau2020}.
In the case of a dust-rich young star cluster like the one of NGC 5253 \cite{turner2015}, assuming a Kroupa (2001) \cite{kroupa2001} IMF
and a WR binary fraction of $40 \%$ \cite{shenar2020},  
a $1.5~10^4$ M$_{\odot}$ dust mass can be accounted for by 
dust-rich Wolf–Rayet winds with a continuous rate of dust production $\dot{M}_d~10^{-6}$M$_{\odot}$ yr$^{-1}$, i.e., by assuming the maximal value of the rate from the current estimates~\cite{lau2020}.  
This still leaves room for a substantial contribution of dust growth in the formation of dust in these systems.  

A separate argument in favor of accretion concerns the high level of depletion ($99\%$) 
normally observed for Fe in various environments \cite{gioannini2017a}.
Considering that Type Ia SNe are the major Fe contributors in the Universe
but are no dust producers, 
a substantial fraction of Fe has to be incorporated in dust by means of some other process \cite{dwek2016,gall2018}.
One possibility is that this occurs via interstellar growth of Fe-rich nanoparticles,
characterised by a particular small size (<0.01~$\upmu$m) and therefore with a small accretion timescale \cite{gioannini2017a}.  

Despite the independent indications supporting a substantial
role of dust growth in a large variety of environments, several unanswered questions undermine our theoretical
understanding of this process at the microphysics level.

\vspace{-6pt}

\subsection{Cosmic Evolution of the Comoving Dust Mass Density}
\label{sec_dmd} 
Performing a direct estimate of how the cosmic dust mass budget
has evolved throughout cosmic history is crucial for constraining a key component of
the cold mass fraction in galaxies.
The quantity suited to this purpose is the comoving dust mass density (DMD), whose evolution has been the focus of a series of investigations in the last couple of decades. 

The most direct way to estimate the DMD is through the dust mass function (DMF), defined as 
the number density of galaxies as a function of dust mass. The DMF is \mbox{defined as} 
\begin{equation}
\Phi(M_d) = \Sigma_i \frac{1}{V_i} \Delta M_d, 
\end{equation}
in which $V_i$ represents the accessible volume in some band in the parent sample of the $i-$th 
source, and with the sum performed over all objects in the mass interval\linebreak  ($M_d$,$M_d +\Delta M_d$)~\cite{dunne2003,dudzeviciute2021}. 
The DMD represents the integral of the DMF over the entire sampled dust mass~range.

So far, only a limited number of studies have focused on the evolution of the DMF.
The first estimate of the DMF at low $z$ is based on the results of the
SCUBA Local Universe Galaxy Survey, the first
attempt at an unbiased submillimetre survey of the local Universe~\cite{dunne2003}.
These authors compare their results at low z with a 
high-redshift ($z\sim 2.5)$ estimate based on submillimetre data from the deep SCUBA submillimetre surveys. 
Essentially, both estimates are based on small samples of ultra-luminous infrared
galaxies. The dust masses were estimated from the submillimetre fluxes as reported in Equation  (\ref{eq_mdust}), assuming
a fiducial value for the average dust temperature. 
 
Later on, Dunne et al. \cite{dunne2011} studied the evolution of the density of galaxies in relation to their dust mass up to a redshift of $z < 0.5$.
Their sample included approximately 2000 sources selected at 250 $\upmu$m
from the Herschel-ATLAS Science Definition Phase, each with a reliable counterpart in the Sloan Digital Sky Survey catalog.
The results indicated an increase in the bright end of the Dust Mass Function (DMF) between z = 0 and $z \sim 0.5$. 
Based on a factor $\sim$10 larger sample consisting of $\sim$15,000 sources,  
Beeston et al. \cite{beeston2018} determined the local dust mass fraction (DMF) by utilizing the combined
data from the Herschel-ATLAS and GAMA surveys, finding a good agreement with previous estimates~
\cite{vlahakis2005,dunne2011}. The substantial improvement of Beeston et al. \cite{beeston2018} was that the sample
probed dust masses as faint as $10^4$ M$_{\odot}$, finding a large abundance of galaxies with low $M_d$ values.
A subsequent, updated estimate based the extended  FIR-selected sample from the Herschel Astrophysical Terahertz Large Area Survey (H-ATLAS)
over a large area of sky featured 
an order of magnitude more galaxies than used in previous studies, confirming the previous results at $z\le 0.5$ \cite{beeston2024}. 

The first study of the evolution of the dust mass function across a comprehensive redshift range, from redshift \( z \sim 0.2 \) to \( z \sim 2.5 \) was
presented by Pozzi et al. \cite{pozzi2020}. This analysis was based on an Herschel-selected FIR catalogue  at 160 $\mu$m 
in the COSMOS field.
The sample consisted of approximately 5500 sources with a flux density greater than 16 mJy,
and it included estimates of either spectroscopic or photometric redshifts.
Traina et al.\cite{traina2024} improved these results performing an ALMA-based exploration using $\sim$20
serendipitous galaxies from the ALMA A$^3$COSMOS database \cite{liu2019} to study the DMF at $0.5 <z<6$.

Besides the direct, DMF-based estimates, other authors have assessed the DMD with alternative approaches.
For instance, the cosmic density of dust can be estimated from 
Mg II absorbers identified by the strong absorption lines in the spectra of distant quasars \cite{menard2012}.
As these strong Mg II absorbers predominantly exist in galactic halos,
their composition is expected to closely reflect the amount of dust that exists outside galaxies. 
Another method to estimate the dust budget evolution is by means of the cosmic far-infrared background.
This background is fueled by the UV and optical emissions from young stars, which is absorbed by the dust 
and subsequently re-emitted at IR  wavelengths \cite{debernardis2012}, accounting for the contribution from 
unresolved sources.

Driver et al. \cite{driver2018} utilized multiwavelength galaxy catalogs,
including GAMA \cite{driver2018} and G10-COSMOS \cite{davies2015}. 
These catalogs encompass panchromatic photometric data ranging from UV to mid-IR.
Using the MAGPHYS code \cite{dacunha2012} and SED-fitting techniques,
the authors aimed to model the spectra of galaxies at various redshifts. 
Their approach was based on the energetic balance between the radiation
that is attenuated by dust in the UV and optical bands and the amount that is re-radiated in the far-IR.

Magnelli et al. \cite{magnelli2020} conducted the first study of the DMD from
redshift \( z \sim 0.5 \) to \( z \sim 5 \). As part of the ALMA Spectroscopic Survey (ASPECS) large program,   
they utilized deep 1.2 mm emission continuum map in the Hubble Ultra Deep Field (HUDF),
with a sensitivity of 9.5 $\upmu$m Jy beam\(^{-1}\), covering an area of about 4 arcmin$^2$.
Other recent indirect estimates from the continuum emission  
do not consider the contribution of individual galaxies, but require an estimate of the temperature of the warm dust \cite{eales2024}. 
Since a given amount of warm dust emit more radiation than the same mass in the form of cold dust,
the luminosity-weighed temperature will always be higher than the more physically-meaningful mass-weighted one
(see also \cite{pozzi2020,pozzi2021}. A way to limit this bias is to estimate the dust mass at
very long rest-frame wavelength, in the Rayleigh Jeans part of the SED, where the approximated emission
spectrum is proportional to $T_d$ \cite{eales2024}.

In Figure \ref{fig:dmd} we show the evolution of DMD from various observational
estimates derived both with direct and indirect methods. 
The data indicate an increase of a factor $\sim$3--4 of the DMD from z = 0 to $z\sim 1$, followed by
a decrease at larger redshifts.
The compilation of data chosen for this review are presented in Table \ref{tab_dmd} and
confirm the presence of a peak of the DMD around $z\sim 1$.
Some authors have noticed the similar behaviour of the DMD evolution and
the cosmic SFR density \cite{driver2018,pozzi2020,eales2024}. 
In principle, a coincidence between the peaks of the two quantities would argue in favour
of a fast dominant dust production mechanism that evolves in lockstep with the star formation.
This could be either from CC SNe or tied with the molecular gas evolution,
as both the comoving CC SNe rate and molecular gas density evolution are known to trace well the one of the
cosmic SFR~\cite{sadat1998,calura2006,magnelli2020,eales2024}. 
The grey area in Figure \ref{fig:dmd}, obtained by simply joining together all the upper and lower endpoints of the
error bars, is aimed to outline the dispersion in the data. 

\vspace{-6pt}
\begin{figure}[H]
  \flushleft
  
\begin{adjustwidth}{-\extralength}{0cm}
\centering 

\includegraphics[width=17 cm]{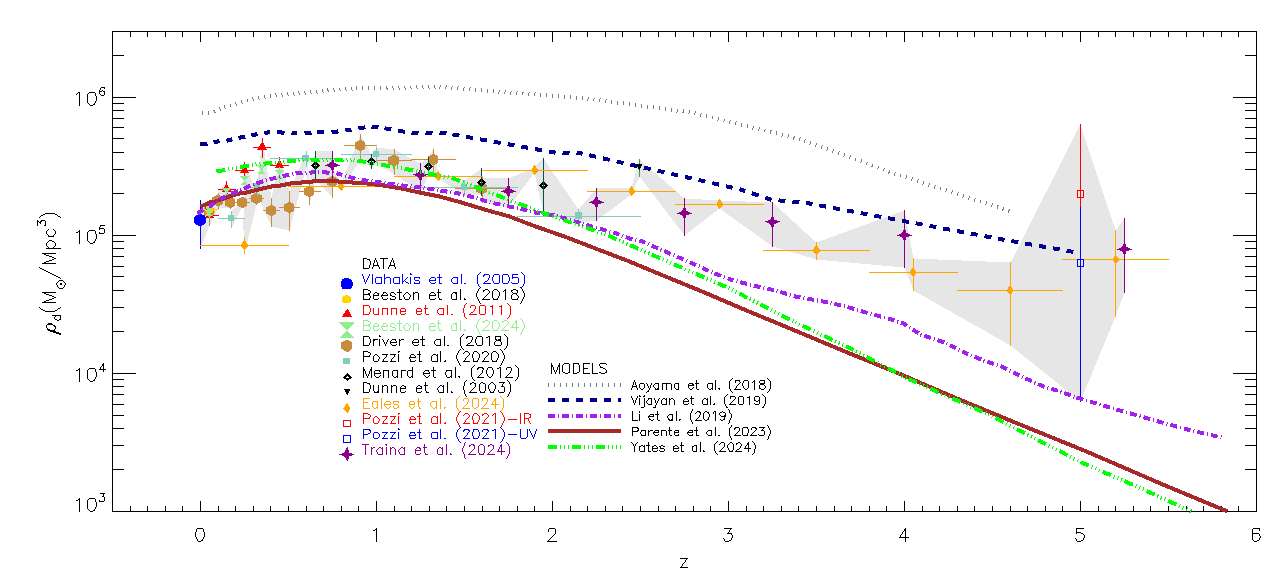}
\end{adjustwidth}

\caption{Redshift evolution of the observed comoving dust mass density compared with theoretical results.
  The observational determinations (symbols with error bars) from various authors are collected in Table \ref{tab_dmd} 
  and include estimates obtained with various methods, both direct (i.e., from the integral of the dust mass function) and
  indirect (without integration from the DMF but using various diagnostics, for further details  see Section \ref{sec_dmd}). 
  The theoretical results are from cosmological models including dust prescriptions, based both on semi-analytic techniques
  and on hydrodynamical simulations.} 
\label{fig:dmd}
\end{figure}

The increasing number of  estimates of the DMD have motivated a series of attempts to
model its behaviour with various approaches.
In particular, some effort was focused on explaining the fall between $z=1$ and $z=0$. 
Non-cosmological approaches conducted with chemical evolution models of galaxies of different morphological
types can account for the observed evolution allowing for an observationally-motivated
number density evolution of spirals and ellipticals \cite{gioannini2017}.
In other cases, the significant decrease observed at $0 \le z \le 1$ can be explained by a
combination of astration and SN destruction \cite{ferrara2021}. 
Other authors have modelled the DMD evolution by means of  ab-initio cosmological models.  
By means of a cosmological SAM that taking into account dust evolution, Parente et al.~\cite{parente2023} explain
the drop of the DMD with efficient supermassive BH growth during disc instabilities, that
causes the quenching of the SF in the most massive galaxies.
In this picture, astration, ejection of dust by outflows and SN destruction have all important roles
in decreasing the overall galactic dust content at $z<1$.
Other works based on cosmological simulations failed  account for the observed DMD evolution \cite{aoyama2018,li2019,vijayan2019}.
However, the largest discrepancies between different models are found at $z>1$.  
In Figure  \ref{fig:dmd} we report the DMD evolution as obtained by means of a few, selected cosmological models.
The model results were chosen to emphasise the dispersion of the results from various authors, significantly larger than the one of the
observational estimates. 
This disagreement concerns both the normalisation and the shape of the DMD evolution.
A broader peak and a higher normalisation at $z\sim 0$ is obtained by some authors \cite{aoyama2018,vijayan2019}, whereas
others find low-redshift values that are more in line with
observations, yet underestimating the observed high-redshift behaviour~\cite{li2019,parente2023,yates2024}.
Although the theoretical frameworks of the models are conceptually similar, the numerical implementations of various processes
can differ significantly, and it may not be easy to identify the causes of such discrepancies. 
Additionally, a major contributing factor may be that the dust model is integrated with various galaxy evolution models.
Each of these models includes sub-grid recipes for processes such as star formation and chemical enrichment,
which are essential for the production and evolution of the dust mass in galaxies. 
Understanding the reasons for such disagreement is beyond the scope of this review. 
However, this is useful to outline the current poor theoretical understanding of the evolution of the
dust mass budget, to stress out the need for pursuing more surveys at high redshift.
This endeavour will be important to confirm the results from extant observational programs and to
constrain further the most fundamental dust production processes implemented in current simulations.

\begin{table}[H]

\caption{Comoving Dust mass function as a function of redshift from various authors.}
\newcolumntype{C}{>{\centering\arraybackslash}X}
\begin{tabularx}{\textwidth}{CCCC}
\toprule
\multirow{2}{*}{\textbf{Reference}}	& \textbf{\textbf{z} $\boldsymbol{\pm}$ Error}	& \multirow{2}{*}{$\boldsymbol{\rho}_\textbf{\emph{D}}$ $\boldsymbol{\pm}$ \textbf{Error}} & \multirow{2}{*}{\textbf{Comments}} \\
                                       &  (\textbf{10}$^\textbf{5}$ \textbf{M$_{\odot}$} \textbf{Mpc}$^{\textbf{$-$3}}$) \\
\midrule
%
%
%
%
 Vlahakis et al. (2005) \cite{vlahakis2005}  & 0 \, - &  1.29$_{-0.49}^{+0.50}$ & Direct method \\ 
 \midrule
 Beeston et al. (2018) \cite{beeston2018}     & 0.05 \, 0.05  &   1.55 \, 0.03  & Direct method, various datasets \\
 \midrule
 Dunne et al. (2011) \cite{dunne2011} &  0.05 \,  0.05 &  1.39 \,  0.21    & Direct method \\
                                   &  0.15 \,  0.05   &  2.14 \,  0.32    &  \\
                                   &  0.25 \,  0.05   &  2.95 \,  0.45    &  \\
                                   &  0.35 \,  0.05   &  4.34 \,  0.66    &  \\
                                   &  0.45 \,  0.05   &  3.20 \,  0.48    &  \\
 \midrule
 Beeston et al. (2024) \cite{beeston2024}     & 0.05 \,  0.05   &  1.47 \,  0.13 &       Direct method    \\
                                             &  0.15 \,  0.05  &  2.04 \,  0.14 &                        \\
                                             &  0.25 \,  0.05  &  2.47 \,  0.04 &                        \\
                                             &  0.35 \,  0.05  &  2.97 \,  0.12 &                        \\
                                             &  0.45 \,  0.05  &  2.76 \,  0.36 &                        \\              
 \midrule
 Driver et al. (2018) \cite{driver2018}  &   0.05 \,  0.03  &   1.48   \, 0.30 & From various datasets\\
                                        &    0.10 \,  0.04 &   1.78  \, 0.22 & \\
                                        &    0.17 \,  0.03 &   1.74  \, 0.17 & \\ 
                                        &    0.24 \,  0.04 &   1.74  \, 0.12 & \\
                                        &    0.32 \,  0.04 &   1.86  \, 0.18 & \\
                                        &    0.41 \,  0.04 &   1.51  \, 0.35 & \\
                                        &    0.50 \,  0.06 &   1.58  \, 0.50 & \\
                                        &    0.62 \,  0.06 &   2.09  \, 0.42 & \\
                                        &    0.75 \,  0.07 &   2.45  \, 0.57 & \\
                                        &    0.91 \,  0.09 &   4.47  \, 0.90 & \\
                                        &    1.10 \,  0.10 &   3.47  \, 0.70 & \\
                                        &    1.33 \,  0.12 &   3.55  \, 0.62 & \\
                                        &    1.60 \,  0.15 &   2.19  \, 0.44 & \\
 \midrule
 Pozzi et al. (2020) \cite{pozzi2020}    &    0.18  \,   0.08 &   1.32 \,  0.17 &  Direct method,\\
                                        &     0.32 \,  0.08 &   2.24  \,  0.29 &  Herschel data\\
                                        &     0.60 \,  0.20 &   3.62  \,  0.46 &  \\
                                        &     1.00 \,  0.20 &   3.85  \,  0.48 &  \\
                                        &     1.50 \,  0.30 &   2.22  \,  0.29 &  \\
                                        &     2.15 \,  0.35 &   1.38  \,  0.17 &  \\
 \midrule
 Menard et al. (2012) \cite{menard2012}  &   0.65 \, - & 3.20$_{-0.65}^{+0.83}$  & Computed from    \\
                                        &   0.97 \, - & 3.44$_{-0.37}^{+0.41}$  & MgII absorbers   \\
                                        &   1.29 \, - & 3.14$_{-0.49}^{+0.52}$  &                  \\
                                        &   1.60 \, - & 2.43$_{-0.49}^{+0.61}$  &                  \\
                                        &   1.95 \, - & 2.28$_{-0.84}^{+1.32}$  &                  \\
 \midrule
 Dunne et al. (2003) \cite{dunne2003}    &  2.49 \, -  & 3.11 \,  0.46  &    DMF estimated from  \\
                                        &             &                &    different surveys   \\

 \bottomrule
\end{tabularx} 
\end{table}

\begin{table}[H]\ContinuedFloat
\caption{\textit{Cont.}}
%
\begin{tabularx}{\textwidth}{CCCC}
\toprule
\multirow{2}{*}{\textbf{Reference}}	& \textbf{\textbf{z} $\boldsymbol{\pm}$ Error}	& \multirow{2}{*}{$\boldsymbol{\rho}_\textbf{\emph{D}}$ $\boldsymbol{\pm}$ \textbf{Error}} & \multirow{2}{*}{\textbf{Comments}} \\
                                       &  (\textbf{10}$^\textbf{5}$ \textbf{M$_{\odot}$} \textbf{Mpc}$^{\textbf{$-$3}}$) \\
\midrule

 Eales et al. (2024) \cite{eales2024}    &     0.25 \,  0.25  &  0.85 \,  0.11  & Indirect method  \\ 
                                        &     0.80 \,  0.30  &  2.27 \,  0.13  &                  \\ 
                                        &     1.35 \,  0.25  &  2.69 \,  0.12  &                  \\
                                        &     1.90 \,  0.30  &  2.96 \,  0.12  &                  \\
                                        &     2.45 \,  0.25  &  2.09 \,  0.12  &                  \\
                                        &     2.95 \,  0.25  &  1.68 \,  0.11  &                  \\
                                        &     3.50 \,  0.30  &  0.78 \,  0.11  &                  \\
                                        &     4.05 \,  0.25  &  0.54 \,  0.14  &                  \\
                                        &     4.60 \,  0.30  &  0.40 \,  0.24  &                  \\
                                        &     5.20 \,  0.30  &  0.67 \,  0.41  &                  \\
   \midrule
 
 Pozzi et al. (2021) \cite{pozzi2021}    &     5 \, -  &   2.00$_{-1.00}^{+4.31}$ &  Direct method, IR estimate \\
                                        &     5 \, -  &   0.63$_{-0.57}^{+0.95}$ &  UV-selected galaxies       \\
 \midrule
 Traina et al. (2024) \cite{traina2024}  &  0.75 \,  0.25  &  3.24$_{-0.72}^{+ 0.84}$     &  Direct Method        \\  
                                        &  1.25 \,  0.25  &  2.75$_{-0.66}^{+ 0.56}$     &                       \\
                                        &  1.75 \,  0.25  &  2.09$_{-0.50}^{+ 0.48}$     &                       \\
                                        &  2.25 \,  0.25  &  1.74$_{-0.45}^{+ 0.45}$     &                       \\
                                        &  2.75 \,  0.25  &  1.45$_{-0.45}^{+ 0.42}$     &                       \\
                                        &  3.25 \,  0.25  &  1.26$_{-0.43}^{+ 0.48}$     &                       \\
                                        &  4.00 \,  0.50  &  1.00$_{-0.41}^{+ 0.51}$     &                       \\
                                        &  5.25 \,  0.75  &  0.79$_{-0.41}^{+ 0.52}$     &                       \\
\bottomrule     
\end{tabularx}                                                                                                        
\label{tab_dmd}
\end{table}

\section{Future Perspectives on Dust Evolution}
\label{sec_future}
The last substantial advancements in our understanding of dust evolution in galaxies 
can be largely attributed to the groundbreaking observations of various space missions, with Herschel playing a primary role. 
Its contributions not only revolutionised our general knowledge of the infrared universe, but also spurred the development
and the use of complementary ground-based instruments.
The synergy between space-borne and terrestrial efforts has been crucial
in overcoming Herschel's inherent limitations---such as its challenges in detecting high-redshift galaxies---thereby enabling more advanced explorations of cosmic dust properties.
Following the conclusion of the Herschel mission, several advanced instrument proposals have emerged, each aiming to extend its groundbreaking 
legacy and deepen our understanding of the infrared universe.
Among the most recent mission proposals, SPICA stood out as a visionary successor to Herschel.
It was designed to transcend Herschel’s capabilities by offering enhanced sensitivity and spectral resolution.
SPICA’s design was anchored by three flagship instruments, 
offering a collecting area of roughly 4.6 m$^2$, while the instruments promised a sensitivity improvement
of about one to two orders of magnitude compared to Herschel, 
potentially with considerable advantages on advancing our understanding of the emission and chemical evolution of dust grains.
The cancellation of SPICA in 2021 was profoundly disheartening for the IR astronomy community.
It left a noticeable void in advanced observational capabilities, effectively rendering the field temporarily ``orphaned", before their
focus gradually shifted towards a new target. 
With no currently operating FIR observing facilities, the IR community's efforts are now focused 
on the PRobe far-Infrared Mission for Astrophysics (PRIMA). 
Currently under evaluation and competing with other advanced facilities,
PRIMA is envisioned to be equipped with two flagship instruments that promise significant advancements
in sensitivity and spectral resolution, heralding a new era of exploration in the infrared universe. 
At the time of this writing, PRIMA has just undergone the NASA `Phase A' study, with final mission selection expected in 2026.
As for the object of the present review, I would like to discuss a few benchmark science cases
in which PRIMA (and its successors in the mid– and far–IR) can uniquely drive forward
our understanding of dust evolution.

The first is the need for significant improvements of our knowledge of the chemical composition of dust. 
From a mineralogical point of view, the composition and structure of dust grain is largely unknown. 
We have mostly indirect constraints regarding elemental depletion from observations of the diffuse ISM
and from the detection of only a handful of known spectral features.
These features include the extinction bump at 2175 \AA, that is the carrier of important information
of carbonaceous dust. 
Solid-state absorption bands due to refractory material occur throughout the spectral range from approximately 2.5~up to $\sim$100 $\upmu$m,
which encompasses the wavelength band where the Earth's atmosphere transitions from being partially obscuring to completely opaque.
Molecular vibrations in solid materials arise as individual atoms oscillate within the constraints of their bonds with neighboring atoms.
When infrared photons interact with cold dust, absorption bands are produced at wavelengths that correspond to the transition
from the ground state to excited vibrational states in the solid. 
Conversely, if the dust is warm or hot, emission bands appear at the same wavelengths as excited molecules relax back to the ground state \cite{snow2004}. 
Examples in this range are absorption features such as the silicate bands at $\sim$8 $\upmu$m and $\sim$10 $\upmu$m and the
hydrocarbons at $\sim$3.4 $\upmu$m and water ice band at 3.08 $\upmu$m. 
A compound or molecule in the gas phase will experience the same vibrational transitions as it does in the solid state.
However, it is typically easier to differentiate the absorption bands of the solid state from those of the gas phase.
This is because the vibrational states in a solid are influenced by neighboring atoms in the lattice, which results in broader and smoother absorption features.
In principle, with high resolution observations, it is possible to distinguish individual rotation lines from those from solid state species. 
Therefore, high-resolution, high S/N observations of local star-forming regions would allow us to 
detect more faint solid-state features and refine our understanding of the chemical composition and structure of interstellar grains \cite{galliano2025}.

The second aspect concerns improving our knowledge of the evolution of the global dust budget with cosmic time.
The best quantity suited for this purpose is the DMF, which is functional to estimate the comoving DMD.   
As outlined above, there are several indirect estimates of the comoving DMD, but very few stem from a direct assessment of the DMF.
While various studies exist of the local DMF, only a handful of studies have been performed of its evolution
with redshift. Estimating the DMF evolution from FIR/submm observations is complicated by various factors.
Estimating it with ground telescopes, such as ALMA, is hard for various reasons.
First of all, ALMA is characterized by a small field of view, which makes wide-area surveys extremely challenging.
Another limitation of ALMA is the limited effective wavelength coverage. 
As it is convenient to derive the dust mass from the Rayleigh-Jeans tail of their dust-continuum emission, ALMA is normally used
mostly for high-redshift sources ($z>1$, e.g., \cite{dacunha2021,traina2024} and complementing the available
data with those obtained with another instrument, typically Herschel.
However, in case archival data are used from inhomogeneous collections of ALMA archival observations or pre-selected targets,
one big issue is to recreate the conditions of a blind survey, therefore using artifacts to eliminate possible biases \cite{traina2024}.
However, the major complication of such a task is that no space instruments is currently running, that can overcome the
limitations of Herschel.  
Past studies based on Herschel data suffered from significant confusion noise and allowed to probe with adequate confidence mostly the bright end of the DMF
at high redshift \cite{pozzi2020}. 
All the extant studies of the DMF could sample only the larger dust masses; as a consequence, a significant fraction (if not the bulk) of the dust mass is
currently missing from the current high-redshift inventories. 
The current limited theoretical understanding of the evolution of the dust mass budget highlights the necessity for conducting more surveys at high redshifts.
This will help better define the shape of the DMF and obtain new estimates during the so-called “Cosmic Noon,”
which refers to the period between z $\sim$ 1 and 3, where the peak of cosmic star formation is observed. 
A high-sensitivity space IR telescope is therefore necessary to perform a blind survey, to overcome the limits of current studies and improve our knowledge 
of the global dust budget at high redshift \cite{traina2025}. 

The last important aspect to probe in the future are the properties and the amount of dust in local, low-metallicity systems.
In general, gaining deeper insights into low-metallicity local galaxies is crucial for interpreting deep surveys,
as they provide valuable information about the faint end of the luminosity function. 
Moreover, such galaxies offer  a unique opportunity to study early cosmic conditions in a more accessible environment. 
The origin of dust in low-metallicity, low-mass galaxies is poorly understood, and studies documenting this aspect are rare, even in the local Universe.
The measures of the dust mass in low-metallicity galaxies require very high telescope sensitivity and are therefore severely 
limited. These include ALMA studies of the blue compact dwarf galaxy SBS 0335-052~\cite{hunt2014,cormier2017} and
Herschel observations of I Zw 18 \cite{fisher2014}. Despite these systems are actively star forming, they stand out as two of the lowest metallicity galaxies of the nearby Universe and
are characterised by 12 + log(O/H) $ \sim  7.2$.
Moreover, these galaxies tend to populate scarcely populated regions of common diagrams used to study local scaling relations.  
The $M_{dust}$–SFR relation in local metal-poor galaxies deviates significantly from the tight, quasi-linear relation observed
in normal star-forming galaxies, showing differences of more than two orders of magnitude \cite{fisher2014}. 
Despite a similar metallicity, in these faint systems the dust-to-gas ratios differ by more than one order of magnitude,
which suggests that there are other factors than metallicity that regulate this quantity and the total dust mass \cite{fisher2014}.
Moreover, these galaxies serve as nearby analogs to distant, primordial galaxies.
Galaxies showing a bursty character and, at the same time, the extreme paucity of heavy elements and dust are presumably very common at high redshift. 
While 'maximum' starbursts (i.e., with SFR $\sim 10^3$ M$_{\odot}$/yr) are very rare even at high redshift,
from the knowledge of the UV luminosity function at high redshift, a density of $\sim 0.001$ Mpc$^3$ can be estimated for
blue, dust-poor systems \cite{fisher2014} and, in very rare cases, constraints on their dust content are accessible only
through gravitational lensing \cite{calura2021}. 
An important standing question is for how much time such systems can remain dust-free, even from a theoretical point of view, if
the buildup of dust occurs through fast processes such as dust growth.

{Observations with  the James Webb Space Telescope (JWST) have identified galaxy candidates at $z\ge 10$ that appear nearly dust-free,
  with extremely low dust attenuation~\cite{tsuna2023}, alongside more dust-obscured systems \cite{rodighiero2023}.  
These findings already point to a complex and potentially rapid build-up of dust in the early Universe. 
In the coming years, JWST will enable tighter constraints on dust attenuation laws and their variation across galaxies through rest-frame
UV-to-optical spectral energy distributions and spectroscopy, including Balmer line measurements.
These observations will provide indirect constraints on dust composition and grain properties, 
offering a crucial complementary framework for interpreting absorption-line studies of high-redshift systems.}

Other peculiar features of the dust in low-Z galaxies have been evidenced in the literature,
which include dust emission spectra broader and peaking at shorter wavelengths than metal-rich galaxies \cite{remyruyer2015},
besides a dust size distribution shifted towards small grains in dwarf galaxies \cite{galliano2005}
and a submillimetre excess despite the moderate observed dust masses~\cite{chang2021} and
a dearth of PAH features \cite{madden2006,hunter2024}. 
Until more advanced, higher sensitivity telescopes can enable the survey of a statistically significant
number of metal-poor galaxies in well-resolved FIR-mm continuum bands, all these phenomena will remain 
a series of unanswered questions.  
At present, samples containing hundreds of low-metallicity galaxies exist such as, e.g., the
ALFALFA [H I] survey \cite{haynes2018}, containing a sizeable amount of systems with M$_{HI}<10^7$M$_{\odot}$,
and the  Sloan Digital Sky Survey (SDSS) with several systems with  12 + log(O/H) $\sim$ 7.20 (Ref. \cite{hsyu2018}, see also \cite{grossi2025}). 
A future IR survey of a statistically significant sample of nearby low-Z galaxies from these catalogues will
allow us to explore the properties of dust in metal-poor systems, providing answers to these questions and advancing
considerably our knowledge of several key topics of modern astronomy. 
This is only a limited list of cases that require advancement by means of future observations.
These testbeds will probe new, elusive regions of the dust evolution parameter space and will serve as pillars for future
dust evolution models, anchoring them with new, observationally driven constraints and highlighting where our current prescriptions must improve.

\section{Summary and Conclusions}
\label{sec_summary}

Despite its tiny mass fraction normally estimated in the ISM, dust is a major galactic constituent 
as it affects a large multitude of properties. 
Of all of these, the ones at the focus of this review are the effects of dust grains on interstellar chemical abundances.

I described the main features of galactic chemical evolution models (Section \ref{sec_chem}). 
These models allow one to compute the time evolution of the elemental abundances in the ISM
starting from a set of basic assumptions to model galactic 
star formation, the relative fraction of stars of various masses and the nucleosynthesis prescriptions, coming
from stellar evolution models,  for various stellar sources. 
These models can include also a detailed treatment of dust evolution.
The main physical processes regulating dust evolution are stardust production, destruction and interstellar growth.
The main sources of stellar dust production considered in most models are AGB stars and CC SNe. 
In particular, the role of SNe as rapid dust producers has been often considered important 
to explain the large dust masses observed at very high redshift, when star formation in early galaxies had presumably just started. 
In the last few years, new estimates of dust production from  other sources, such as Wolf-Rayet stars, acting on smaller timescales
than SNe, lead these systems to gain considerable interest. 
Also the role of the production from QSOs is discussed, in general regarded as minor with respect to
the other sources.

Dust destruction is thought to occur primarily as due to interstellar SN shocks.
In chemical evolution models, this process is regulated by a characteristic timescale that depends (i) 
on the total SN rate, including both contributions from CC and type Ia; 
(ii) on the average mass swept by a SN remnant and (iii) the dust fraction present in this material.
Other destruction or removal processes were mentioned, including astration (the dust removal due to star formation)
and the dust expelled via galactic outflows.

Also dust growth is regulated by a typical timescale, assumed to depend on the amount of dust already present
in the ISM and, in some cases, on the metallicity. 
The overall dust budget is therefore regulated by a significant amount of parameters, accompanied by a substantial lack of knowledge
of the roles of each one of them.
This stresses out the urgent need for more observational constraints from future IR facilities.

In Section \ref{sec_howdust}, I discussed the main effects of  interstellar dust on the abundances measured in high-redshift galaxies. 
Past studies of the chemical abundance pattern measured in the
absorption spectra of distant QSOs turned out to carry precious information on dust at high redshift. 
A significant deal  of work in this field have evidenced the presence of considerable amounts of dust in high-redshift systems
found along the lines of sight of QSOs. This achievement was mostly due to the collection and study of a large number of 
interlopers known as Damped Lyman $\alpha$ absorbers, recognizable for the deep absorption lines visible in the spectra, 
indicating the presence of large HI reservoirs.  
These systems were normally identified as `normal' (i.e., not too much intense or, in some cases, similar to the Milky Way)
star-forming galaxies that were intercepted by the radiation emitted by \linebreak the QSOs. 

Through a detailed characterisation of the abundance pattern of the gas in distant galaxies, DLAs enabled
to confirm the refractory nature of several elements (such as Fe and Si) for which precise estimates of their column density were achievable.
The conclusion that some elements were depleted by dust grains was possible thanks to the evaluation of the abundance ratios between 
refractory elements and non-refractory ones, such as Zn, another fundamental element for which precise column density measures were accessible. 

The nature of the DLAs was unveiled thanks to the interpretation of the observed abundance ratios by means of chemical evolution models.

The comparison of depletion-corrected abundances with the theoretical ones confirmed that the DLA absorption originated
mostly in spiral and irregular systems, and that some categories of galaxies were excluded or very unlikely, mostly because too much dusty,
or because they could retain the neutral gas reservoirs for a too small amount of time, such as very intense starbursts.

While it was becoming clear that metal-rich systems with intense star formation were largely absent from QSO DLAs, 
a new category of DLAs was emerging from the study of the absorption spectra of GRB afterglows. 
Also in this case, the background sources were intense enough to generate a large multitude of absorption in the gas present along
their line of sight. 
While QSO DLAs are useful to probe the gas in galactic outskirts, 
GRB DLAs occur in dense star-forming regions associated with the innermost, most metal-rich regions of galaxies. 
For these reasons, they represent complementary probes with respect to QSO DLAs. 
Depletion corrections based on the measured abundance ratios were applied also to analyse the nature of GRB DLAs. 
In the meantime, galactic chemical evolution models incorporating dust depletion were used to interpret
their abundance ratios. 
This effort allowed us to realise that the SFHs of GRB DLAs are different than those of QSO DLAs, in that the samples of GRB host galaxies may well
include also intense starbursts. Therefore, in principle, GRB-DLAs allow one to probe also the interstellar gas of proto-spheroids.
It is however important to note that GRB-DLAs samples are not entirely bias-free, in particular as far as the effects of
dust and metallicity are concerned.

In Section \label{sec_ontheredshift} I reviewed  the evolution of the dust budget in galaxies. 
I discussed the measure of the dust mass in galaxies, and how it rests upon the assumption that the thermal emission of
grains heated by stellar radiation can be approximated by means of a black-body model.
The optically thin approximation is a further simplification to estimate the dust mass, along with the Rayleigh-Jeans one, 
often adopted in particular in studies of high-redshift galaxies.
The discovery of copious dust reservoirs dust in the most distant galaxies triggered several theoretical attempts to explain how dust can form so rapidly in the early Universe, highlighting the role of various  production mechanisms. 
Although core-collapse supernovae are often invoked as primary dust producers, a growing body of observational
and theoretical evidence demonstrates  that the role of interstellar dust growth cannot be neglected.

The measure of the dust mass has been used to estimate the global evolution of the dust mass budget,
expressed by the comoving dust mass density.
This quantity has been estimated by various authors. A few studies rely on a direct estimate based on the
integral of the comoving dust mass function, measured directly in the IR and submillimetre bands. 
Other authors use indirect probes, such as Mg II absorbers or surveys at other wavelength, estimating 
the dust luminosity via SED fitting techniques, or without considering the contribution
from resolved galaxies, i.e., from the submm or IR background. 
Most studies indicate a comoving dust density increasing progressively from $z=0$ to z $\sim$ 1, then presenting a broad
peak followed by a slow decrease towards higher redshifts. 
At the present day, the direct evaluation of both the dust mass function and density  is problematic
as current ground IR/submm facilities do not allow one to perform easily wide, \linebreak unbiased surveys.

Current cosmological models fail to account for this behaviour, as some predict an excessively flat trend
as a function of redshift, a too broad peak and a too much steep decrease at high redshift. 

Moreover, the predictions from different models are much more dispersed than the
observational estimates. The current poor theoretical knowledge outlines the need for a high-sensitivity 
IR instrument to improve our understanding of the evolution of the global dust budget. 
In the post-Herschel era, the community is still awaiting an 
instrument to probe adequately and with an improved sensitivity the IR luminosity function,
the dust composition in local galaxies from a sufficient set of spectral features and the amount of dust
in the local most metal-poor galaxies —among other critical diagnostics that remain to be explored.  
This is a pivotal moment for the infrared astronomical community, as crucial decisions will soon be made on the approval
of a future mission that promises to address a series of outstanding key questions.
The decisions made now will shape our understanding and exploration of the infrared universe for years to come,
offering a unique opportunity to push the boundaries of our knowledge.

\vspace{6pt} 




\funding{I acknowledge support from the INAF Mini Grant 2024 program ``DustPedia meets Metal-THINGS:
Dust-METAL'' and from the INAF Theory Grant 2024 program ``Magnetohydrodynamic Simulations of Galactic
Molecular Clouds: Resolving Stellar Birth and Proto-planetary Discs with an Enhanced Chemical Network''.}
\dataavailability{This review did not involve the generation or analysis of new data.
  All information discussed is derived from previously published literature, which is cited throughout the manuscript.}

\acknowledgments{I acknowledge many collaborators and friends in various astronomical institutes in Bologna and Trieste,
  with whom I had many interesting and inspiring discussions on dust evolution, including 
  F. Pozzi, R. Gilli, F. Matteucci, A. Pipino, G. Vladilo, G. Granato, L. Gioannini, V. Casasola, M. Palla, C. Vignali, M. Parente 
  and others met through the years at conferences and meetings. \linebreak Two anonymous referees are also acknowledged for valuable comments.}     

\conflictsofinterest{The author declares no conflict of interest. 
} 

\abbreviations{Abbreviations}{~The following abbreviations are used in this manuscript:\vspace{6pt}\\
\noindent 
\begin{tabular}{@{}ll}

  AGB &  ~~Asymptotic Giant Branch\\
  AGN &  ~~Active Galactic Nuclei\\
  BEL &  ~~Broad Emission Lines\\
  BH  &  ~~Black Hole\\
  CC  &  ~~Core Collapse\\
  CNM &  ~~Cold Neutral Medium \\
  DLA &  ~~Damped Lyman $\alpha$\\
  DM  &  ~~Dark Matter\\
  DMD &  ~~Dust mass density\\
  DMF &  ~~Dust mass function\\
  DTM & ~~Dust-to-metal ratio \\
  DTG &  ~~Dust-to-gas ratio\\
  DTS &  ~~Dust-to-stellar mass ratio\\
  GRB &  ~~Gamma-Ray Bursts \\
  IGIMF & ~~integrated galactic initial mass function \\
  IGM  & ~~intergalactic medium \\
  IMF &  ~~Initial Mass Function \\
  ISM & ~~Interstellar Medium \\
  LBG &  ~~Lyman-Break Galaxy\\
  MC  & ~~Molecular Cloud\\
  PAH & ~~polycyclic aromatic hydrocarbon\\
  QSO &  ~~Quasi-Stellar Objects \\
    SED &  ~~Spectral Energy Distribution \\
  \end{tabular}}

  \noindent 
\begin{tabular}{@{}ll}

  SN  &  Supernova  \\
  SNe &  Supernovae \\
  S/N &  Signal-to-Noise  \\
  SF  &  Star Formation \\
  SFE &  Star Formation Efficiency \\
  SFH &  Star Formation History \\
  SFR &  Star Formation Rate\\
  SNR &  Supernova Remnant \\
  THIMF & Top-heavy IMF \\
  WR  &  Wolf-Rayet  \\
\end{tabular}}

%
%
%
%
%
%

\externalbibliography{yes}


\reftitle{References}

\PublishersNote{}

\end{document}